\documentclass{ieeeaccess}
\usepackage{cite}
\usepackage{algorithm}      % algorithm
\usepackage[noend]{algpseudocode} % algorithm
\usepackage{graphicx}
\usepackage{textcomp}
\usepackage{amsmath,amssymb,amsfonts}
\usepackage{graphicx}
\usepackage{subcaption}
\usepackage{color}

\newcommand{\algrule}[1][.2pt]{\par\vskip.1\baselineskip\hrule height #1\par\vskip.1\baselineskip}

\def\BibTeX{{\rm B\kern-.05em{\sc i\kern-.025em b}\kern-.08em
    T\kern-.1667em\lower.7ex\hbox{E}\kern-.125emX}}
\begin{document}
\history{Date of publication xxxx 00, 0000, date of current version xxxx 00, 0000.}
\doi{XXXX}

\title{Estimating the Magnitude and Phase of Automotive Radar Signals under Multiple Interference Sources with Fully Convolutional Networks}

\author{\uppercase{Nicolae-C\u{a}t\u{a}lin Ristea}\authorrefmark{1},
\uppercase{Andrei Anghel}\authorrefmark{1}\IEEEmembership{Senior Member, IEEE}, and \uppercase{Radu Tudor Ionescu}\authorrefmark{2}, \IEEEmembership{Member, IEEE}}

\address[1]{Department of Telecommunications, University Politehnica of Bucharest, Romania}
\address[2]{Department of Computer Science and the Romanian Young Academy, University of Bucharest, Romania}

% \tfootnote{}

\markboth
{Ristea \headeretal: Estimating the Magnitude and Phase of Automotive Radar Signals with Fully Convolutional Networks}
{Ristea \headeretal: Estimating the Magnitude and Phase of Automotive Radar Signals with Fully Convolutional Networks}

\corresp{Corresponding author: Andrei Anghel (e-mail: andrei.anghel@munde.pub.ro)}

\begin{abstract}
Radar sensors are gradually becoming a wide-spread equipment for road vehicles, playing a crucial role in autonomous driving and road safety. The broad adoption of radar sensors increases the chance of interference among sensors from different vehicles, generating corrupted range profiles and range-Doppler maps. In order to extract distance and velocity of multiple targets from range-Doppler maps, the interference affecting each range profile needs to be mitigated. In this paper, we propose a fully convolutional neural network for automotive radar interference mitigation. In order to train our network in a real-world scenario, we introduce a new data set of realistic automotive radar signals with multiple targets and multiple interferers. 
To our knowledge, we are the first to apply weight pruning in the automotive radar domain, obtaining superior results compared to the widely-used dropout. While most previous works successfully estimated the magnitude of automotive radar signals, we propose a deep learning model that can accurately estimate the phase. For instance, our novel approach reduces the phase estimation error with respect to the commonly-adopted zeroing technique by half, from 12.55 degrees to 6.58 degrees. Considering the lack of databases for automotive radar interference mitigation, we release as open source our large-scale data set that closely replicates the real-world automotive scenario for multiple interference cases, allowing others to objectively compare their future work in this domain. Our data set is available for download at: http://github.com/ristea/arim-v2.
\end{abstract}


\titlepgskip=-15pt

\maketitle

\section{Introduction}
\label{sec:introduction}
% \PARstart{T}{his} 

\begin{figure}[!t]
\centering
%\hspace{-0.6cm}
\includegraphics[width=1.07\linewidth]{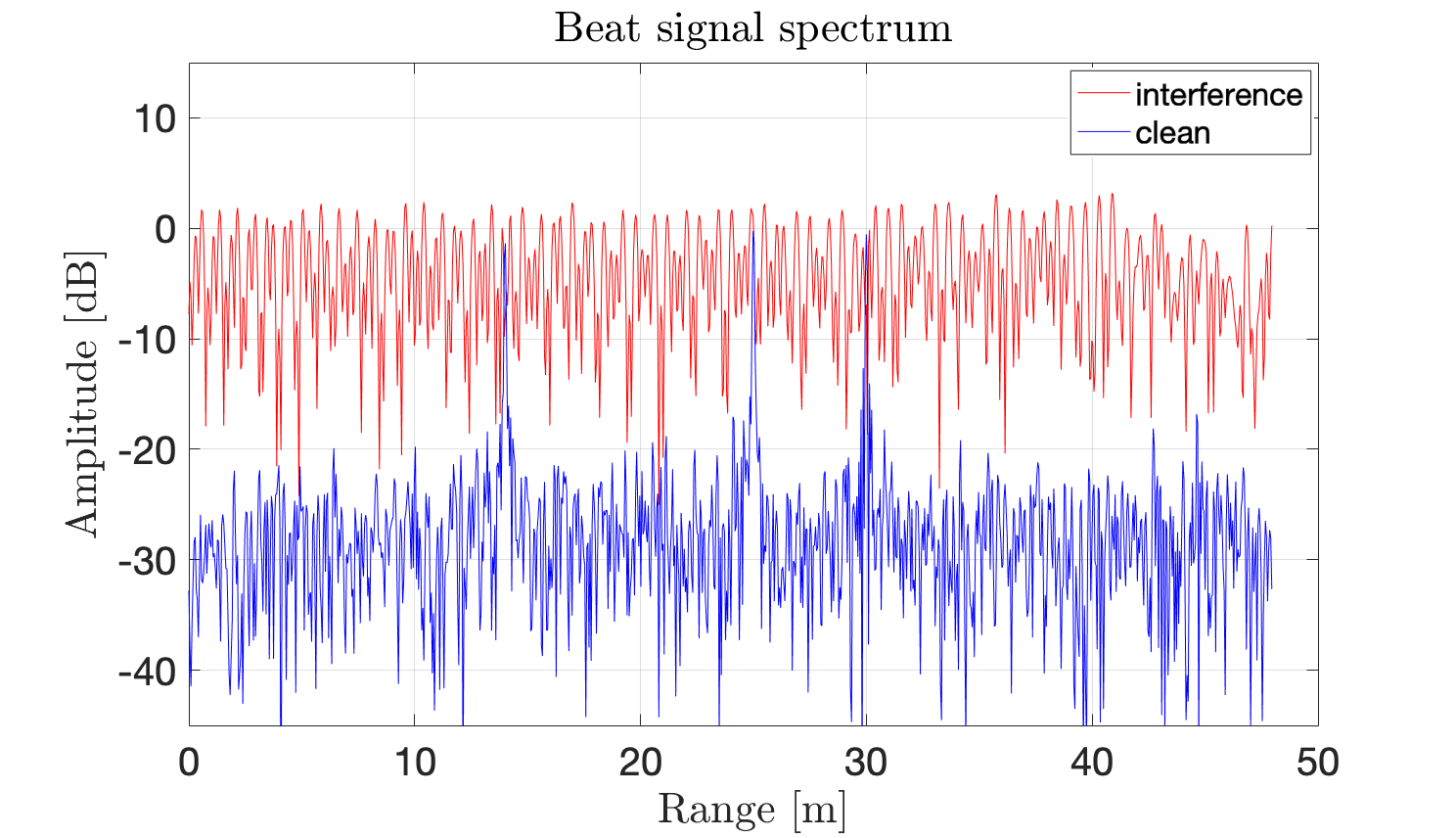}
%\hspace{-0.6cm}
\includegraphics[width=1.07\linewidth]{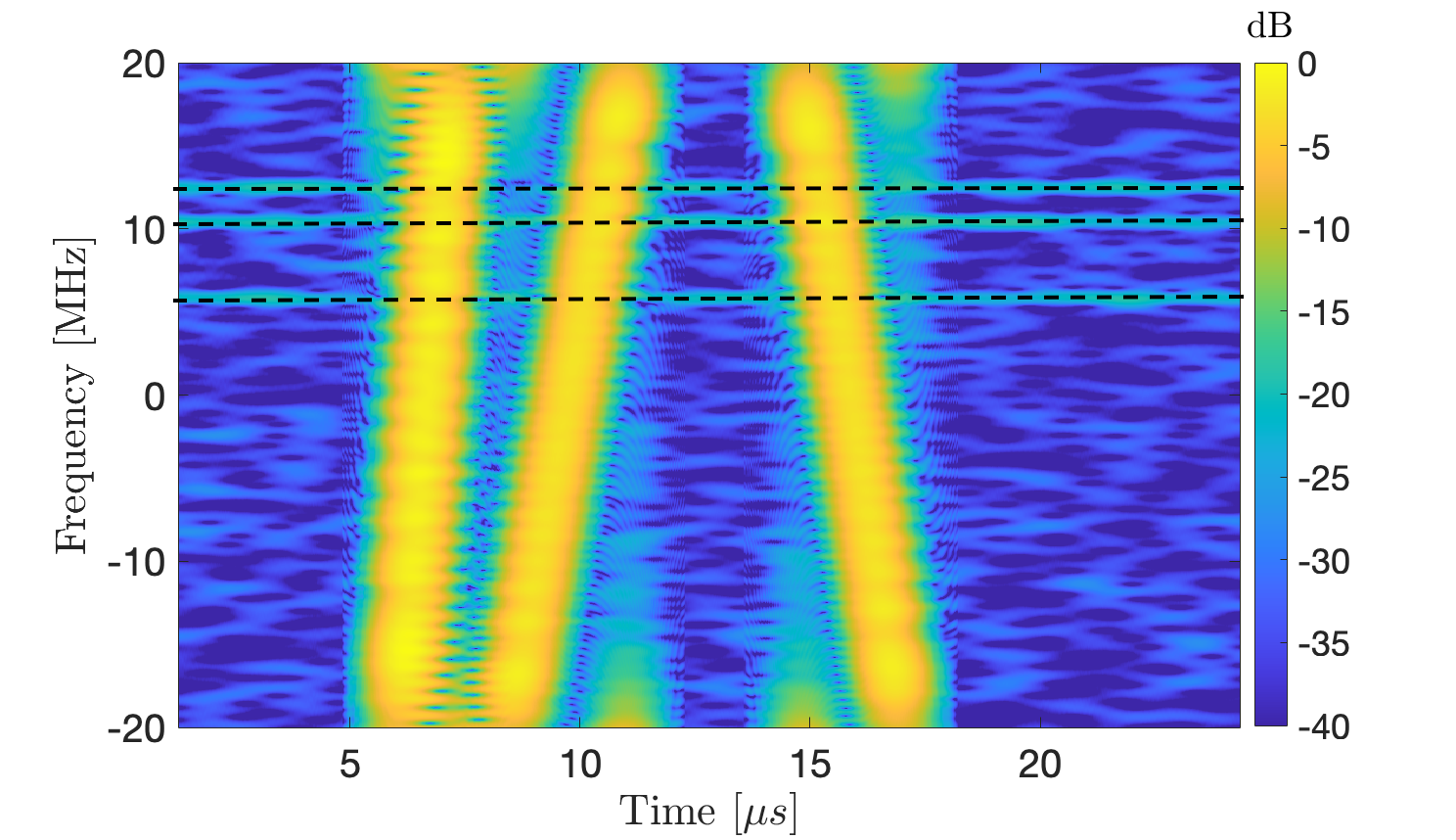}
\vspace{-0.3cm}
\caption{Top: Range profile magnitude of an FMCW radar sensor. The clean profile is shown in blue, while the profile with interference is shown in red. Bottom: The magnitude of the STFT of the corresponding range profile. The targets are the thin horizontal lines and the interference sources are the thick (more pronounced) diagonal lines. Best viewed in color.}
\vspace{-0.2cm}
\label{fig_1}
\end{figure}

\PARstart{A}{utonomous} driving and road safety are very important topics in order to reduce the number of traffic accidents and the number of deaths on the road. One of the solutions proposed by automotive companies to build autonomous and safer vehicles is based on scanning the surrounding environment using radar sensors. The most common radar senors used in the automotive industry are frequency modulated continuous wave (FMCW) / chirp sequence (CS) radars, which transmit sequences of linear chirp signals. The signals transmitted and received by such sensors provide the means to estimate the distance and the velocity of nearby targets (e.g., vehicles, pedestrians or other obstacles). For instance, automotive radar sensors have even been used to detect very small objects (e.g., road debris~\cite{Shibao-VTC-2019}). However, the growing adoption of radar sensors \cite{kunert2012eu} increases the probability of interference among sensors from different vehicles, generating corrupted and unusable signals. Indeed, radio frequency interference can raise the noise floor by a large margin, to the point where potential targets are completely hidden by noise, thus reducing the sensitivity of target detection methods \cite{brooker2007mutual}. In Figure \ref{fig_1}, we present a range profile of a radar signal with and without interference, in which some of the targets visible in the clean range profile are absorbed by the risen noise floor caused by multiple interference sources. In order to be able to detect such targets, the radar interference has to be mitigated. To address this problem, researchers have proposed various techniques ranging from conventional approaches \cite{mosarim2010d15, uysal2019sync, kim2018peer, xu2018orthogonal_noise, bechter2017dbfmimo, laghezza2019nxp, neemat2018interference, mun2018deep} to deep learning methods \cite{Ristea-VTC-2020,rock2019complex,fan2019interference, mun2020automotive, rock2021resource}. 

%Range profile magnitude of an FMCW radar sensor. The clean profile is shown in blue, while the profile with interference is shown in red. Because of interference, the noise floor has risen with over $30\;dB$ and the targets become undetectable. Best viewed in color.

In this paper, we extend our prior work~\cite{Ristea-VTC-2020} by designing a novel fully convolutional network (FCN) \cite{Long-CVPR-2015} that $(i)$ can recover the phase along with the magnitude of radar beat signals and $(ii)$ can cope with multiple non-coherent radio-frequency (RF) interference sources. Our network takes as input the real part, imaginary part and magnitude of the Short-Time Fourier Transform (STFT) of the beat signal with interference, providing as output the real part, imaginary part and magnitude of the range profile, respectively. Although our network does not directly estimate the phase, it can be trivially computed from the real and imaginary parts. To our knowledge, we are among the few to propose a deep learning model that can accurately estimate the phase, this being a well-known problem, which is often left as future work in related articles~\cite{Fuchs2020}. While most deep learning approaches studied radar interference mitigation with a single interference source \cite{Ristea-VTC-2020,fan2019interference,mun2018deep}, we aim to address the task under multiple interference sources. To achieve this goal, we generate a large-scale data set that closely replicates the real-world automotive scenario for multiple interference sources, considering up to three interference sources during training and up to six interference sources during inference. We compare our approach with three state-of-the-art methods, one based on zeroing \cite{mosarim2010d15,laghezza2019nxp} and two based on deep neural networks \cite{Ristea-VTC-2020,rock2019complex}, reporting superior results for various evaluation metrics. In this paper, we also apply weight pruning \cite{han2015learning,liu2018rethinking}, improving the signal-to-noise ratio contained in the weights of our neural network models. We compare our weight pruning to the widely-adopted dropout~\cite{Srivastava-JMLR-2014}, showing that the former approach helps the neural model to reach a better convergence point. Furthermore, we release our novel data set as open source, allowing other researchers and engineers to objectively compare their future work on radar interference mitigation. Our data set is available for download at: http://github.com/ristea/arim-v2. Along with the data set, we also release the code to reproduce our results.

In summary, our contribution is threefold:
\begin{itemize}
    \item We propose a deep learning model able to mitigate non-coherent RF interference from multiple sources.
    \item We design a fully convolutional network architecture that outputs clean range profiles, estimating both the magnitude and, indirectly, the phase.
    % \item We introduce a new training regime, termed Weight Noise Reduction Training (WeNoRT), that reduces the amount of noise stored in the neural network weights.
    \item We introduce a radar interference data set with a wide and realistic range of signal parameter variations as well as multiple interference sources.
\end{itemize}

We organize the rest of this paper as follows. We present related work on radar interference mitigation in Section~\ref{sec_Related_Work}. We describe our method based on fully convolutional networks in Section~\ref{sec_Method}. We present our data set composed of generated range profiles in Section~\ref{sec_Data_Set} and we provide a comprehensive set of experimental results in Section~\ref{sec_Experiments}. Finally, we draw our conclusions in Section~\ref{sec_Conclusion}.

% \vspace{-0.1cm}
\section{Related Work}
\label{sec_Related_Work}

% \vspace{-0.1cm}
\subsection{Conventional methods}

State-of-the-art interference mitigation methods are usually classified according to the domain in which the interference is mitigated \cite{mosarim2010d15, uysal2019sync, kim2018peer, xu2018orthogonal_noise, bechter2017dbfmimo, laghezza2019nxp}: polarization, time, frequency, code and space. Polarization-based methods assume the use of cross-polarized antennas between the two interfering radars and the mitigation margin is around 20 dB, but ground reflections or other surrounding targets can severely reduce this margin. Time domain methods include the following approaches: using low transmit duty cycles (to reduce the probability of hitting other receivers), using short receive windows (to reduce the probability of being hit by an interferer), or employing a variable pause between transmitted chirps or a variable chirp slope (to avoid periodic interference). Frequency domain methods involve a division of the authorized operating bandwidth into several sub-bands, such that nearby systems operate in different sub-bands. Radio frequency interference (RFI) mitigation in the coding domain implies the modulation of radar wave forms with a device-specific code (to minimize cross-talk between radars, the codes of different devices should be orthogonal), whereas in the case of space domain techniques, the antenna radiation pattern is adaptively configured to avoid interfering signals. 

A particular class of methods are the strategic RFI mitigation techniques \cite{mosarim2010d15}, which require additional hardware and/or software, yet rely on some of the basic techniques. The classical strategic approaches are: ``communicate and avoid'' (requires inter-vehicle communication to avoid simultaneous transmission), ``detect and avoid'' (e.g., detects the interference in a sub-band and changes the operating sub-band of the radar), ``detect and repair'' (after detection, the measurement with interference is reconstructed), ``detect and omit'' (after detection, the measurements affected by interference are removed) and ``listen before talk'' (the radar transmits only when no other transmitting device is detected). 

Currently, mitigation of a single FMCW interferer on an FMCW victim radar is quite well understood \cite{alland2019spm}, ongoing research in the field being focused on other scenarios that involve multiple interference sources due to the increasing number of vehicles equipped with radars and the increase in the number of radar systems per vehicle.

Different from all these methods, which rely on algorithms handcrafted by researchers, we propose an approach based on end-to-end learning from data. In order to obtain our approach, we extended the data set and the method proposed in our previous work~\cite{Ristea-VTC-2020} in order to learn deep neural networks for RFI under multiple interference sources.

%\vspace{-0.3cm}
\subsection{Deep learning methods}

Deep learning techniques have been applied in a vast diversity of tasks with remarkable results, including object detection \cite{Ren-NIPS-2015,Girshick-ICCV-2015,soviany2018optimizing}, speech separation \cite{wang2018supervised} and medical image super-resolution \cite{Chen-MICCAI-2018,georgescu2020convolutional,Yu-ICIP-2017}. One such task is image denoising, where deep learning achieved state-of-the-art results \cite{bricman2018coconet}, outperforming classical filtering approaches (i.e., median or bilateral filtering). 
By transforming an arbitrary signal with STFT, we obtain an image-like representation that can eliminate the gap between the task of signal denoising and that of image denoising. Indeed, the interference becomes a noise pattern that is overlapped over the signal's STFT, opening up the possibility to employ novel ways for interference mitigation, previously applied only on images. In this context, we propose to apply fully convolutional networks, a deep learning technique, to transform a noisy STFT image into a clean range profile of an FMCW radar sensor.

To our knowledge, there are only a handful of related works \cite{Ristea-VTC-2020,rock2019complex,fan2019interference,mun2018deep, Fuchs2020, mun2020automotive} that employ deep learning models for radar interference mitigation. Most of the existing deep RFI mitigation approaches consider only scenarios with one source of interference. Complex scenarios with multiple sources of interference, which are very likely to happen in daily driving, were only considered by Rock et al.~\cite{rock2019complex}. In \cite{rock2019complex}, the authors proposed a convolutional neural network (CNN) to address RFI, aiming to reduce the noise floor while preserving the signal components of detected targets. Their CNN architecture can be trained using either range processed data or range-Doppler (RD) spectra as inputs. The authors reported promising results, but they still had concerns regarding the generalization capacity on real data. In the experiments, we show that our approach outperforms the model of Rock et al.~\cite{rock2019complex} by a significant margin. Additionally, we demonstrate that our method can generalize to real data.

Another approach that relies on CNNs is proposed in \cite{Fuchs2020}. The authors employed an auto-encoder based on the U-Net architecture \cite{ronneberger2015u}, performing interference mitigation as a denoising task directly on the range-Doppler map. They surpassed classical approaches, but their method fails to fully preserve the phase information. Similarly, in \cite{fan2019interference}, the network architecture is build upon CNNs, but the authors added residual connections, inspired from the ResNet model \cite{he2016deep}.
A different interference mitigation method is proposed in \cite{mun2018deep}, which is based on applying a recurrent neural network model with Gated Recurrent Units (GRU) \cite{chung2014empirical} on the time domain signal. The authors reported better performance and lower processing times compared to previous signal processing methods. Similarly, Mun et al. \cite{mun2020automotive} proposed an approach that is also based on GRU, but they add a novel attention block. This approach attains better results than classical methods and the authors empirically prove that the attention block brings a performance boost. Nevertheless, the algorithm is not tested on real data or on a large test collection, so there are concerns regarding its generalization capacity.

\noindent
{\bf Relation to preliminary VTC-Fall 2020 version \cite{Ristea-VTC-2020}.}
We recently  proposed two novel FCN models in our preliminary work~\cite{Ristea-VTC-2020}, which are able to transform an STFT sample affected by interference into the corresponding clean range profile. The models have the capacity to generalize on real data, but they are not designed to estimate the \emph{phase} of beat signals. In the current work, we extended our preliminary work presented in~\cite{Ristea-VTC-2020} by designing a method able $(i)$ to recover the phase of beat signals and $(ii)$ to cope with multiple interference sources. In addition, we employ a new training regime based on weight pruning \cite{han2015learning,liu2018rethinking}, which is aimed at improving the signal-to-noise ratio of the neural network weights. Moreover, the performance of our novel neural model is highlighted by the experimental results. Indeed, we achieved the best performance on the ARIM-v2 data set and we empirically proved the model's generalization capability. Additionally, we extended the data set proposed in our preliminary work~\mbox{\cite{Ristea-VTC-2020}} to multiple sources of interference, releasing the first freely available interference mitigation data set for multiple sources of interference.

%Once with the development of FCNs, more and more machine learning based approaches have been proposed in signal processing related tasks, ranging from denoising to synthesis. Generally, signals and specific signal transforms (such as Fourier Transform) are defined in the complex set of numbers, therefore models architectures were adapted in order to handle properly the definition data domain.
%Several theories have been developed, but the most common one is to neglect the phase of the complex numbers and consider just the magnitude. This way, a spectrogram could be computed from the signal and feed into a conventional FCN with a single input channel. Other theory have been developed by Trabelsi et al. \cite{trabelsi2017deep}, which designed neural network blocks capable of handling complex value numbers, by exploiting the formula of convolution for complex value numbers. They showed that those blocks attains better performances in signal related tasks.

%\vspace{-0.1cm}
\section{Method}
\label{sec_Method}

% \vspace{-0.1cm}
\subsection{Radar Signal Model}

\begin{figure}[t]
\centering
\includegraphics[width=1.0\linewidth]{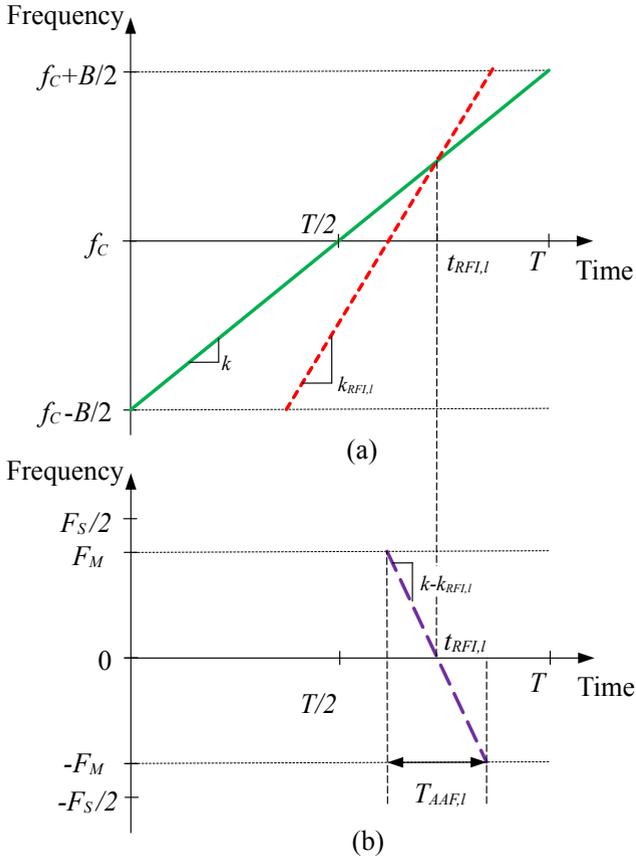}
\caption{Qualitative time-frequency diagrams of (a) the transmitted signal (green) and an un-correlated interference (red) and (b) the resulting baseband interference signal after mixing with the transmitted waveform. $B$ is the band of the transmitted signal, T is the time of chirp, $f_c$ is the radar central frequency, $F_S$ is the sampling frequency and $F_M$ is maximum frequency of the baseband interference signal.}
\label{fig_tf_interf_diagram} 
\end{figure}

In FMCW radar solutions, the transmitted signal $s_{TX}(t)$ is a chirp sequence, whose frequency usually follows a sawtooth pattern. In the presence of mutual interferences, the receiving antenna collects a mix from two signals, the reflected signal and the interference signal. Consequently, the received signal is defined as follows:
\begin{equation}\label{int_sign}
s_{RX}(t) = \displaystyle\sum_{i=0}^{N_{t}-1} \underline{A}_{i} \cdot s_{TX}(t-\tau_{i}) + \displaystyle\sum_{l=0}^{N_{int}-1} s_{RFI,l}(t),
\end{equation}
where $\underline{A}_{i} = A_{i} \cdot e^{j\phi_i}$ is the complex amplitude, $\tau_{i}$ is the propagation delay of target $i$, $N_{t}$ is the number of targets, and $N_{int}$ is the number of interferers. The receive antenna collects the reflected signal $s_{RX}(t)$, which is further  mixed  with  the  transmitted signal and low-pass filtered, resulting in the beat signal $s_{b}(t)$.
After mixing the signal reflected by a point-like target with the transmitted signal, we obtain a signal with constant frequency, whereas by mixing an uncorrelated interference with the transmitted chirp, we obtain a baseband chirp signal (as depicted in a qualitative manner in Figure \ref{fig_tf_interf_diagram}). 

The slope of the interference chirp (after mixing) is equal to the difference between the slope of the transmitted signal $k$ and the slope of the interference $k_{RFI,l}$, while its zero-frequency point $t_{RFI,l}$ corresponds to the intersection between the instantaneous frequency laws (IFL) of the transmitted and interference chirps. Based on the time-frequency diagram from Figure~\ref{fig_tf_interf_diagram}b, the IFL of the interference chirp in baseband can be written as:
\begin{equation}
\begin{split}
& f_{RFI,l}(t) = (k - k_{RFI,l})  \cdot (t - t_{RFI,l}) \cdot \\
&  \cdot p \left( \frac{t-t_{RFI,l}}{T_{AAF,l}} \right)  \cdot p \left( \frac{t-\frac{T}{2}}{T} \right)\!,
\end{split}
\label{sig_tf_tx_chirp}
\end{equation}
where $T_{AAF,l} = \frac{2 F_M}{|k - k_{RFI,l}|}$ is the duration of the interference, which is limited by the anti-aliasing filter used before sampling. If the slope of the transmitted signal is close to the interference slope, $T_{AAF,l}$ can get much longer than the chirp duration $T$ and the actual time extent of the baseband interference limited by $T$. A similar effect occurs if $t_{RFI,l}$ is near the ends of the repetition interval. Using the introduced notations, the resulting analytical beat signal in the presence of interferences is expressed as:
\begin{equation}
\begin{split}
& s_b(t)= \Bigg\{\sum_{i=0}^{N_{t}-1}\underline{A}_{i} \cdot \textrm{exp}(j 2 \pi k \tau_i t) + \\ 
& +\!\! \sum_{l=0}^{N_{int}-1} \underline{A}_{RFI,l}  \cdot\textrm{exp}\left[j \pi (k - k_{RFI,l} )(t - t_{RFI,l})^2\right]  \cdot \\ 
& \cdot p \left( \frac{t-t_{RFI,l}}{T_{AAF,l}} \right) \Bigg\}  \cdot p \left( \frac{t-\frac{T}{2}}{T} \right)\!,
\end{split}
\label{sig_rec_eq}
\end{equation}
where $\underline{A}_{RFI,l}$ is the complex amplitude of interference signal $l$ and $p(t)$ is the window function described below:
\begin{equation}\label{eq_adaf1}
\begin{split}
p\Bigg(\frac{t}{a} \Bigg) = \left\{\begin{array}{ll} 1, & \mbox{if} \; -\frac{a}{2} \leq t \leq \frac{a}{2} \\ 0, & \mbox{otherwise} \; \end{array}\right.\!\!, \forall a \in \mathbb{R}.
\end{split}
\end{equation}

Hence, $s_{b}(t)$ consists of a sum of complex exponentials (representing the targets) and a sum of interfering signals (baseband chirps). Therefore, the uncorrelated interference appears as a highly non-stationary component on the beat signal's spectrogram, being spread across multiple frequency bins, as opposed to the signal received from targets, which is present only at some frequency values \cite{alland2019spm}. This explains the general aspect of the magnitude of STFT presented in Figure~\ref{fig_1}.

%In this paper we extend the prior work of Ristea et al.~\cite{Ristea-VTC-2020} by considering more realistic scenarios with multiple sources of interference.

\begin{figure*}[!th]
\centering
  \includegraphics[width=1.0\linewidth]{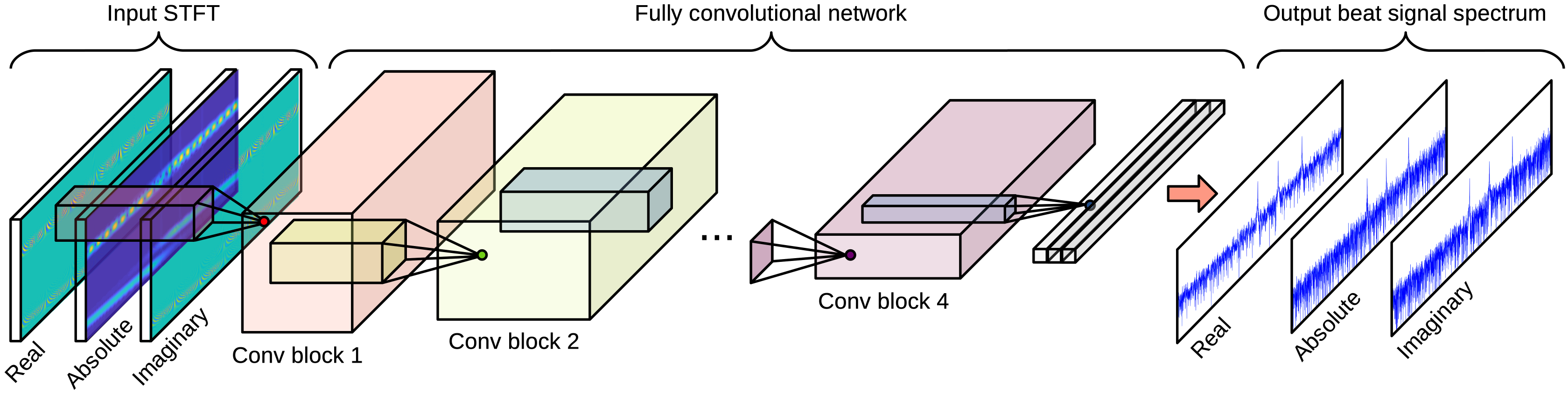}
  \vspace{-0.3cm}
  \caption{The architecture of our FCN model. The input STFT is processed through a series of four conv blocks (composed of conv and pooling layers) until the vertical dimension is reduced to 1, while preserving the horizontal dimension. The output is a beat signal spectrum without the interference removed by the FCN. Best viewed in color.}
  \vspace{-0.2cm}
  \label{fig_2}
\end{figure*}

%\vspace{-0.2cm}
\subsection{Data Preprocessing}

As shown in Figure~{\ref{fig_1}}, we need to compute the discrete STFT in order to disentangle the targets from the inference sources. The following equation shows how to transform a time domain signal into an image using the discrete STFT:
\begin{equation}
\label{stft}
    STFT\{x[n]\}(m, k) = \sum_{n=-\infty}^{\infty} x[n] \cdot w[n-m R] e^{-j \frac{2 \pi}{N_x}k n},
\end{equation}
where $x[n]$ is the discrete input signal (the sampled version of $s_b(t)$), $w[n]$ is a window function, $N_x$ is the STFT length and $R$ is the hop/step size \cite{allen1977stft}. There are a multitude of window functions proposed in literature, such as \textit{hann}, \textit{blackman} and others. We chose to perform the STFT with \textit{hamming} window. Additionally, we scale the STFT (by dividing it with $\alpha=40$, which was obtained statistically on the entire training set) in order to have the input data approximately within the range of $[ -1, 1]$. 

Since our data samples are now represented as images, we consider convolutional neural networks (CNNs) to model the mapping between input images and clean range profiles, noting that CNNs attain state-of-the-art results on natural images~\cite{LeCun-IEEE-1998,Krizhevsky-NIPS-2012,georgescu2019recognizing}, medical images~\cite{georgescu2020convolutional} and artificial images resulted after transforming time domain signals~\cite{Ristea-INTERSPEECH-2020}.

Our goal is to obtain clean range profiles from the STFT of the beat signal affected by noise and uncorrelated interference. We design a custom FCN architecture to provide the clean range profiles as output (during training, the FCN has to learn to reproduce the ground-truth clean range profiles). For this reason, we perform a Fast Fourier Transform (FFT) of our time domain labels (to obtain the ground-truth clean range profiles) and train our network to map the STFT input to the FFT output (computed in $N_x$ points, as the number of STFT frequency bins).
The intuition behind choosing the input (STFT) and the output (FFT) domains of our network as presented above is that the spectrum of a beat signal affected by interference is covered in noise and the targets are almost undetectable, as could be seen in Figure~\mbox{\ref{fig_1}}. The advantage of using STFT is that there are portions in the representation where the targets are visible (the thin horizontal lines in the STFT in Figure~\mbox{\ref{fig_1}}), even if the signal is affected by multiple sources of interference. We empirically tested our intuition by training the same network architecture (redesigned for the one-dimensional FFT input) and discovered that this approach has convergence issues and unusable results.

%Formally, the FFT is computed as follows:
%\begin{equation}
%\label{fft}
%    \textrm{FFT}\{x[n]\}(k) = \sum_{n=0}^{N_x-1} x[n] \cdot e^{-j \frac{2 \pi}{N_x}k n},
%\end{equation}
%\noindent
%where $N_x$ is the number of FFT points (same value as the number of STFT range bins).

%\vspace{-0.2cm}
\subsection{Neural Network Model}

Our goal is to create a neural network that can mitigate RFI and is able to map a noisy STFT input to the clean FFT output for any given signal, in terms of both magnitude and phase. Therefore, we propose a novel FCN architecture that can meet the above requirement. There are related works which take an STFT as input and give an FFT as output, such as \cite{neemat2018interference}, but these are not based on deep learning techniques. To the best of our knowledge, there are no approaches based on deep learning models that transform an STFT input sample affected by interference into a clean FFT range profile.

The novelty of our neural architecture is mainly related to the input and output structures, each consisting of a representation composed of three different channels. The first and third channels of the input are the real and imaginary parts of the STFT, while the channel in the middle is the magnitude of the STFT. In terms of information theory, the second channel is redundant information, which could be mathematically determined from the real and imaginary parts. The motivation behind adding the magnitude of the STFT as input is given by our preliminary FCN models \cite{Ristea-VTC-2020}, which successfully used it to predict the magnitude of the FFT. Furthermore, the magnitude of an STFT has the most meaningful visual information and can be seen as an attention map \cite{woo2018cbam, wang2017deep}, which, in our case, is not computed by the network, but offered as an input channel. The output follows a similar design in terms of the number of channels, the only difference being its spatial dimension, as described next. Although our network does not directly compute the phase, it can be computed from the real and imaginary parts. We hereby underline that we have tried various architectures to explicitly output the phase of the FFT, such as having as input channels only the magnitude and phase, but we never obtained convergence. However, it appears that modeling the phase indirectly is achievable. Regarding the magnitude, we can choose between taking the middle output channel directly predicted by the network or computing the magnitude from the real and imaginary parts, as a post-processing step. The results are very close, but slightly better for the former approach. In summary, we take the magnitude predicted as output and compute the phase from the real and imaginary parts. We also noticed that without the magnitude of the STFT as input channel, our model achieves significantly lower performance (see Table~{\ref{tab_results_arim}}). Hence, the seemingly redundant magnitude channel is actually of utter importance.

Our neural model is designed to process an input tensor of size $154 \times 2048\times3$ and give an output tensor of size $1 \times 2048 \times 3$. The network progressively reduces the dimension on the vertical axis ($154$), which corresponds to the number of time bins in which the STFT is computed, to the size $1$, while keeping the dimensions on the other axes constant (the number of FFT points, $N_{x}$, and the number of channels, respectively).

Our architecture, illustrated in Figure~\ref{fig_2}, consists of $10$ convolutional (conv) layers organized into $4$ convolution blocks. Each of the first $2$ blocks are composed of 3 conv layers, followed by a max-pooling layer. The third block is formed of $2$ conv layers and a max-pooling layer, while the last block has the same number of convolutions as the third, but without any pooling layer. 
%(### Care este paper-ul pentru leaky relu?###)
Additionally, each conv layer is followed by leaky Rectified Linear Units (ReLU) \cite{Maas-WDLASL-2013}, except for the last $2$ layers. The number of convolutional filters (kernels) is independently established for each block. The number of kernels starts from $32$ in the first block, growing by $32$ with each subsequent block, ending up at $128$ in the last one. Exceptionally, the very last conv layer has only $3$ kernels in order to fit to the desired number of output channels. We also reduce the kernel size from $13\times 13$ in the first block to $9 \times 9$ in the second block and $5\times5$ in the third block. Regarding the last conv block, we set the kernel size to $5\times5$ in the first conv layer and to $1\times1$ in the last conv layer, respectively. The conv filters are always applied at stride $1$, a circular padding being added to preserve the horizontal dimension of the activation maps. The pooling filters are always of size $2\times1$, reducing the size of the activation maps by half on the vertical axis only. Zero padding for the max-pooling layers is added only when we need to make sure that the input activation maps have an even size. 

%\vspace{-0.2cm}
\subsection{Loss Function}

The procedure of learning a neural network model $f$ is cast as an optimization problem, which is typically solved using a gradient-based algorithm that navigates the space of possible sets of weights $W$ the model may use in order to attain a convergence point. Typically, a neural network model is trained using the stochastic gradient descent optimization algorithm, the weights being updated using back-propagation \cite{Rumelhart-N-1986}.
In the context of an optimization problem, the function used to evaluate a candidate solution (i.e., a set of weights) is referred to as the objective function, or the loss function.

In our case, we employ a custom loss function based on the mean squared error (MSE), in order to properly train the model and achieve optimal results. Our main goal is to recover the targets, which are typically at the upper extremity of the amplitude interval. To make sure that our model gives proper attention to such extreme values, we favor MSE instead of the mean absolute error (MAE). Furthermore, our loss function is designed to adjust the importance of the FFT magnitude in relation to its real and imaginary parts, because estimating the real and the imaginary parts of a complex number is more difficult to achieve compared to estimating its magnitude \cite{Fuchs2020}. We therefore introduce the hyperparameter $\lambda \in \mathbb{R}^+$ to control this importance. Our loss function is formally defined below:
\begin{equation}
\begin{split}
\label{loss_function}
 \mathcal{L}(y, \hat{y}) &= \mathcal{L}_{abs}(y, \hat{y}) + \lambda \cdot ( \mathcal{L}_{re}(y, \hat{y}) + \mathcal{L}_{im}(y, \hat{y})) ,
\end{split}
\end{equation}
\noindent where $y$ is the true label, $\hat{y}=f(x,W)$ is the label predicted by the model $f$ for the input $x$ associated with label $y$, and the loss function $\mathcal{L}_{\{abs,re,im\}}$ is the MSE applied to the corresponding parts of $y$ and $\hat{y}$, respectively. As explained earlier, the factor $\lambda$ adjusts the importance of the magnitude with respect to the real and the imaginary parts. We stress out that the label $y$ is actually the FFT of the clean range profile, being composed of the magnitude, the real part and the imaginary part, respectively.

%\vspace{-0.2cm}
\subsection{Weight Pruning}

Convolutional neural networks have shown major performance improvements in a broad range of domains~\cite{georgescu2020convolutional,Krizhevsky-NIPS-2012,georgescu2019recognizing}, once the training on powerful graphical processing units was made possible by the technological advancements in parallel processing, gaining orders of magnitude in terms of training time~\cite{Krizhevsky-NIPS-2012}. This also allowed researchers to explore deeper and deeper models~\cite{Simonyan-ICLR-14,Szegedy-CVPR-2015,He-CVPR-2016}, requiring appropriate changes to avoid vanishing and exploding gradients after a certain point, for example by introducing residual blocks~\cite{He-CVPR-2016}. %The main benefit of CNNs is that they can capture high-order relationships between input and output, thus being able to generalize better and take more complex decisions. 
However, a downside of such large models is that they are also likely to capture noise from training data, easily falling into the pitfall of overfitting. The noise learned by a CNN through its weights is not representative for the generic data distribution, inherently leading to high variance and poor performance. Nonetheless, simply reducing the model's capacity would not be a proper solution, because it will lead to the other extreme, underfitting. This problem may occur when the high-order relationships between input and output cannot be captured by a model with reduced capacity. Since it is already proven that CNNs attain better results as the models grow deeper~\cite{Krizhevsky-NIPS-2012,Simonyan-ICLR-14,Szegedy-CVPR-2015,He-CVPR-2016}, the main focus in this area of research is to find ways to avoid overfitting for models with higher capacity. One such example is dropout~\cite{Srivastava-JMLR-2014}. Moreover, a well-known fact is that noise reduction is a hot topic in the field of signal processing, therefore, a lot of approaches have been proposed by researchers~\cite{bentler2006digital}. Both classic signal processing algorithms~\cite{boudraa2004emd, klasen2007binaural, hu2007comparative} and machine learning methods~\cite{rudy2019deep, chiang2019noise, zhang2013denoising} have been developed in order to mitigate the noise from signals.
The noise problem is even more relevant when we refer to denoising solutions based on deep learning, because models have a large capacity and tend to also replicate the noise from label signals, preventing the network from achieving a global optimum. 
%This may eventually leads to inferior results.

\begin{algorithm}[!t]
\caption{Training with Weight Pruning}\label{alg_1}
\textbf{Input:}{ $f$ - a neural network, $X$ - a training set, \\ $r \in [0, 1]$ - ratio of noise reduction.}\\
\textbf{Initialization:}{ $W^{(0)} \sim \mathcal{N}(0, \Sigma)$}\\
\textbf{Output:}{ $W^{(t)}$}
\vspace{0.2em}
\algrule
\vspace{0.2em}
\textbf{Stage 1:}{ Conventional training regime}
\vspace{0.2em}
\algrule
\begin{algorithmic}[1]
\While{not converge}
 \State $W^{(t+1)} = W^{(t)} - \eta^{(t)} \nabla{f(W^{(t)}, x^{(t)})}$
 \State $t = t + 1$
\EndWhile

\vspace{0.2em}
\algrule
\vspace{0.2em}
\noindent\hspace{-2em}
\textbf{Stage 2:}{ Weight pruning training regime}
\vspace{0.2em}
\algrule
\vspace{0.2em}
\State $S = sort(||W^{(t-1)}||)$
\State $k = \lfloor r/|S| \rfloor$
\State $\epsilon = S_k$
\For{$w \in W^{(t)}$} 
\If{$w < \epsilon$}
\State $M_w = 0$
\Else
\State $M_w = 1$
\EndIf
\EndFor

\While{not converge}
 \State $W^{(t+1)} = W^{(t)} - \eta^{(t)} \nabla{f(W^{(t)}, x^{(t)})}$
 \State $W^{(t+1)} = W^{(t+1)} \cdot M$
 \State $t = t + 1$
\EndWhile 
\end{algorithmic}
\end{algorithm}

To solve this problem, we apply a weight pruning~\cite{han2015learning,liu2018rethinking} method that starts with a conventional training phase, followed by a noise-constrained training phase with the aim of pruning the inner network noise, thus improving its signal-to-noise ratio. The network architecture is perfectly consistent, assuming no further modifications at testing time. The steps required by weight pruning are formally described in Algorithm~\ref{alg_1}. The training starts with the random initialization of the weights $W$ from a normal distribution, as usual.

In the first stage, we apply the standard training procedure based on gradient descent, until we reach an optimal convergence point. During this stage, the weights $W$ are updated in the negative direction of the gradient $\nabla f$, the update step being controlled through the learning rate $\eta$. After the first training stage, we observed that our neural networks contains many weights that are close to zero. When put together, these very small weights can affect the model, acting as some kind of noise learned from the training data. In the next phase, we compute a binary mask $M$ with the aim of clipping the less important weights to zero. In step 4, the weights are first sorted by their magnitude in ascending order. The index of the largest ``noisy'' weight to be used as threshold is computed in step 5, based on the ratio of noise reduction $r$ given as input. The actual value of the threshold weight is stored into the parameter $\epsilon$. In steps 7 to 11, we build the mask $M$, assigning $0$ for every weight lower than $\epsilon$ and $1$ for every other weight. After obtaining the mask $M$, we can further proceed by training the model using gradient descent. After each weight update, the training algorithm introduces step 14, which removes the weights close to zero. We note that the mask $M$ can also be recomputed at every iteration, but we did not observe any significant improvement in terms of convergence during our preliminary experiments. To save computational time during training, we decided to compute the mask $M$ only at the beginning. Last, we note that, although Algorithm~\ref{alg_1} is based on the standard stochastic gradient descent, the weight update steps are independent of the training regime. Hence, weight pruning~\cite{han2015learning,liu2018rethinking} can be applied on top of any modern optimization algorithm for neural networks, e.g.~Adam~\cite{Kingma-ICLR-2015}.

\subsubsection{Relation to dropout}
Being designed as a method to prevent overfitting, we note that weight pruning can be seen as a competitor to dropout \cite{Srivastava-JMLR-2014}, which may lead to superior results \cite{xiao2019autoprune}. Dropout is a regularization technique that drops out a certain percentage of neural units randomly, at each iteration. Weight pruning works in a similar way, but instead of dropping units randomly, it chooses the units that have the weights closer to zero. We expect such units to contain noise rather than useful information. We therefore believe that weight pruning is able to preserve (or even improve) the signal-to-noise ratio inside the neural network. Another difference from dropout is that our training regime based on weight pruning is divided into two stages. In the first stage, we allow the network to converge to an optimal point, using only early stopping to prevent overfitting. Weight pruning is applied only in the second stage, enabling convergence to a more generic solution. We compare dropout and weight pruning experimentally, showing that the latter training regime provides superior results.

%\vspace{-0.3cm}
\section{Data Set}
\label{sec_Data_Set}

One of the key factors in the training process of deep models is the data set. It was empirically shown many times before, e.g.~\cite{Krizhevsky-NIPS-2012,He-CVPR-2016}, that a large database, e.g.~ImageNet~\cite{Russakovsky-IJCV-2015}, is essential to enable deep models to attain state-of-the-art results. Therefore, we extend the automotive radar interference mitigation (ARIM) data set proposed in~\cite{Ristea-VTC-2020} in order to cover complex real-world automotive scenarios that include multiple sources of interference. To the best of our knowledge, there are no other public databases for the interference mitigation task with multiple sources of interference. Our data set is created in the fast FMCW hypothesis, where the beat frequency is usually much larger than the Doppler shift. Each range profile can include both static and dynamic targets, which will practically appear as straight lines in the beat signal's spectrogram.

In this paper, we introduce a novel and complex large-scale database, called ARIM-v2, consisting of 144,000 synthetically generated samples, replicating realistic automotive scenarios. We generated each sample using randomly selected values from the set of realistic parameters enumerated in Table~\ref{tab1}. The number of interference sources and the signal-to-noise ratio (SNR) values are selected using a fixed step between the minimum and the maximum values. The other parameters from Table~\ref{tab1} are interpreted as random variables that follow an uniform distribution between the minimum and the maximum values. The amplitude of each target is proportional with the power expected from that particular target. Moreover, we added a random phase to each target to obtain more realistic radar signals.

\begin{table}
\centering
\caption{Minimum and maximum values for each parameter in our joint uniform distribution used for generating the samples in our database.}
\label{tab1} 
\vspace{-0.1cm}
 \begin{tabular}{|l|ccc|}
 \hline
 Parameter & Minimum & Maximum & Step \\ 
 \hline\hline
 Interference sources & 1 & 3 & 1 \\ 
 \hline
 SNR [dB] & 5 & 40 & 5 \\ 
 \hline
 SIR [dB] & -5 & 40 & - \\
 \hline
 Relative interference signal slope & 0 & 1.5 & - \\
 \hline
 Number of targets & 1 & 4 & - \\
 \hline
 Target amplitude & 0.01 & 1 & - \\
 \hline
 Target distance [m] & 2 & 95 & - \\ 
 \hline
 Target phase [rad] & $-\pi$ & $\pi$ & - \\ 
 \hline
\end{tabular}
%\vspace{-0.3cm}
\end{table} 

\begin{table}
\centering
\caption{Fixed parameters for simulating a realistic radar sensor.}
\label{tab2}
\vspace{-0.1cm}
 \begin{tabular}{|clc|}
 \hline
 Parameter & Description & Value\\
 \hline\hline
 B & Bandwidth & 1.6 GHz\\ 
 \hline
 $T$ & Time of chirp & 25.6 $\mu s$\\ 
 \hline
 $f_{s}$ & Sampling frequency & 40 MHz\\ 
 \hline
 $f_{c}$ & Radar central frequency & 78 GHz\\ 
 \hline
\end{tabular}
%\vspace{-0.3cm}
\end{table}

Real data acquisition involves capturing signals with radar sensors that have specific parameters. Even when we consider a particular application, the deployed radar sensor could have distinct values of parameters, which may lead to differences between captured data.
For this reason, in our data generation procedure, we considered a set of parameters (i.e., bandwidth, sweep time, sampling frequency and central frequency) that can be adjusted for a specific radar sensor. In this way, our database could be adapted and used in various circumstances.
Without loss of generality, we developed the database by setting the acquisition parameters to typical values used in automotive radar sensors. The exact values used for these parameters are listed in Table~\ref{tab2}. We underline that, for a different sensor, the database can be regenerated with the parameters of the specific radar system.

One of the greatest advantages regarding synthetically generated data is that we can control the entire process with the purpose of obtaining more complete and relevant information, which may help to develop a better solution. In our case, we have access to the signal with interference as well as access to the clean signal. Hence, we can properly evaluate any interference mitigation algorithm. For example, the clean signals can be used as ground-truth labels when training a machine learning model. Moreover, access to the clean signals provides the means to conduct an objective performance assessment, by comparing the output predicted by the model with the corresponding ground-truth (expected) output. In ARIM-v2, a data sample is composed of:
\begin{itemize}
\item a time domain signal without interference; 
\item a time domain signal with interference;
\item a label vector with complex amplitude values in target locations;
\item a label vector with the following information: number of sources of interference, SNR, SIR and interference slopes. 
\end{itemize}

We randomly split our data samples into a training set of 120,000 samples and a test set of 24,000 samples. The generated data set will allow future works on RFI mitigation to objectively compare newly developed methods with the state of the art, provided that our data set  is freely available for download at: http://github.com/ristea/arim-v2. 

%\vspace{-0.2cm}
\section{Experiments}
\label{sec_Experiments}

Since both ARIM and ARIM-v2 databases consist of multiple radar signals (with and without interference) referring to different range profiles, in our experiments, the interference mitigation is performed individually, on each range profile. We consider as label the amplitude and phase of targets from the range profiles obtained by applying FFT on signals without interference.

%\vspace{-0.2cm}
\subsection{Performance Measures}

Usually, the goal in radar signal processing is to maximize the detection performance. Therefore, a rather intuitive measure is the area under the Receiver Operating Characteristics (ROC) curve, known as AUC for short, which describes the ability to disentangle targets from noise at various thresholds. When computing the AUC, the target detection threshold slides iteratively from the lowest value to the largest value in the range profile, modifying the probability of false alarms. Another performance indicator is the mean absolute error (MAE) in decibels (dB) between the range profile amplitude of targets computed from label signals and the amplitude of targets from predicted signals. However, in radar signal processing not only the target's amplitude is important, but also its phase, because the latter is necessary to estimate other essential parameters (e.g., target velocity) or to perform beamforming. Thus, we also report the MAE in degrees between the range profile phase of targets computed from label signals and the phase of targets from predicted signals. 

In summary, we employ the AUC, amplitude MAE, phase MAE and mean SNR improvement ($\overline{\Delta \mbox{SNR}}$), which is computed for the target with the highest amplitude as the difference between the SNR before and after interference mitigation in the range profile.

%\vspace{-0.2cm}
\subsection{Hyperparameter Tuning}

Hyperparameter tuning is performed on ARIM-v2, employing the same hyperparameters on ARIM without further tuning. In order to minimize the chance of overfitting in hyperparameter space, we split the ARIM-v2 training set into training and validation, keeping $20\%$ (24,000 samples) for validation, the rest (96,000 samples) being used for training. Regarding the our regime based on weight pruning, we trained our model for $100$ epochs in the conventional training regime, followed by $20$ epochs with weight pruning. The ratio of noise reduction $r$ was validated considering values in the set $\{0.15, 0.3, 0.45\}$, the best performance gains being obtained for $r=0.3$. We compared the weight pruning regime with conventional training for $120$ epochs and dropout for $120$ epochs, respectively. The dropout rate was validated in the range $[0.1, 0.5]$, the best rate being $0.25$. In all cases, we used mini-batches of $16$ samples using the Adam optimizer \cite{Kingma-ICLR-2015} with a learning rate of $5 \cdot 10^{-5}$ and a weight decay of $10^{-5}$. Regarding the parameter $\lambda$ in the loss function, we tried out several values ranging from $1$ to $20$, the best solution being achieved for $\lambda=10$.

\begin{table}[t]
\centering
\noindent
\caption{Results on the ARIM-v2 test set obtained by our FCN under different training regimes: conventional, dropout and weight pruning. Regarding conventional training, we considered a network with half capacity (HC) along with the full network. For weight pruning, we report results for all considered reduction ratios. For the reported metrics, we use $\uparrow$ to denote that higher values are better and $\downarrow$ to denote that lower values are better, respectively. Best scores are highlighted in bold.} 
\begin{tabular}{|l|cccc|}
\hline
 &  & & {MAE}$\downarrow$ & {MAE}$\downarrow$ \\
Training Method & $\overline{\Delta \mbox{SNR}}\uparrow$                                      & {AUC}$\uparrow$                  & Amplitude     & Phase\\
                &                                       &                   & {(dB)}        & {(degrees)} \\ 
\hline
\hline
Conventional                        & 15.28 & 0.959 & 1.40 & 7.44\\ 
Conventional + HC                   & 12.44 & 0.953 & 1.48 & 9.88\\ 
Dropout~\cite{Srivastava-JMLR-2014} & 14.42 & 0.958 & 1.68 & 8.90 \\ 
\hline
Pruning ($r=0.15$)                   & 15.32 & 0.960 & 1.33 & 6.96\\ 
Pruning ($r=0.3$)                    & 15.36 & \textbf{0.961} & \textbf{1.27} & \textbf{6.58} \\ 
Pruning ($r=0.45$)                   & \textbf{15.42} & 0.960 & 1.37 & 7.12 \\ 
\hline
\end{tabular}
\label{tab_results_nrt}
%\vspace{-0.3cm}
\end{table}

\begin{table*}[t]
\renewcommand{\arraystretch}{1.3} 
\centering
\noindent
\caption{Validation and test results on the ARIM data set (containing only one source of interference per range profile) attained by our model (trained with both conventional and weight pruning regimes) versus the oracle (based on ground-truth labels), the zeroing approach, a state-of-the-art deep learning method \cite{rock2019complex} and our earlier FCN models~\cite{Ristea-VTC-2020}. The best results (excluding the oracle) are highlighted in bold. The symbol $\uparrow$ means higher values are better and $\downarrow$ means lower values are better.} 
\setlength\tabcolsep{4.5pt}
\begin{tabular}{|l|cccc|cccc|}
\hline
%\multicolumn{1}{N }{\textbf{}} & \multicolumn{4}{c }{\textbf{Evaluation}} & \multicolumn{4}{c }{} \\  
 & \multicolumn{4}{|c|}{{Validation set}} & \multicolumn{4}{|c|}{{Test set}} \\ 
\cline{2-9}
{{Method}} & $\overline{\Delta \mbox{SNR}}\uparrow$ & {AUC}$\uparrow$ & {MAE Amplitude}$\downarrow$& {MAE Phase}$\downarrow$ & $\overline{\Delta \mbox{SNR}}\uparrow$ & {AUC}$\uparrow$ & {MAE Amplitude}$\downarrow$& {MAE Phase}$ \downarrow$\\
 &  & & {(dB)} & {(degrees)} & & & {(dB)} & {(degrees)} \\ 
\hline
\hline
Oracle (true labels) & 12.92 & 0.978 & 0 & 0 & 13.08 & 0.978 & 0 & 0 \\ 
\hline
{Zeroing} & 5.27 & 0.951 & 1.26 & 6.80 & 5.44 & 0.951 & 1.27 & 6.79 \\ 
{Shallow FCN~\cite{Ristea-VTC-2020}} & 10.34 & 0.965 & 2.20 & - & 10.49 & 0.965 & 2.21 & -\\ 
{Deep FCN~\cite{Ristea-VTC-2020}} & \textbf{12.90} & 0.972 & 1.21 & - & \textbf{13.06} & 0.972 & 1.22 & - \\
{CNN \cite{rock2019complex}} & 10.87 & 0.970 & 1.83 & 7.13 & 10.93 & 0.970 & 1.81 & 7.15 \\ 
\hline
{FCN (no magnitude)} & 4.32 & 0.893 & 3.54 & 30.23 & 4.66 & 0.879 & 3.87 & 35.23\\ 
{FCN (ours)} & 11.28 & 0.973 & 1.06 & 5.33 & 11.65 & 0.974 & 1.06 & 5.39\\ 
{FCN + pruning (ours)} & 11.31 & \textbf{0.973} & \textbf{1.02} & \textbf{5.15} & 11.67 & \textbf{0.974} & \textbf{1.02} & \textbf{5.30} \\ 
\hline
\end{tabular}
\label{tab_results_arim}
%\vspace{-0.2cm}
\end{table*}

\begin{table*}[t]
\renewcommand{\arraystretch}{1.3} 
\centering
\noindent
\caption{Validation and test results on the ARIM-v2 data set (containing up to three sources of interference per range profile) attained by our model (trained with both conventional and weigth pruning regimes) versus the oracle (based on ground-truth labels), the zeroing approach and a state-of-the-art deep learning method \cite{rock2019complex}. The best results (excluding the oracle) are highlighted in bold. The symbol $\uparrow$ means higher values are better and $\downarrow$ means lower values are better.} 
\setlength\tabcolsep{4.5pt}
\begin{tabular}{|l|cccc|cccc|}
\hline
%\multicolumn{1}{N }{\textbf{}} & \multicolumn{4}{c }{\textbf{Evaluation}} & \multicolumn{4}{c }{} \\  
 & \multicolumn{4}{|c|}{{Validation set}} & \multicolumn{4}{|c|}{{Test set}} \\ 
\cline{2-9}
{{Method}} & $\overline{\Delta\mbox{SNR}}\uparrow$ & {AUC}$\uparrow$ & {MAE Amplitude}$\downarrow$ & {MAE Phase}$\downarrow$ & $\overline{\Delta\mbox{SNR}}\uparrow$ & {AUC}$\uparrow$ & {MAE Amplitude}$\downarrow$ & {MAE Phase}$\downarrow$ \\
 &  & & {(dB)} & {(degrees)} & & & {(dB)} & {(degrees)} \\ 
\hline
\hline
Oracle (true labels) & 16.87 & 0.971 & 0 & 0 & 17.15 & 0.970 & 0 & 0 \\ 
\hline
{Zeroing} & 8.64 & 0.930 & 2.11 & 12.63 & 8.94 & 0.929 & 2.13 & 12.55 \\ 
{CNN \cite{rock2019complex}} & 12.64 & 0.952 & 2.14 & 8.25 & 12.94 & 0.953 & 2.17 & 8.13 \\ 

\hline
% {FCN* (ours)} & 3.62 & 0.878 & 5.14 & 38.50 & 3.99 & 0.881 & 5.91 & 39.77\\
{FCN (ours)} & 15.02 & 0.961 & 1.39 & 7.32 & 15.28 & 0.959 & 1.40 & 7.44\\ 
{FCN + pruning (ours)} & \textbf{15.09} & \textbf{0.963} & \textbf{1.25} & \textbf{6.41} & \textbf{15.36} & \textbf{0.961} & \textbf{1.27} & \textbf{6.58} \\ 
\hline
\end{tabular}
\label{tab_results_arim_v2}
%\vspace{-0.2cm}
\end{table*}

%\vspace{-0.2cm}
\subsection{Results of Weight Pruning versus Competing Methods}

In order to prove that weight pruning attains better performance due to its inner network noise reduction principle, we present the results obtained on the ARIM-v2 test set in comparison with a set of competing training regimes. We consider as competing methods the conventional training regime applied on a network with full capacity, the conventional regime applied on a network with half capacity (HC), and the regime known as dropout~\cite{Srivastava-JMLR-2014}. The corresponding results are presented in Table~\ref{tab_results_nrt}.

We first observe that dropout offers the lowest results in terms of all performance metrics, even compared to the conventional training regime. The poor results attained by dropout actually motivated us to seek an alternative training regime, this being the main driver behind using weight pruning. 
In order to establish the optimal noise reduction ratio $r$ for weight pruning, we performed several validation experiments. However, we observed that weight pruning produces better results than conventional training, irrespective of the considered reduction ratio. We therefore present test results using three different reduction ratios in Table~{\ref{tab_results_nrt}}. Although weight pruning is generally better than dropout and conventional training, it seems that the best results are achieved for the ratio $r=0.3$. Even if the noise reduction ratio of $0.45$ gives better results in terms of $\overline{\Delta \mbox{SNR}}$, the other performance metrics are in favor of the ratio $r=0.3$.

\begin{figure*}[!t]
\begin{subfigure}{.33\textwidth}
  \centering
  \includegraphics[width=1.0\linewidth]{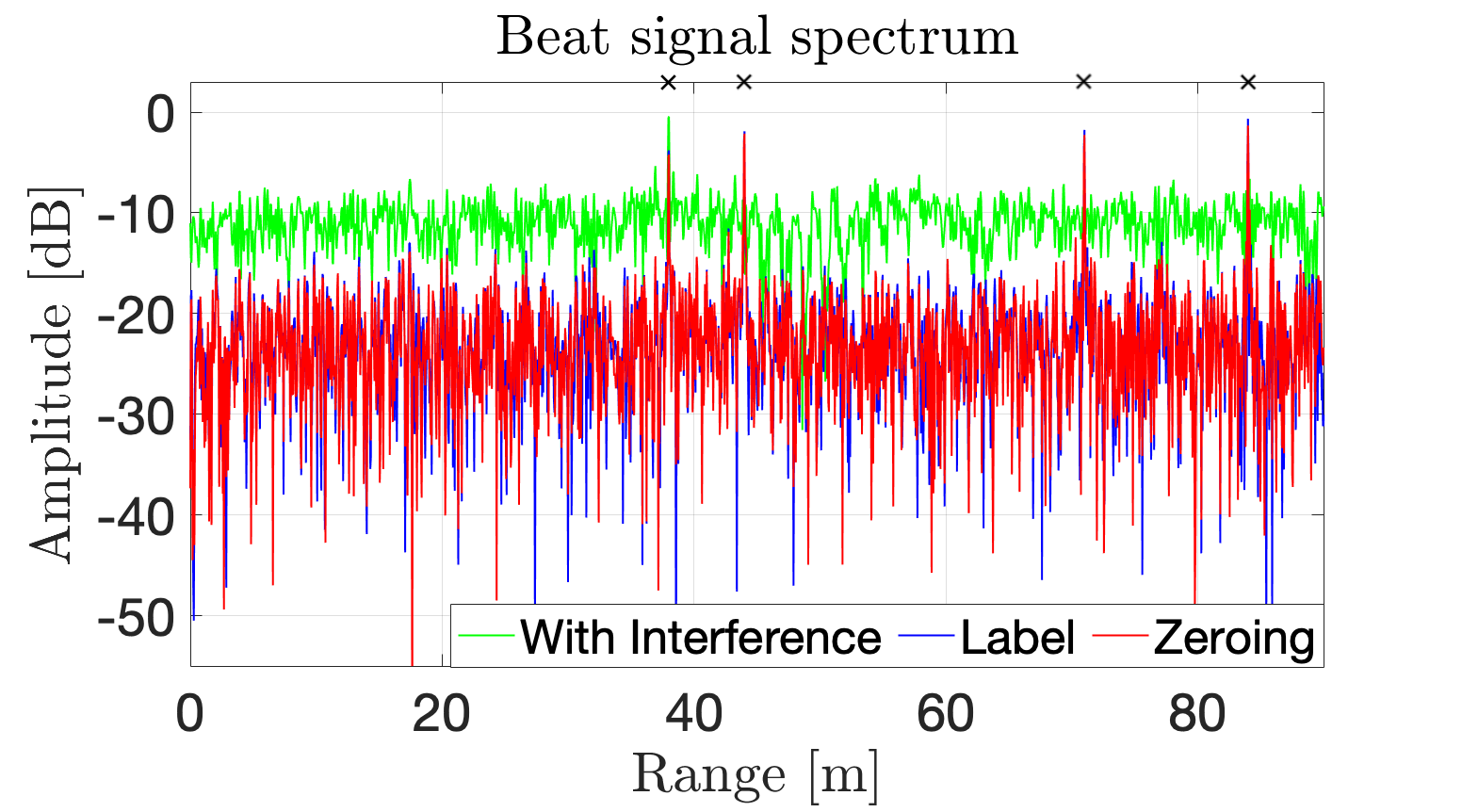} 
  \vspace{-0.55cm}
  \caption{$\mbox{SNR}\!=\!20dB$, $\mbox{SIR}\!=\!10dB$, $k\!=\!0.5$.}
  \label{arim_subfig1}
\end{subfigure}
\begin{subfigure}{.33\textwidth}
  \centering
  \includegraphics[width=1.0\linewidth]{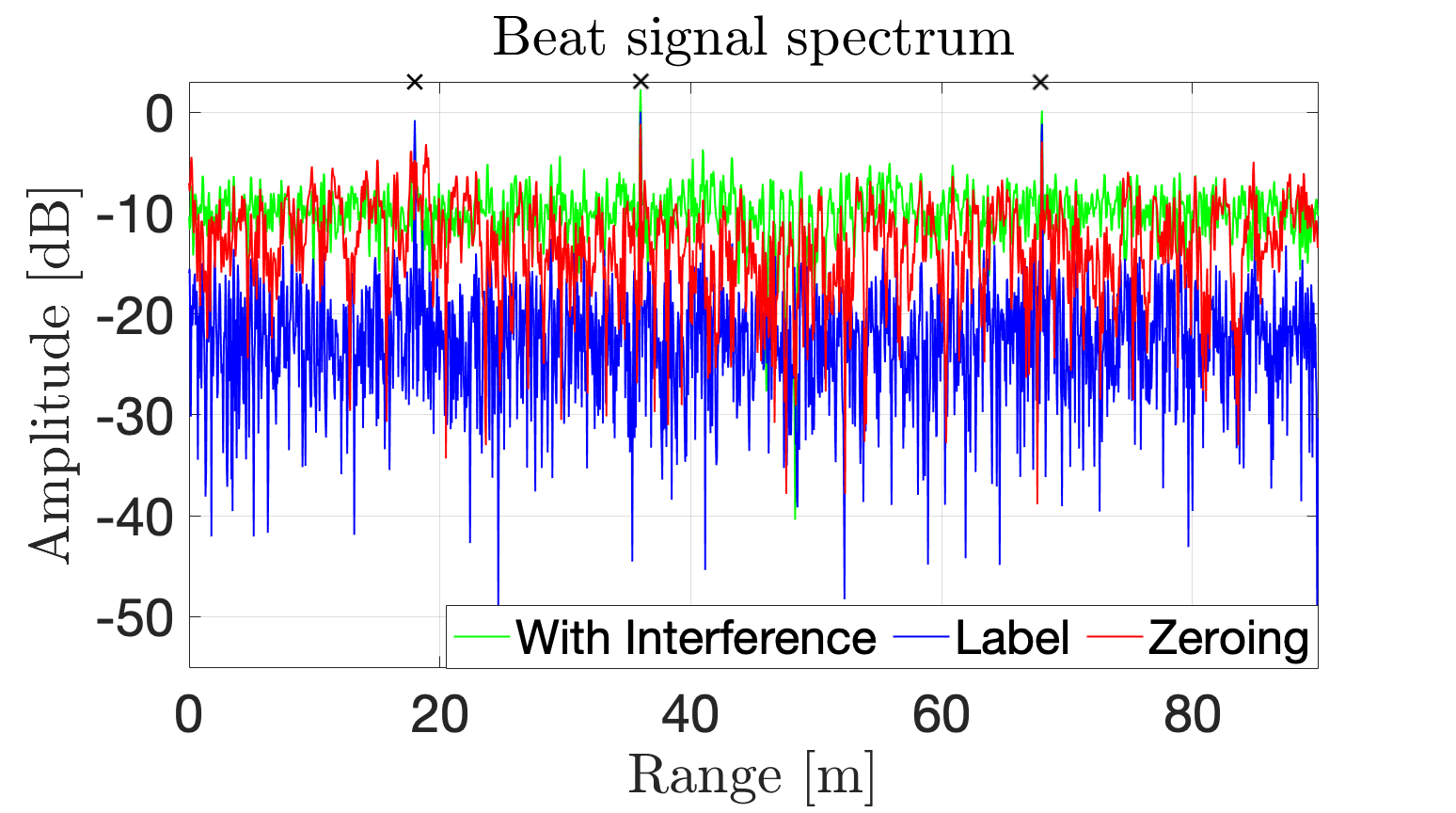} 
  \vspace{-0.55cm}
  \caption{$\mbox{SNR}\!=\!20dB$, $\mbox{SIR}\!=\!10dB$, $k\!=\!0.7$.}
  \label{arim_subfig2}
\end{subfigure}
\begin{subfigure}{.33\textwidth}
  \centering
  \includegraphics[width=1.0\linewidth]{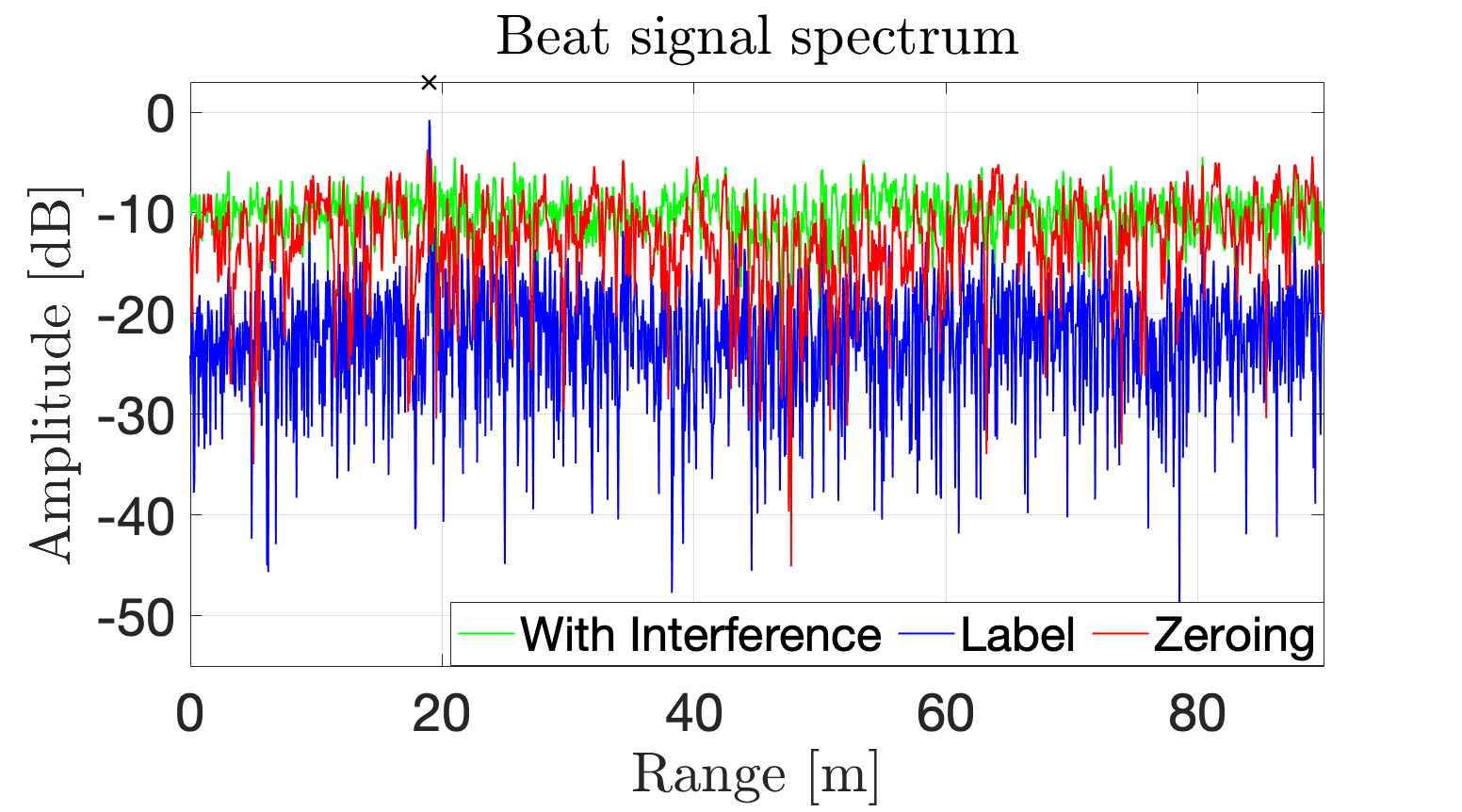}
  \vspace{-0.55cm}
  \caption{$\mbox{SNR}\!=\!20dB$, $\mbox{SIR}\!=\!10dB$, $k\!=\!0.9$.}
  \label{arim_subfig3}
\end{subfigure}
\begin{subfigure}{.33\textwidth}
  \centering
  \includegraphics[width=1.0\linewidth]{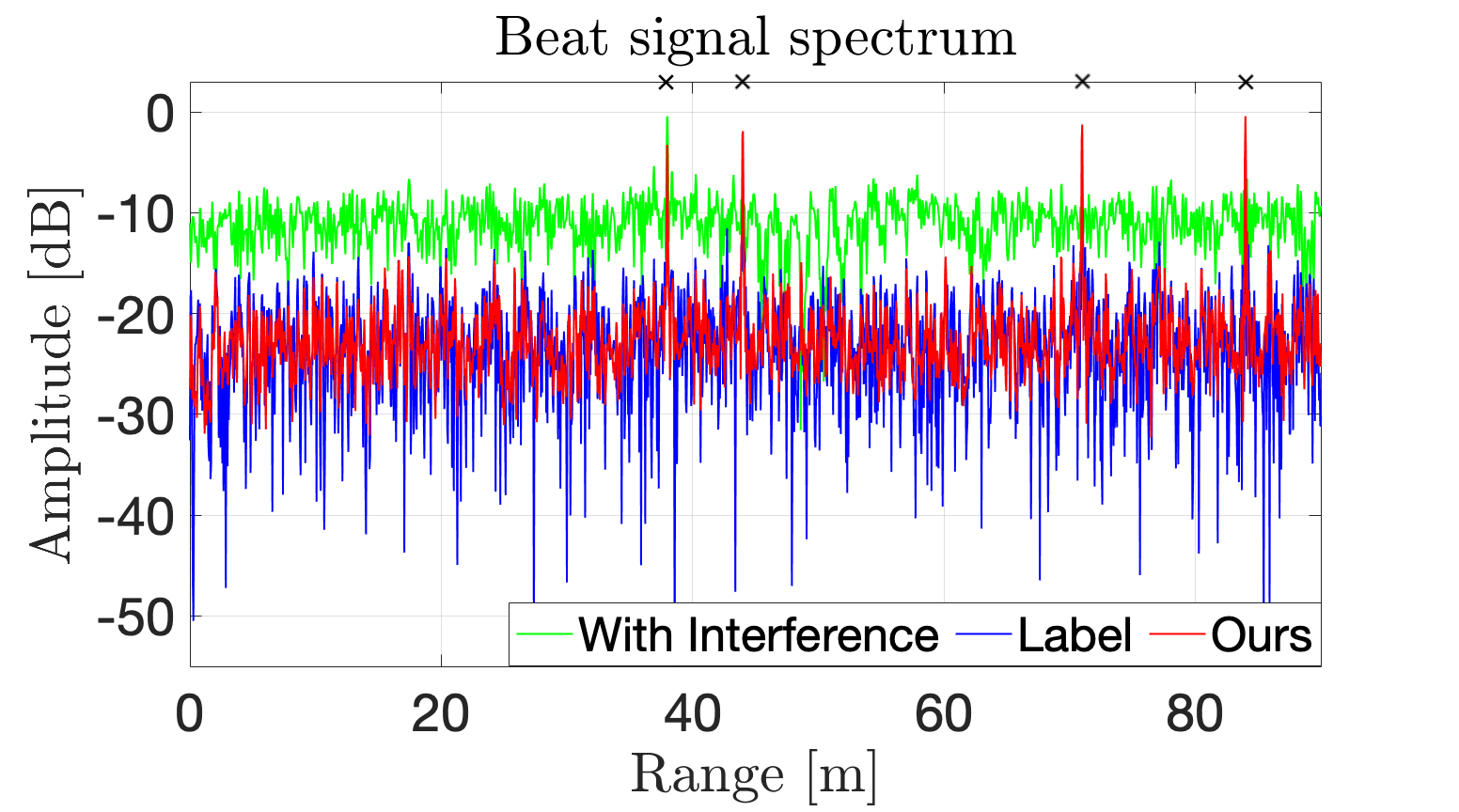}  
  \vspace{-0.55cm}
  \caption{The same parameters as above.}
  \label{arim_subfig4}
\end{subfigure}
\begin{subfigure}{.33\textwidth}
  \centering
  \includegraphics[width=1.0\linewidth]{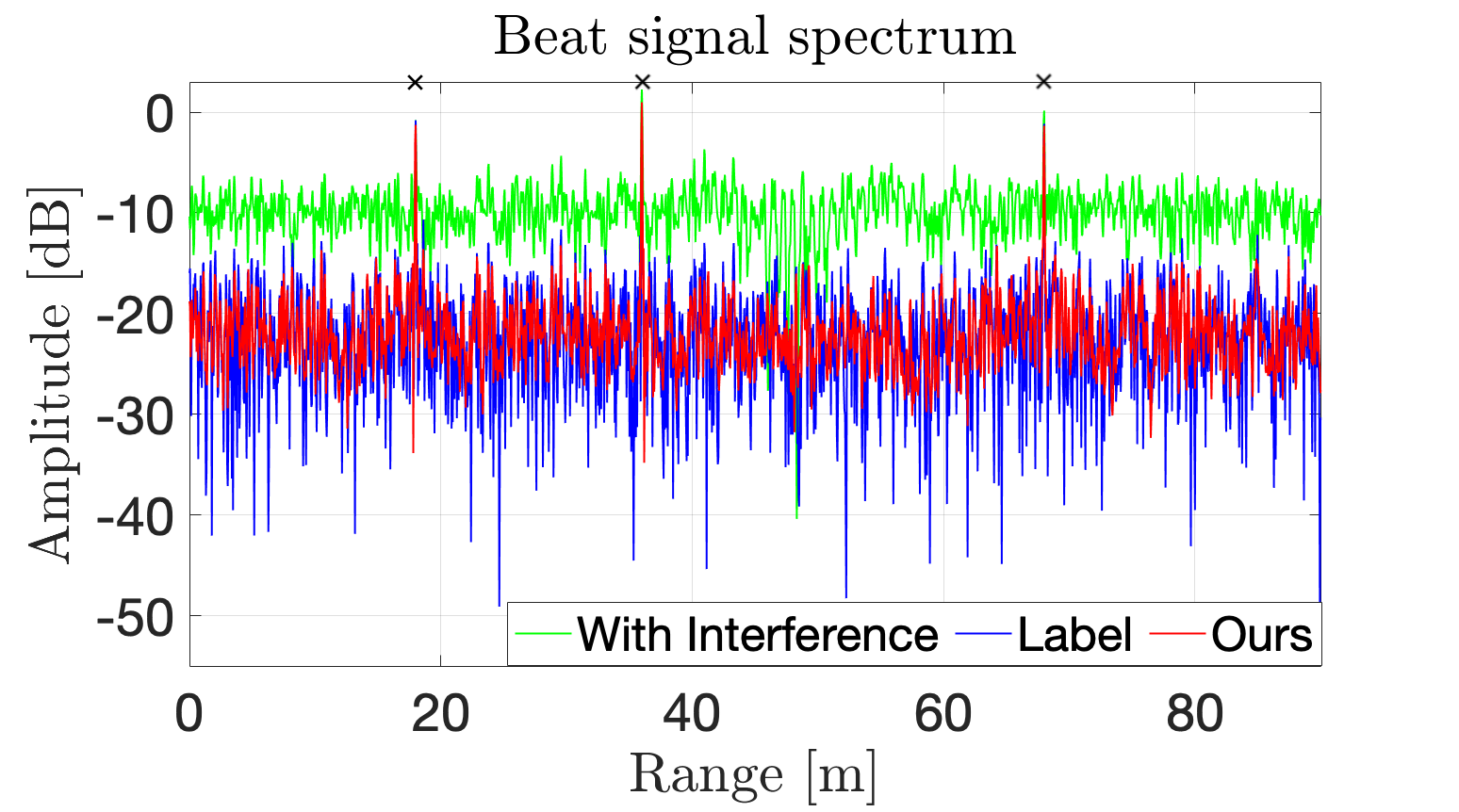}  
  \vspace{-0.55cm}
  \caption{The same parameters as above.}
  \label{arim_subfig5}
\end{subfigure}
\begin{subfigure}{.33\textwidth}
  \centering
  \includegraphics[width=1.0\linewidth]{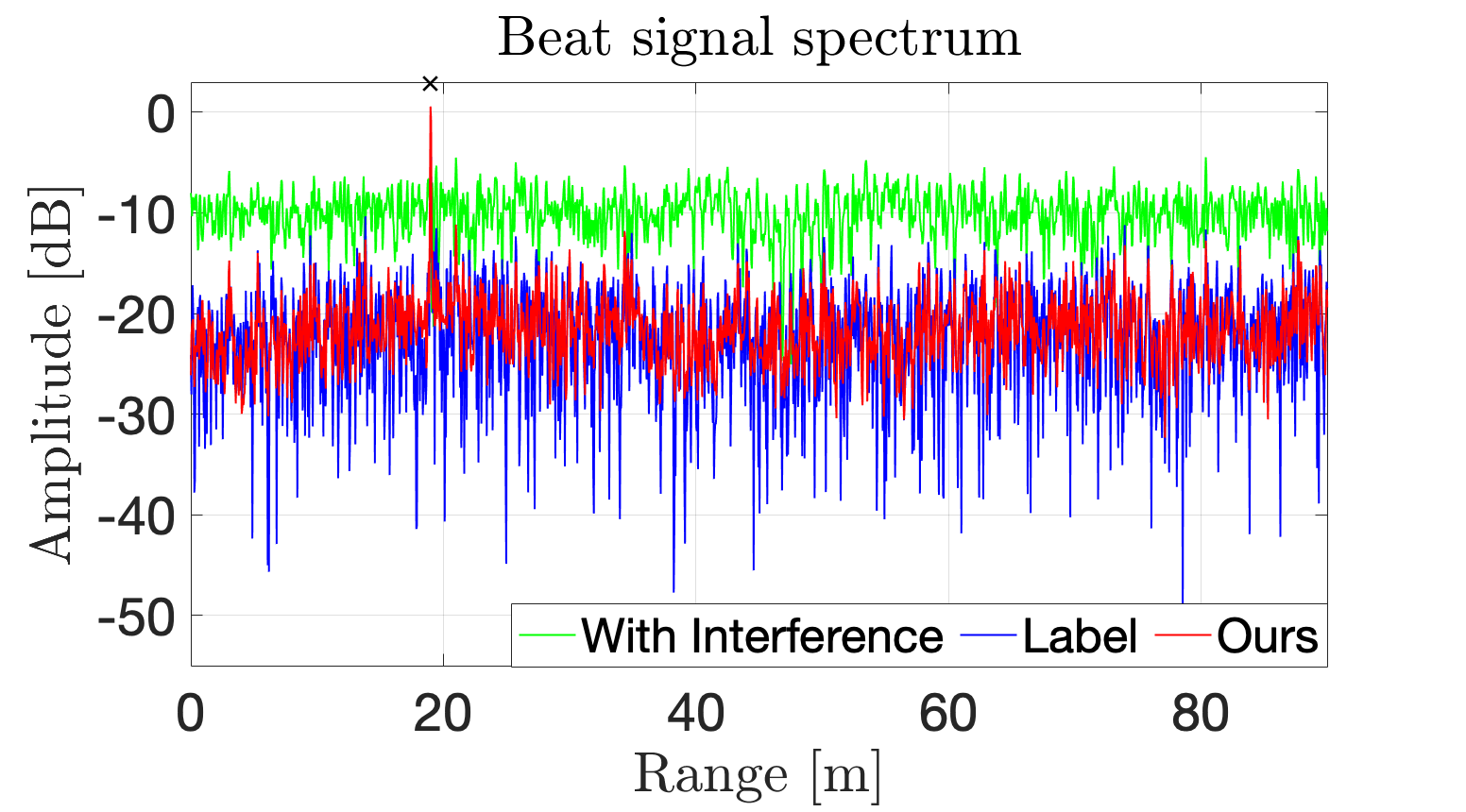}  
  \vspace{-0.55cm}
  \caption{The same parameters as above.}
  \label{arim_subfig6}
\end{subfigure}
\vspace{-0.1cm}
\caption{Qualitative results provided by our FCN+pruning model in comparison with the zeroing method. Examples are selected from the ARIM test set, each having one source of interference. Both plots on each column illustrate the same reference signal, with and without interference. On the top row, the interference is mitigated by zeroing, while on the bottom row, the interference is mitigated by our FCN+pruning model. The parameter $k$ refers to the ratio between signal and interference slopes. Best viewed in color.}
\vspace{-0.3cm}
\label{arim_data_fig}
\end{figure*}

\begin{figure*}[!t]
\begin{subfigure}{.33\textwidth}
  \centering
  \includegraphics[width=1.0\linewidth]{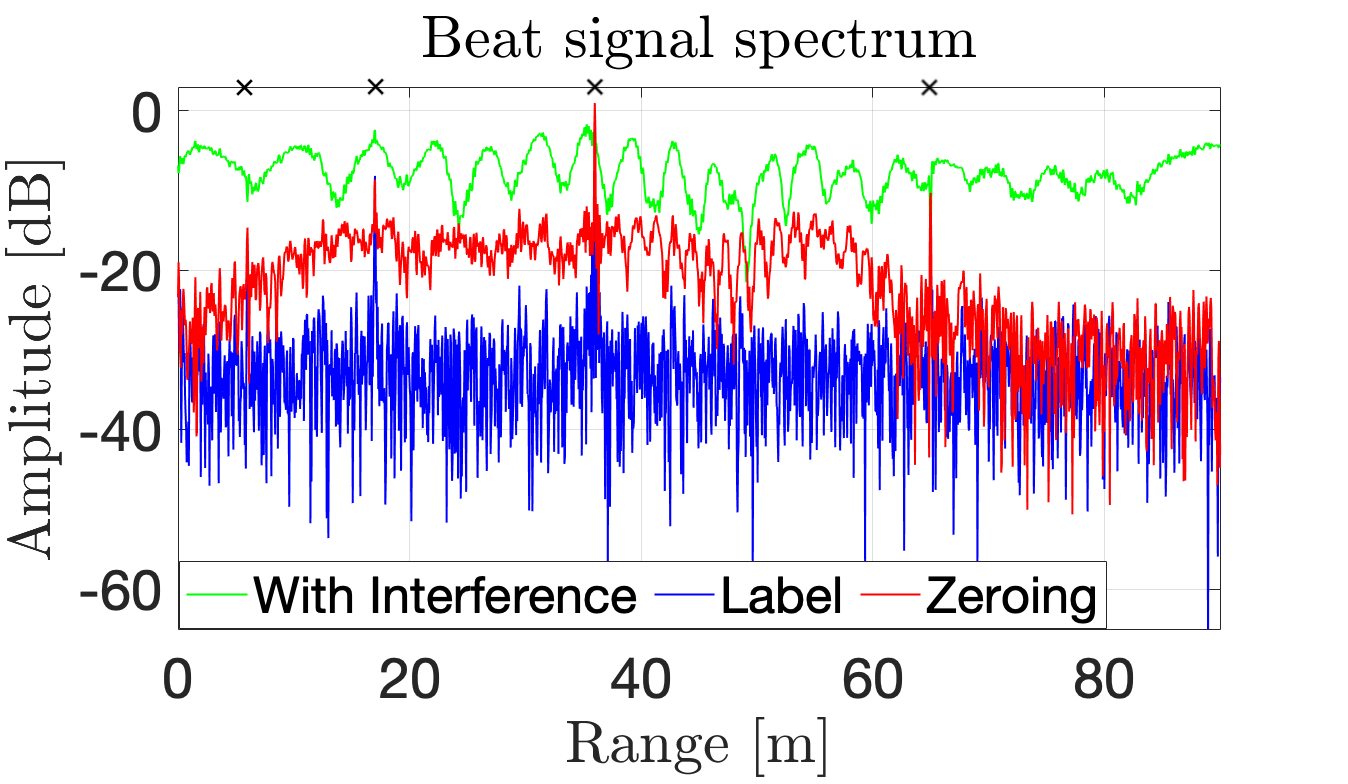} 
  \vspace{-0.55cm}
  \caption{$N_{int}=2$, $\mbox{SIR}_{min}=0.5dB$.}
  \label{arim2_subfig1}
\end{subfigure}
\begin{subfigure}{.33\textwidth}
  \centering
  \includegraphics[width=1.0\linewidth]{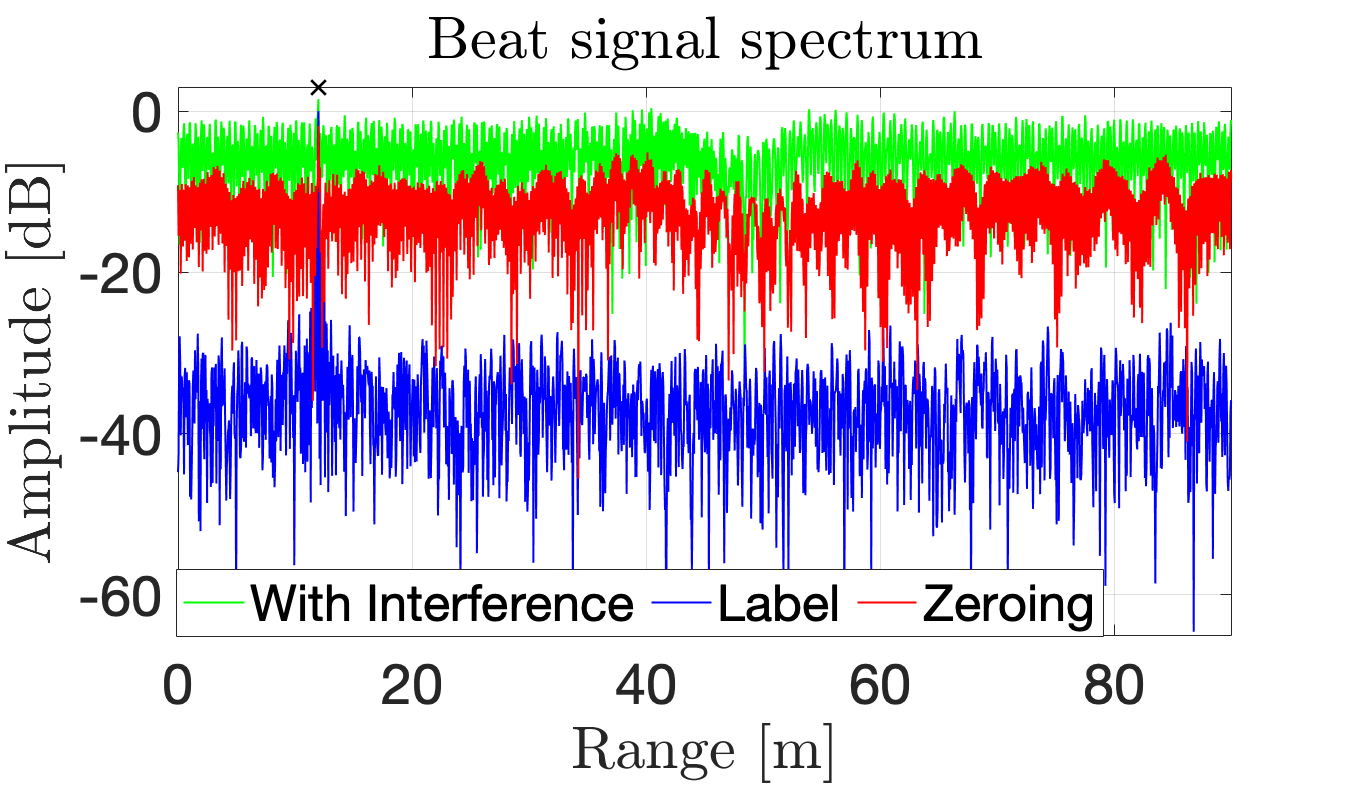} 
  \vspace{-0.55cm}
  \caption{$N_{int}=3$, $\mbox{SIR}_{min}=3dB$.}
  \label{arim2_subfig2}
\end{subfigure}
\begin{subfigure}{.33\textwidth}
  \centering
  \includegraphics[width=1.0\linewidth]{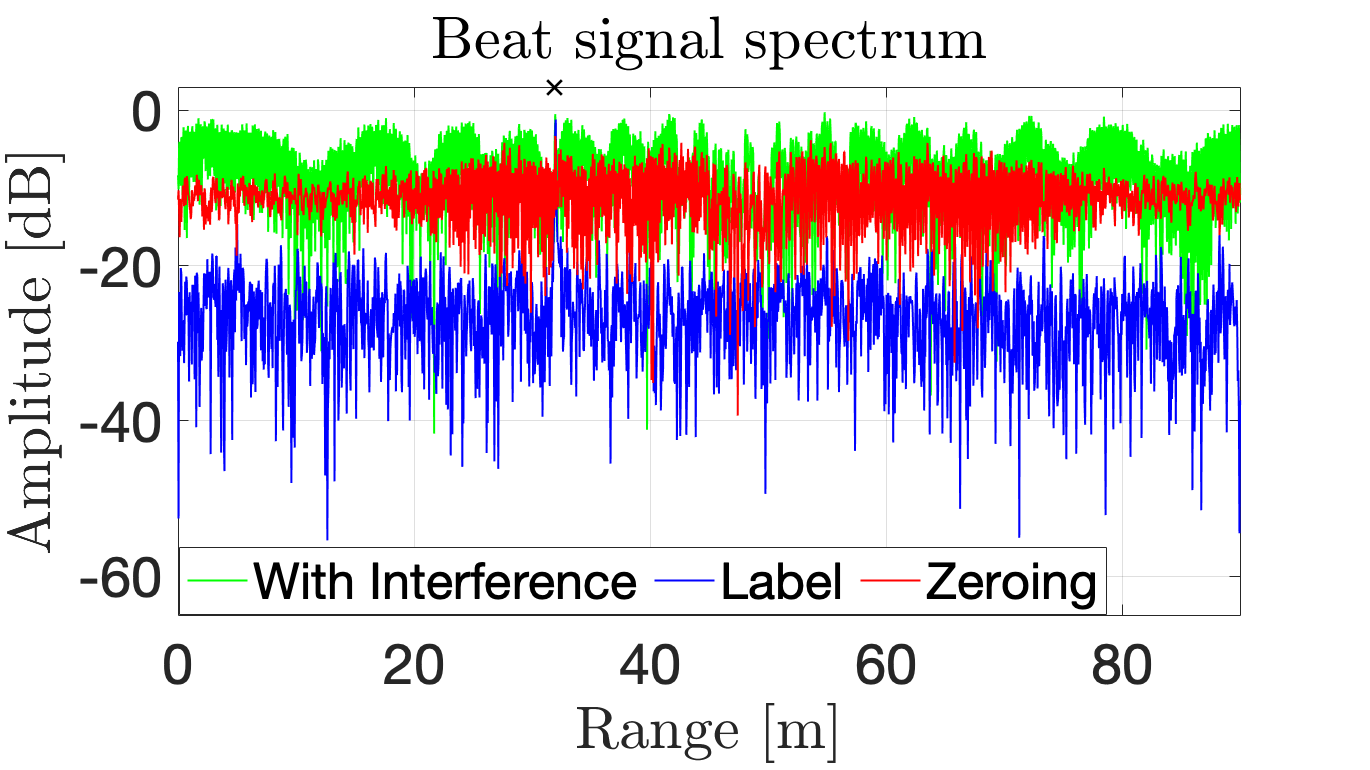}  
  \vspace{-0.55cm}
  \caption{$N_{int}=3$, $\mbox{SIR}_{min}=5dB$.}
  \label{arim2_subfig3}
\end{subfigure}
\begin{subfigure}{.33\textwidth}
  \centering
  \includegraphics[width=1.0\linewidth]{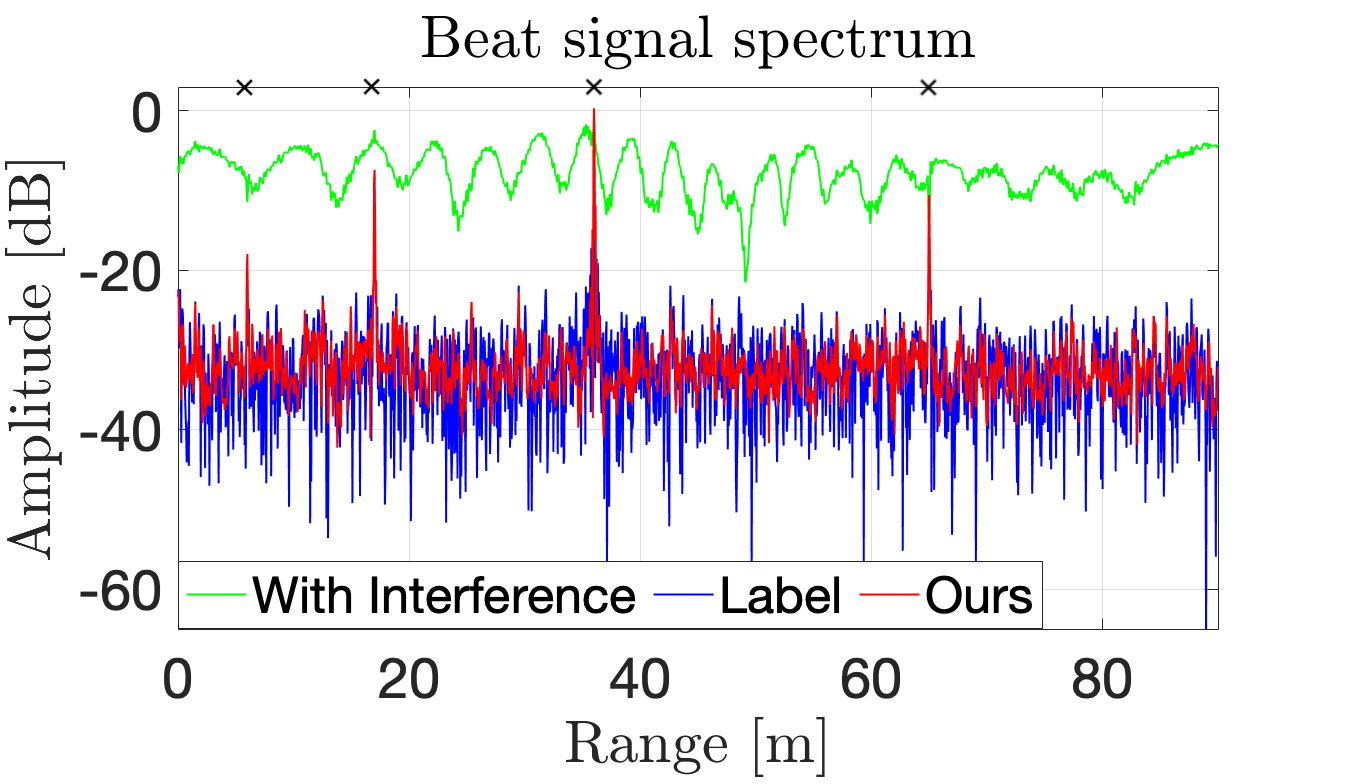}  
  \vspace{-0.55cm}
  \caption{The same parameters as above.}
  \label{arim2_subfig4}
\end{subfigure}
\begin{subfigure}{.33\textwidth}
  \centering
  \includegraphics[width=1.0\linewidth]{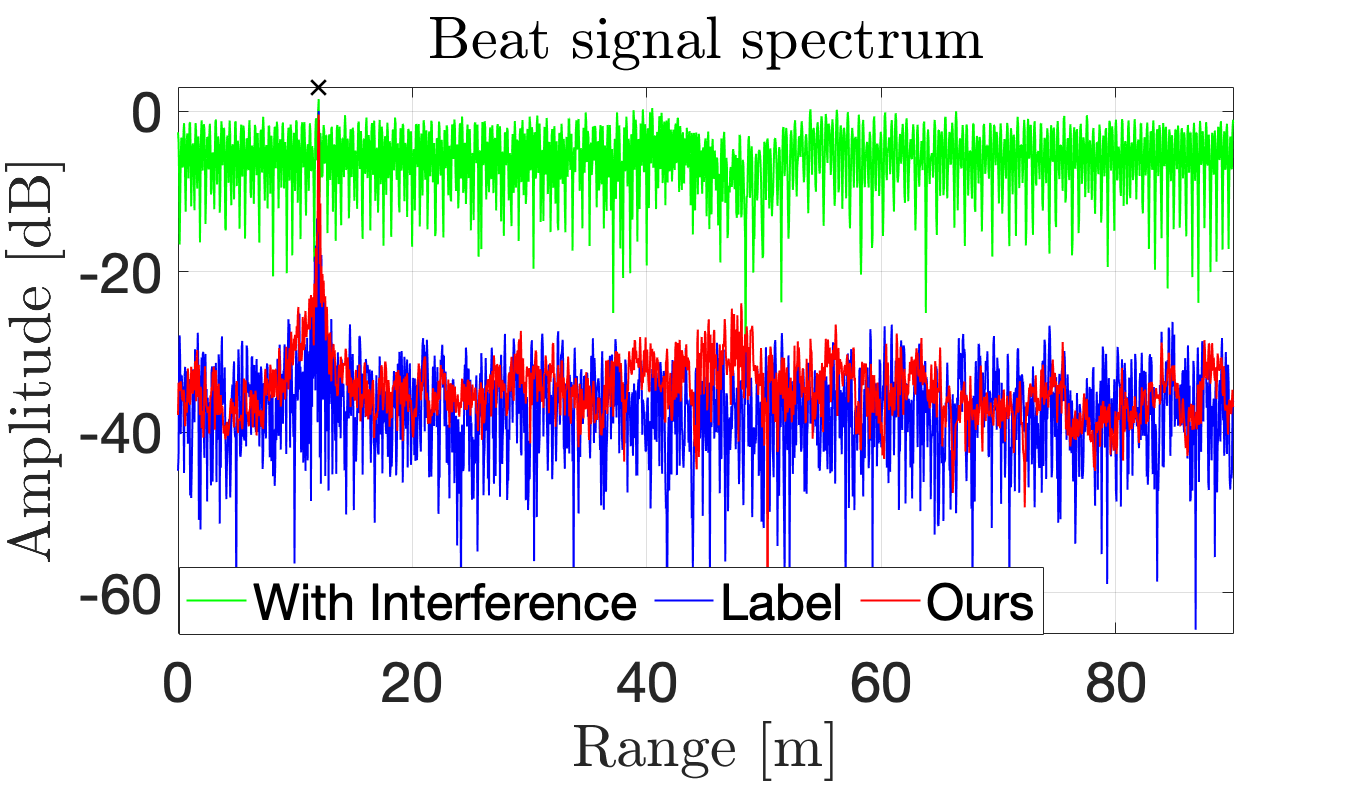}  
  \vspace{-0.55cm}
  \caption{The same parameters as above.}
  \label{arim2_subfig5}
\end{subfigure}
\begin{subfigure}{.33\textwidth}
  \centering
  \includegraphics[width=1.0\linewidth]{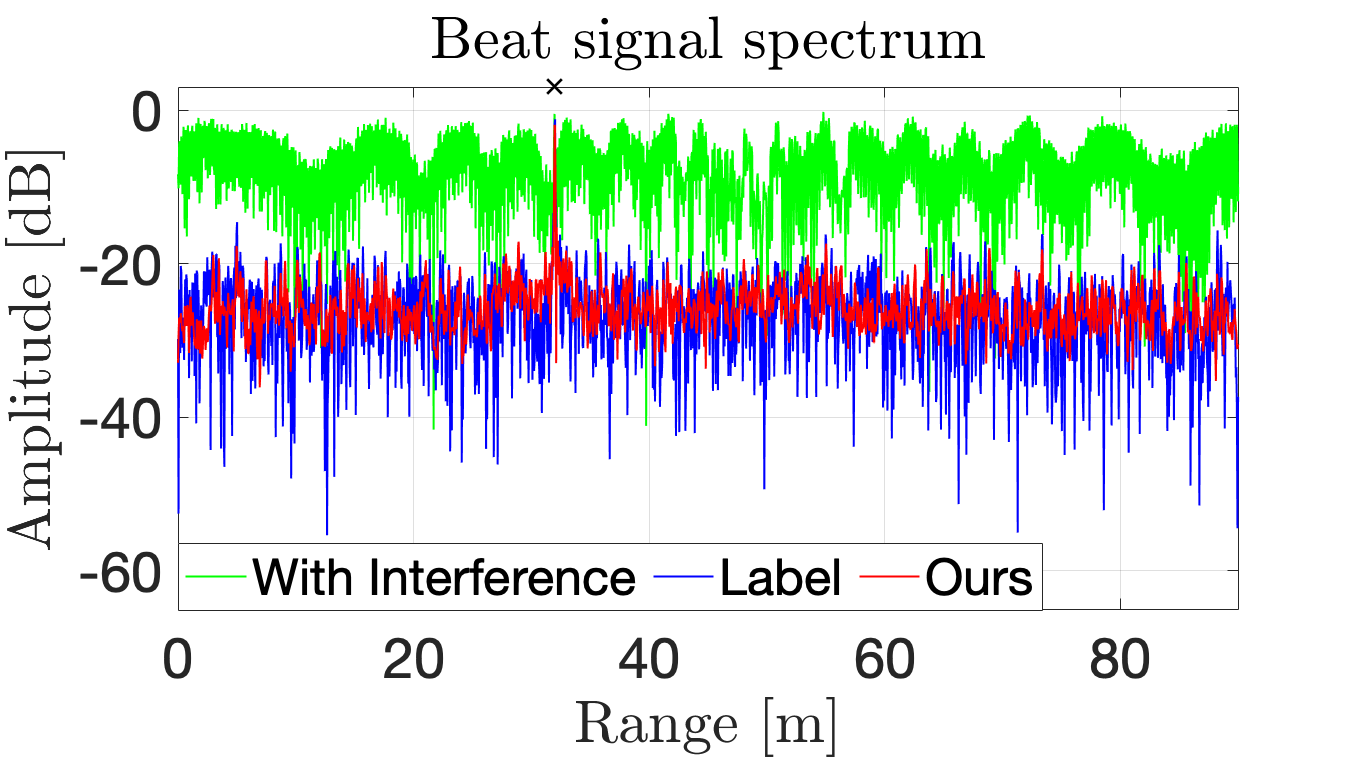}  
  \vspace{-0.55cm}
  \caption{The same parameters as above.}
  \label{arim2_subfig6}
\end{subfigure}
\vspace{-0.1cm}
\caption{Qualitative results provided by our FCN+pruning model in comparison with the zeroing method. Examples are selected from the ARIM-v2 test set, each having multiple sources of interference. Both plots on each column illustrate the same reference signal, with and without interference. On the top row, the interference is mitigated by zeroing, while on the bottom row, the interference is mitigated by our FCN+pruning model. The $N_{int}$ parameter refers to the number of interference sources and $\mbox{SIR}_{min}$ refers to the minimum value of SIR for every interference source. Best viewed in color.}
\label{arim-v2_data_fig}
\vspace{-0.3cm}
\end{figure*}

Since weight pruning replaces a certain percentage of weights with zero, it can be argued that it can be equivalent to simply reducing the model's capacity. We therefore present results using conventional training, while considering a model having $50\%$ of the original FCN capacity. As shown in Table~{\ref{tab_results_nrt}}, reducing the network's capacity is a sub-optimal solution. In terms of $\overline{\Delta SNR}$, the difference between our best pruning variant and conventional training with half capacity is $2.92$ dB in favor of the former approach. Additionally, an important difference can be observed for the MAE computed on the phase of targets, where the score of the best pruning solution is $6.58$ degrees, while the score of the model with half capacity is $3.30$ degrees higher. We thus emphasize that weight pruning is not equivalent to reducing the network's capacity, as it attains superior results.

\begin{figure*}
\begin{subfigure}{.33\textwidth}
  \centering
  \includegraphics[width=1.0\linewidth]{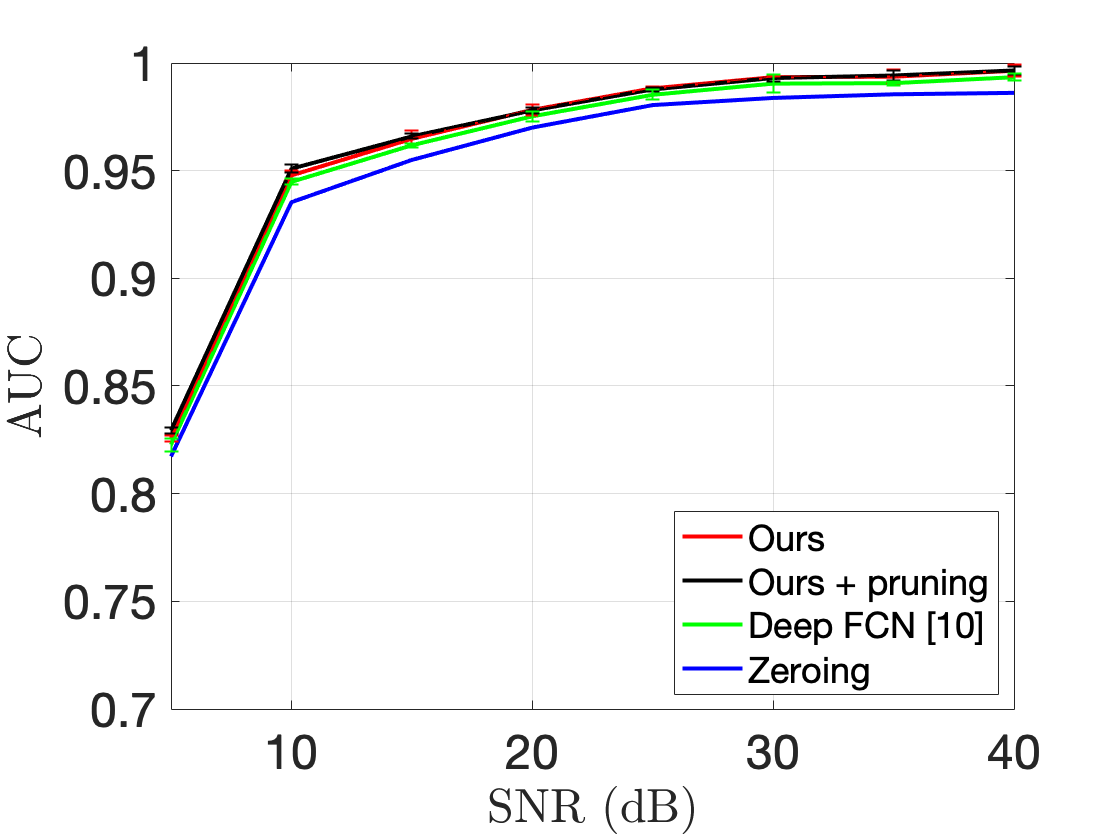}  
  \vspace{-0.55cm}
  \caption{}
  \label{subfig1}
\end{subfigure}
\begin{subfigure}{.33\textwidth}
  \centering
  \includegraphics[width=1.0\linewidth]{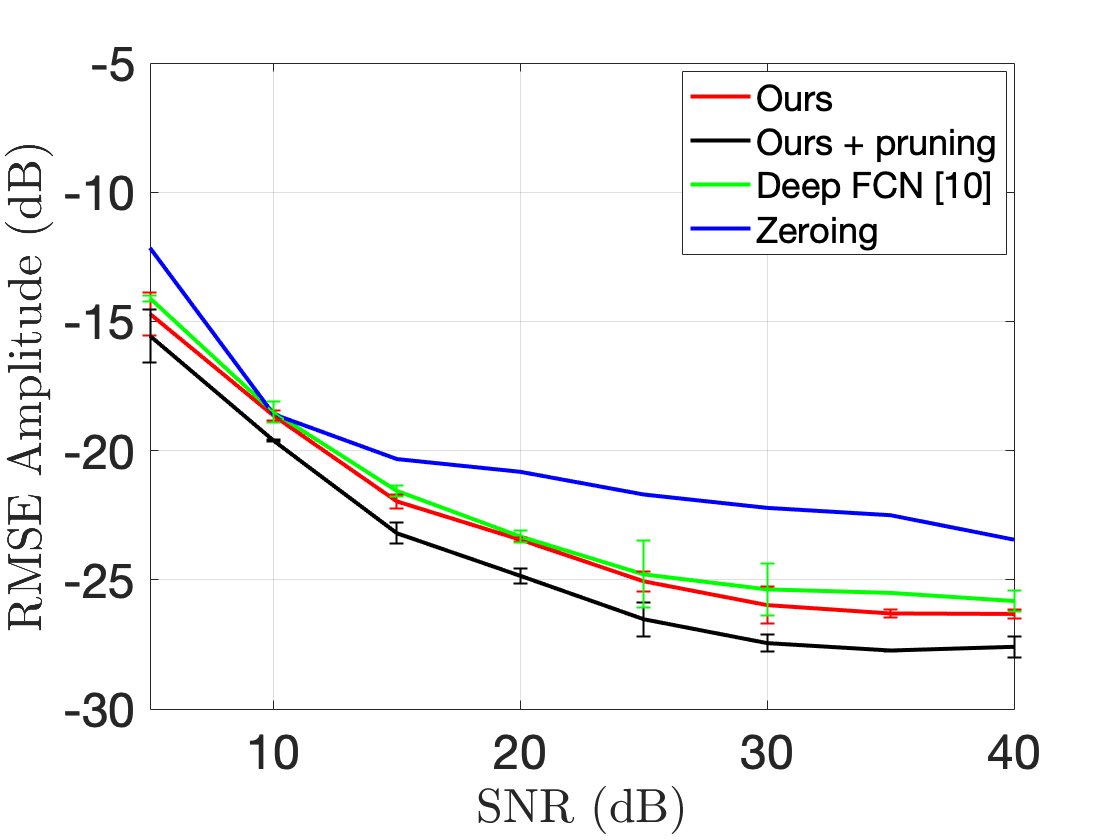} 
  \vspace{-0.55cm}
  \caption{}
  \label{subfig2}
\end{subfigure}
\begin{subfigure}{.33\textwidth}
  \centering
  \includegraphics[width=1.0\linewidth]{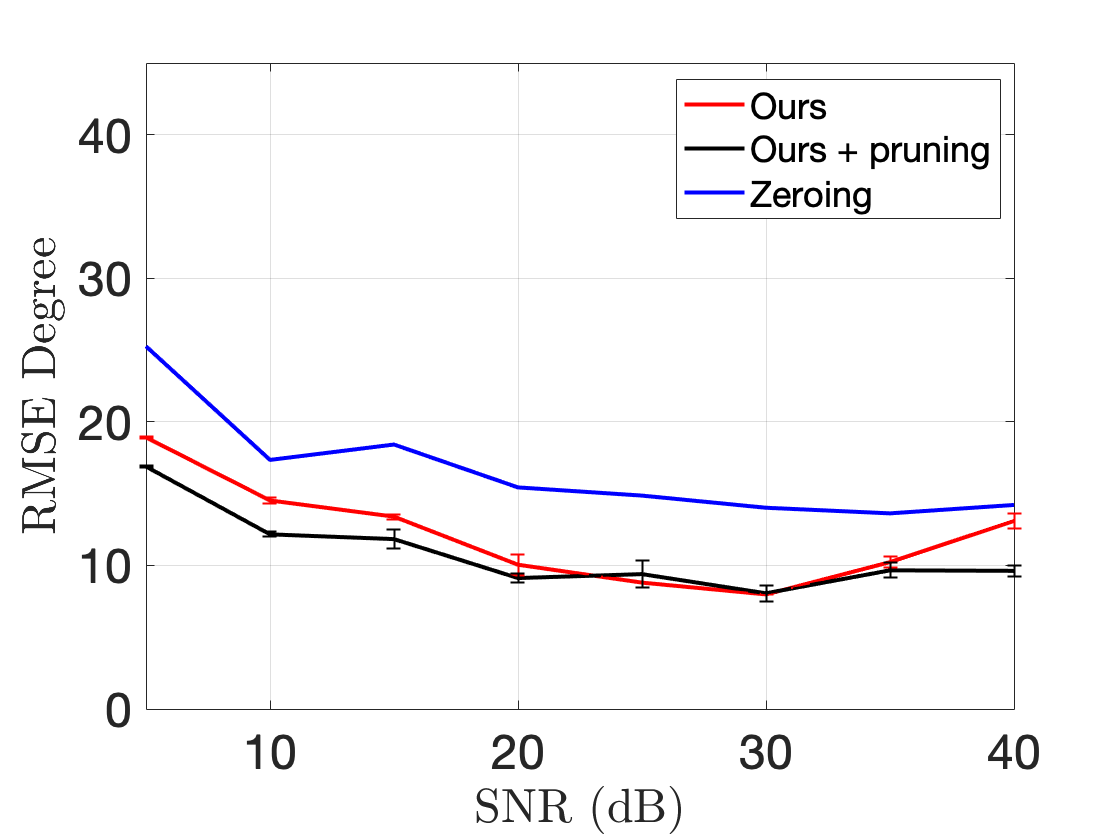} 
  \vspace{-0.55cm}
  \caption{}
  \label{subfig3}
\end{subfigure}
\begin{subfigure}{.33\textwidth}
  \centering
  \includegraphics[width=1.0\linewidth]{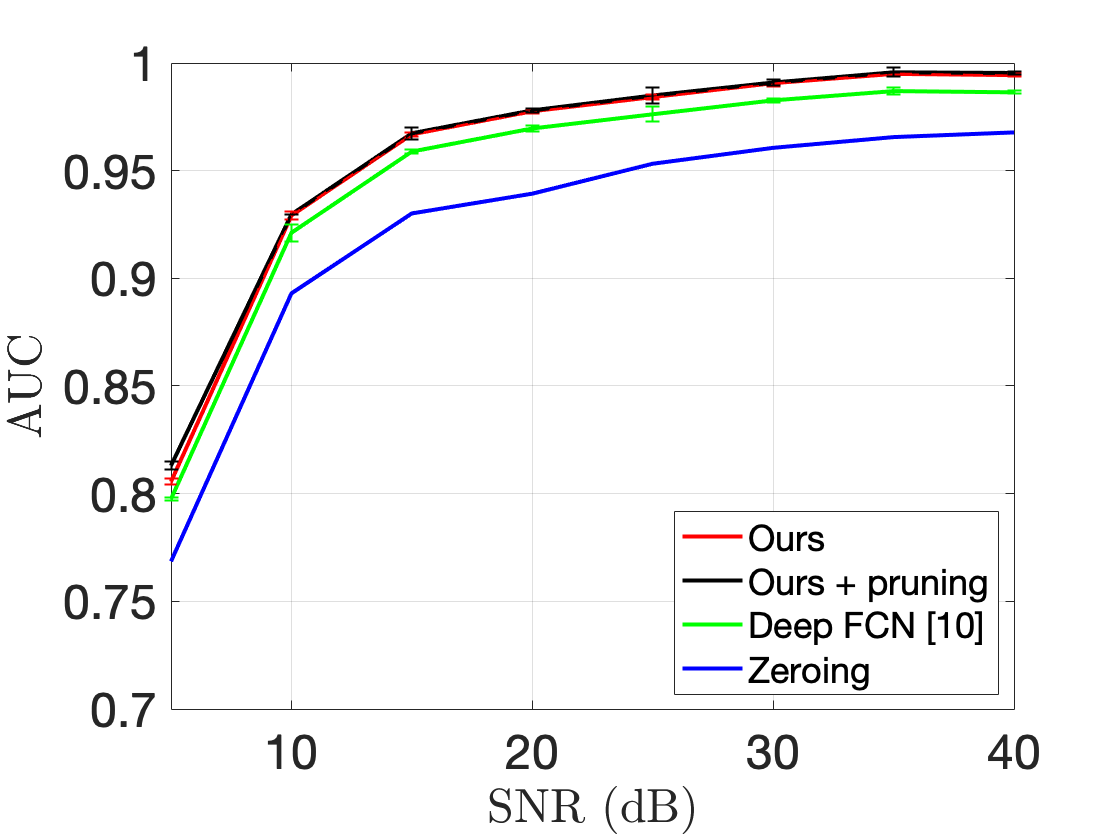}  
  \vspace{-0.55cm}
  \caption{}
  \label{subfig4}
\end{subfigure}
\begin{subfigure}{.33\textwidth}
  \centering
  \includegraphics[width=1.0\linewidth]{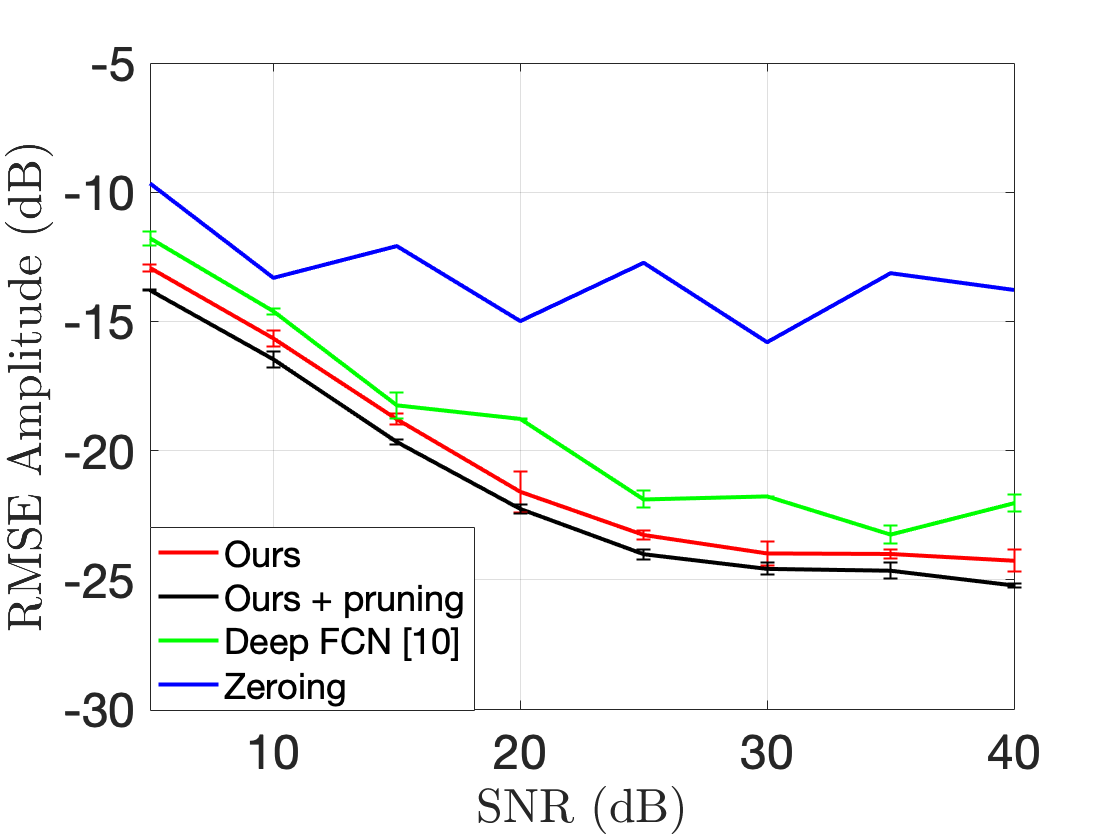} 
  \vspace{-0.55cm}
  \caption{}
  \label{subfig5}
\end{subfigure}
\begin{subfigure}{.33\textwidth}
  \centering
  \includegraphics[width=1.0\linewidth]{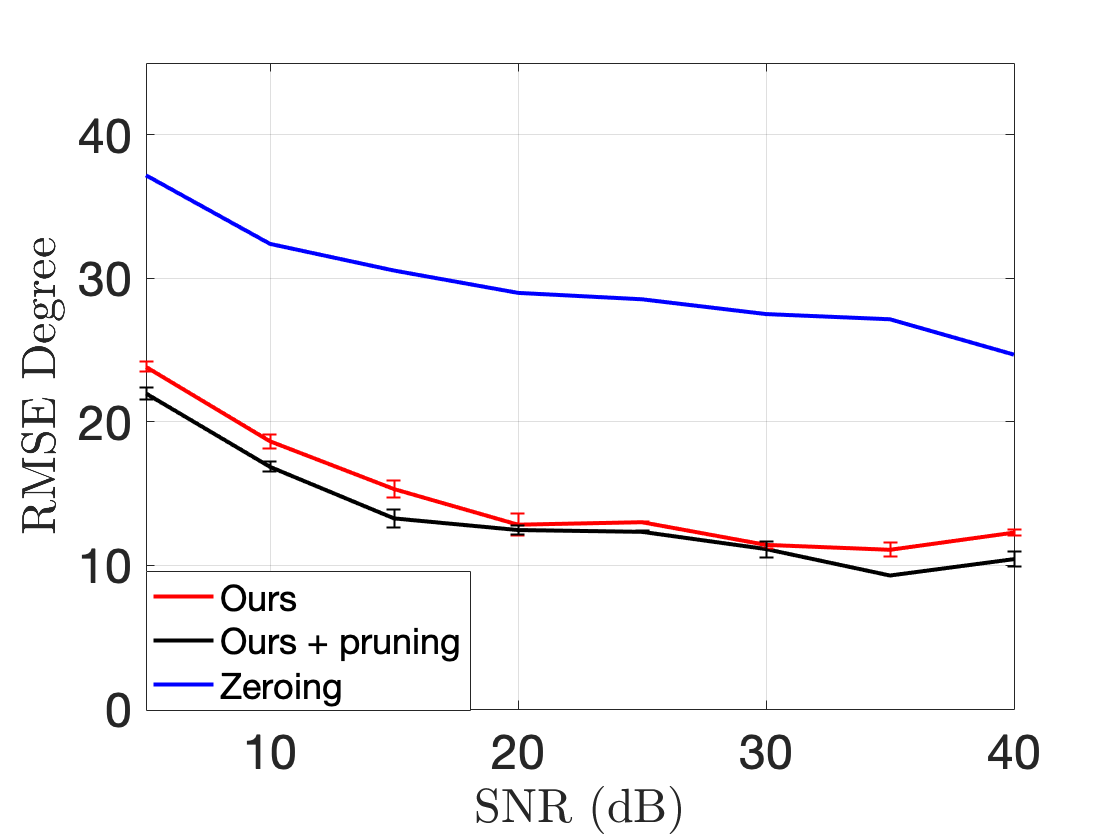} 
  \vspace{-0.55cm}
  \caption{}
  \label{subfig6}
\end{subfigure}
\begin{subfigure}{.33\textwidth}
  \centering
  \includegraphics[width=1.0\linewidth]{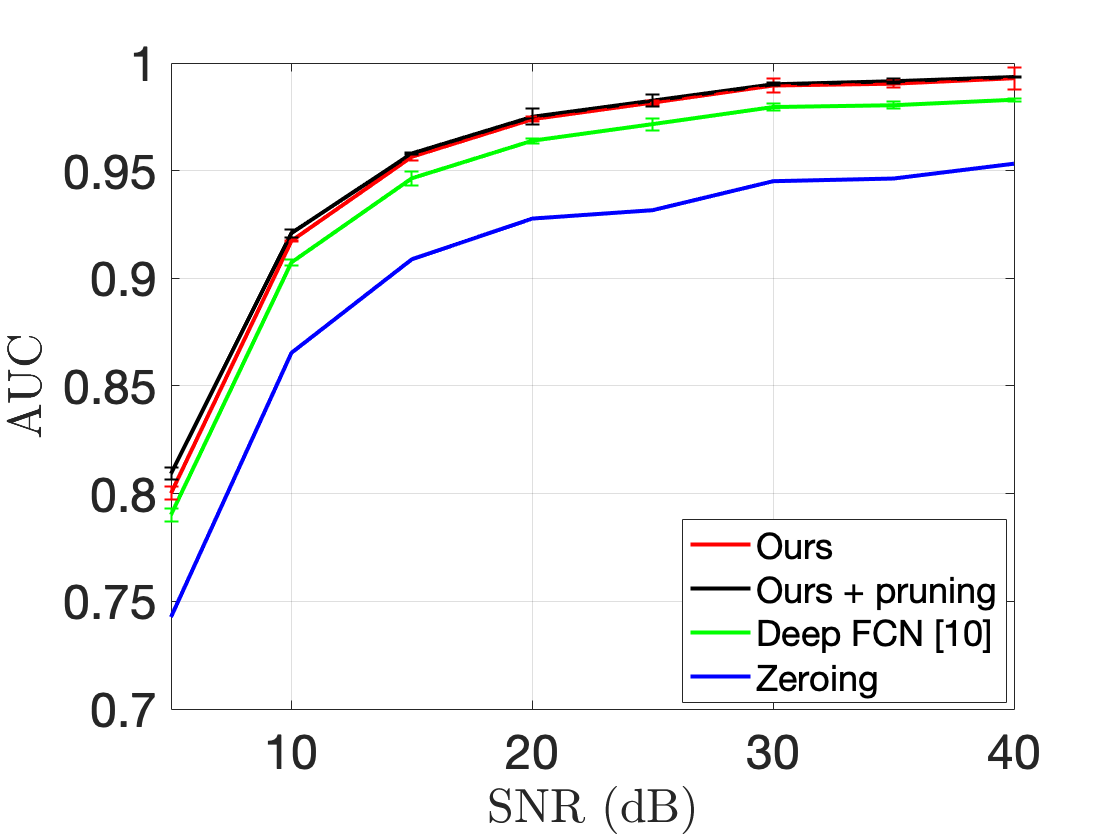} 
  \vspace{-0.55cm}
  \caption{}
  \label{subfig7}
\end{subfigure}
\begin{subfigure}{.33\textwidth}
  \centering
  \includegraphics[width=1.0\linewidth]{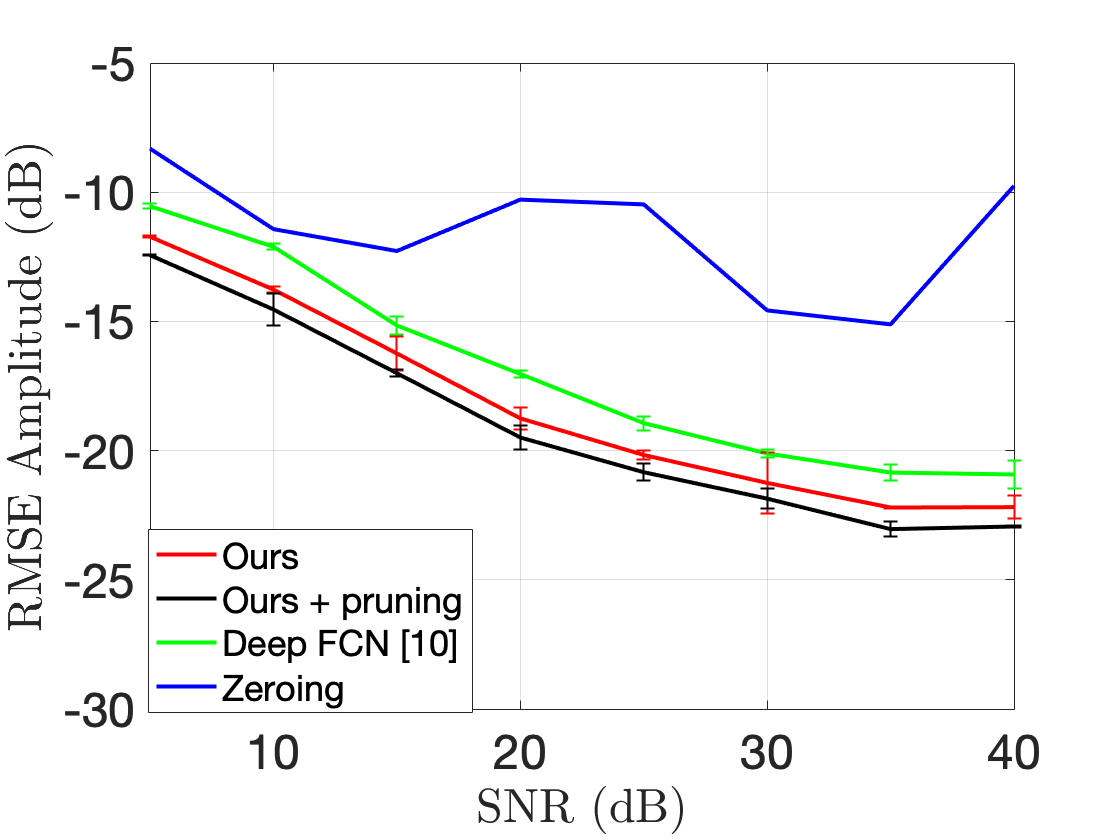}
  \vspace{-0.55cm}
  \caption{}
  \label{subfig8}
\end{subfigure}
\begin{subfigure}{.33\textwidth}
  \centering
  \includegraphics[width=1.0\linewidth]{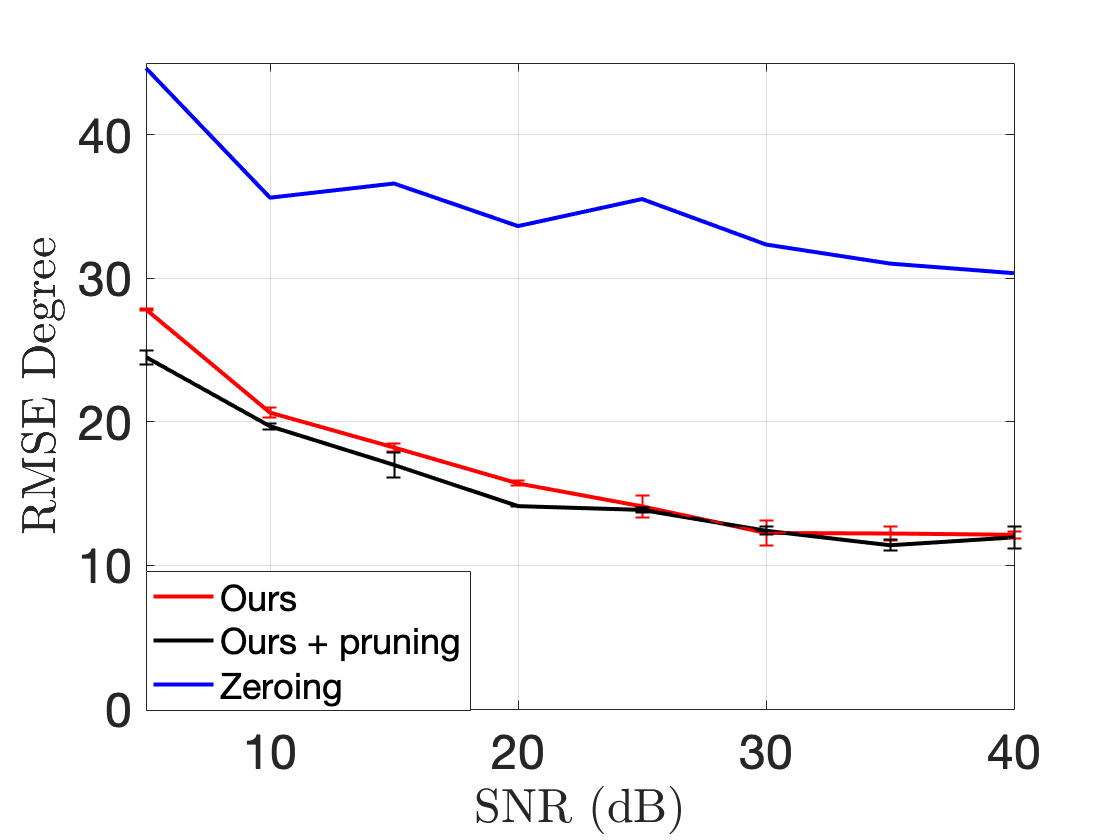}
  \vspace{-0.55cm}
  \caption{}
  \label{subfig9}
\end{subfigure}
\vspace{-0.1cm}
\caption{Results provided by our model trained with both conventional and weight pruning regimes against the zeroing baseline and the deep FCN from our preliminary work \cite{Ristea-VTC-2020}. Figures (a), (b) and (c) show results for one source of interference, figures (d), (e) and (f) show results for two sources of interference and figures (g), (h) and (i) show results attained for three sources of interference. Three different performance measures are presented: AUC, RMSE for the amplitude and RMSE for the phase. We added confidence intervals (based on the standard deviations for three runs) for the neural models. Best viewed in color.}
\label{statistics_fig}
\vspace{-0.3cm}
\end{figure*}

%\vspace{-0.2cm}
\subsection{Results on ARIM}

%We report results on both ARIM and ARIM-v2 data sets in order to prove that our approach attains better results than the baseline methods.

On the ARIM data set, which contains one interference source per data sample, we compared our FCN (considering both conventional and weight pruning regimes) with the oracle, the zeroing approach, the CNN proposed by Rock et al.~\cite{rock2019complex} and the FCN models proposed in our earlier work~\cite{Ristea-VTC-2020}. The results are presented in Table~\ref{tab_results_arim}. The oracle is a model based on ground-truth labels, which represents an upper bound for the other models. We used the same network architecture proposed by Rock et al.~\cite{rock2019complex}, but instead of range-Doppler maps, we trained the network with the STFT of radar signals.

A major drawback of the FCN models proposed in~\cite{Ristea-VTC-2020} is their inability to estimate the phase of signals, which is a mandatory quality, the phase being necessary in subsequent radar signal processing blocks. Even if our method attains a poor performance in terms of $\overline{\Delta \mbox{SNR}}$ compared with the Deep FCN, we outperform all methods regarding the other performance measures. On the test set, our FCN model trained with weight pruning surpasses the Deep FCN with 0.20 dB in terms of target amplitude MAE, as well as the zeroing baseline, with 1.49 degrees in terms of target phase MAE. In addition, we observe that weight pruning leads to slightly better results, sustaining the idea that noisy weights may alter the overall performance of the neural model.

%In order to show the necessity and the effectiveness of adding the magnitude channel besides the real and imaginary parts of the input, we removed the 
%absolute channel from both the input and the output domains and trained our network approach (FCN* from Table~\ref{tab_results_arim}). The results are illustrated in Table~\ref{tab_results_arim}. We can observe that the network performance has significantly dropped, attaining the weaker results compared with all methods. This enforces the idea that the magnitude channel could be seen as an attention mechanism which helps the network to focus on certain input locations. This experiment is not repeated on the ARIM-v2 data set since the results are predictable.

In order to show the necessity and the effectiveness of adding the magnitude channel besides the real and the imaginary parts of the input, we removed the absolute channel from the input and the output, obtaining an ablated FCN network (no magnitude). Its results are included in Table~{\ref{tab_results_arim}}. Without the magnitude channel, we observe that the network's performance drops by a significant margin, attaining weaker results compared with the other methods (even weaker than zeroing). This enforces the idea that the magnitude channel is a useful input channel, acting as an attention mechanism that helps the network to focus on relevant input locations.

In addition to the quantitative results presented so far, we illustrate a series of qualitative results on the ARIM test set in Figure~\ref{arim_data_fig}, comparing our approach against the zeroing method. The examples are vertically corespondent and they demonstrate that in certain conditions, for example when wide-length interference affects the signal, classical approaches, such as zeroing, fail to mitigate the interference and provide unsatisfying results. In Figure~\ref{arim_data_fig}, we observe that our model successfully produces signals that are very similar with the labels, while the zeroing method cannot perform the interference mitigation. All parameters are identical except for the parameter $k$ (the ratio between signal and interference slopes), which quantifies the length of interference with respect to the length of signal. More exactly, the closer $k$ is to $1$, the longer the interference is, i.e.~$k=1$ refers to a coherent interference.

\begin{figure*}[!t]
\begin{subfigure}{.33\textwidth}
  \centering
  \includegraphics[width=1.0\linewidth]{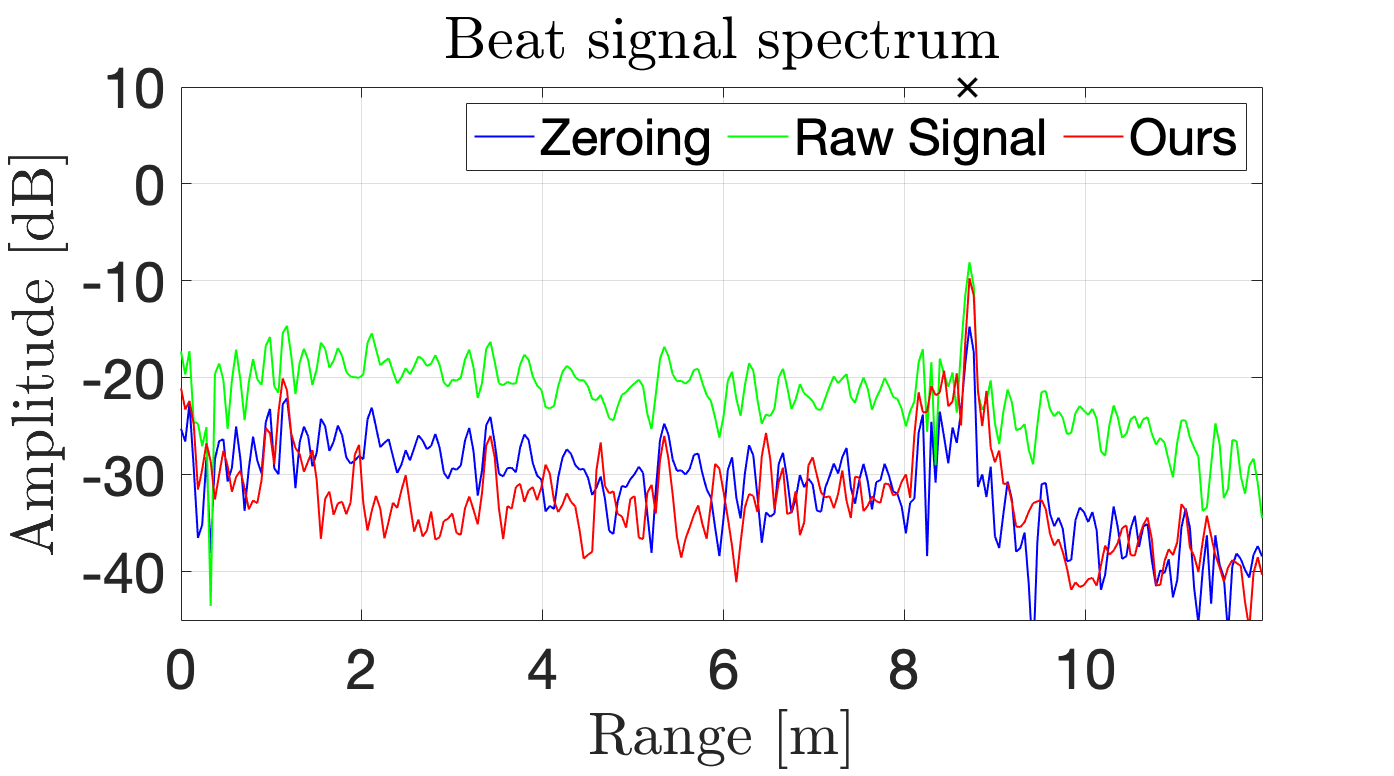}  
  \vspace{-0.55cm}
  \caption{$f_{0} = 76.25 GHz, B=1GHz$.}
  \label{real_subfig1}
\end{subfigure}
\begin{subfigure}{.33\textwidth}
  \centering
  \includegraphics[width=1.0\linewidth]{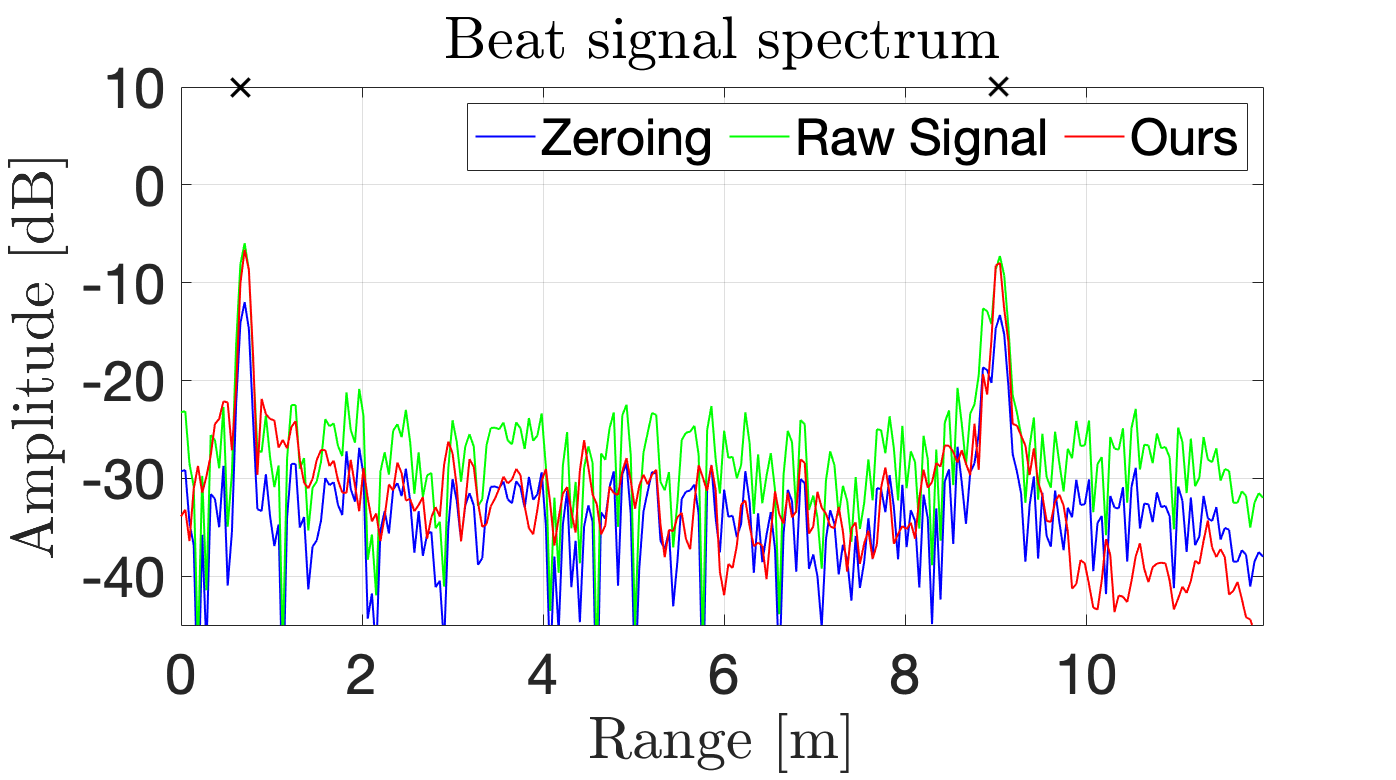}  
  \vspace{-0.55cm}
  \caption{$f_{0} = 76.50 GHz, B=1GHz$.}
  \label{real_subfig2}
\end{subfigure}
\begin{subfigure}{.33\textwidth}
  \centering
  \includegraphics[width=1.0\linewidth]{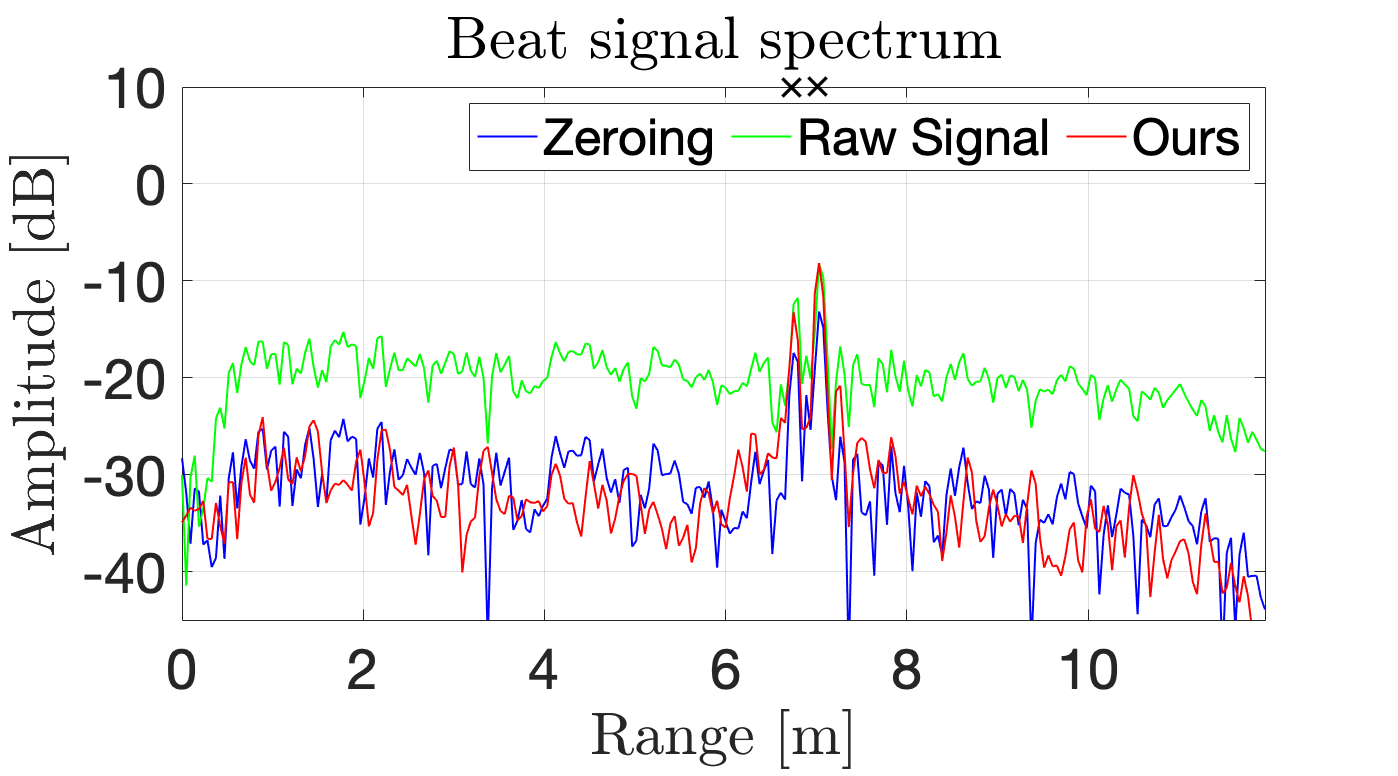}
  \vspace{-0.55cm}
  \caption{$f_{0} = 76.50 GHz, B=1GHz$.}
  \label{real_subfig3}
\end{subfigure}
\begin{subfigure}{.33\textwidth}
  \centering
  \includegraphics[width=1.0\linewidth]{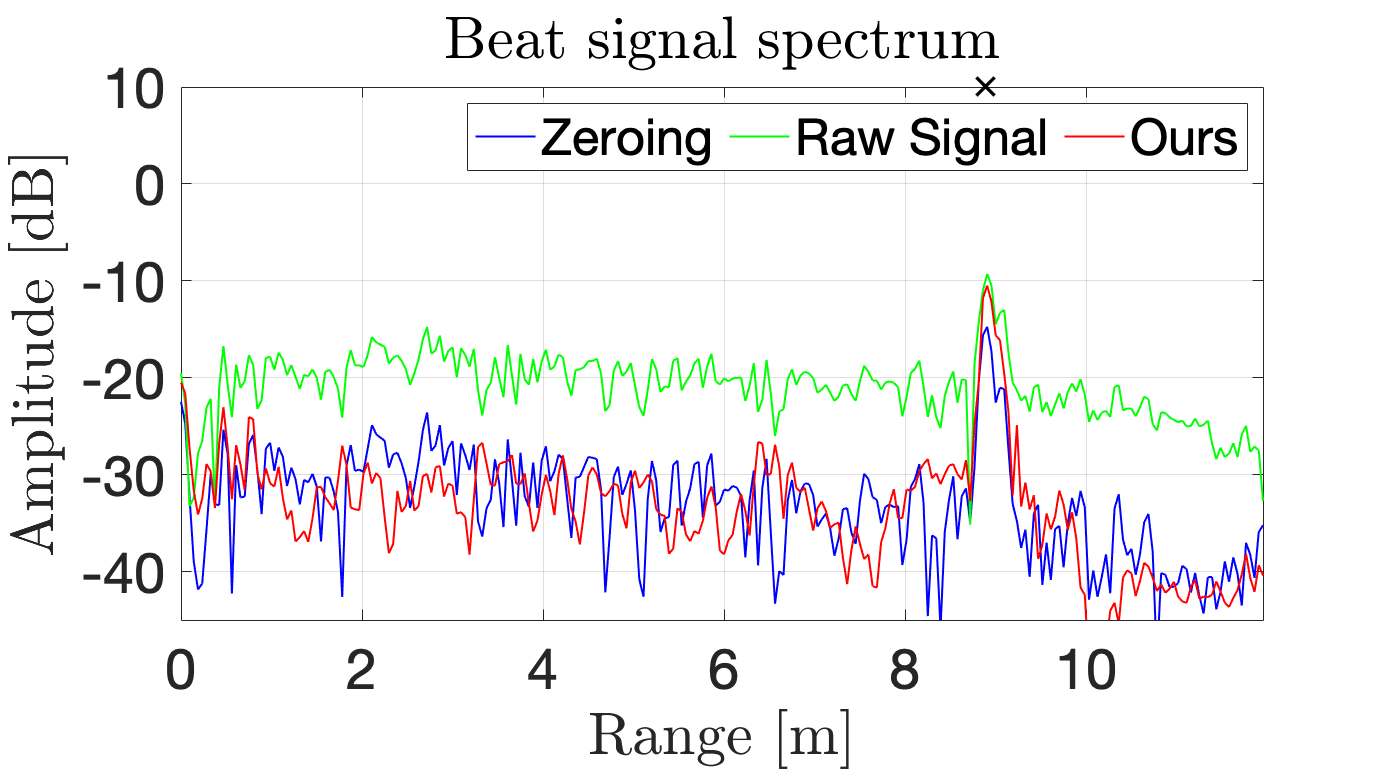} 
  \vspace{-0.55cm}
  \caption{$f_{0} = 76.75 GHz, B=1GHz$.}
  \label{real_subfig4}
\end{subfigure}
\begin{subfigure}{.33\textwidth}
  \centering
  \includegraphics[width=1.0\linewidth]{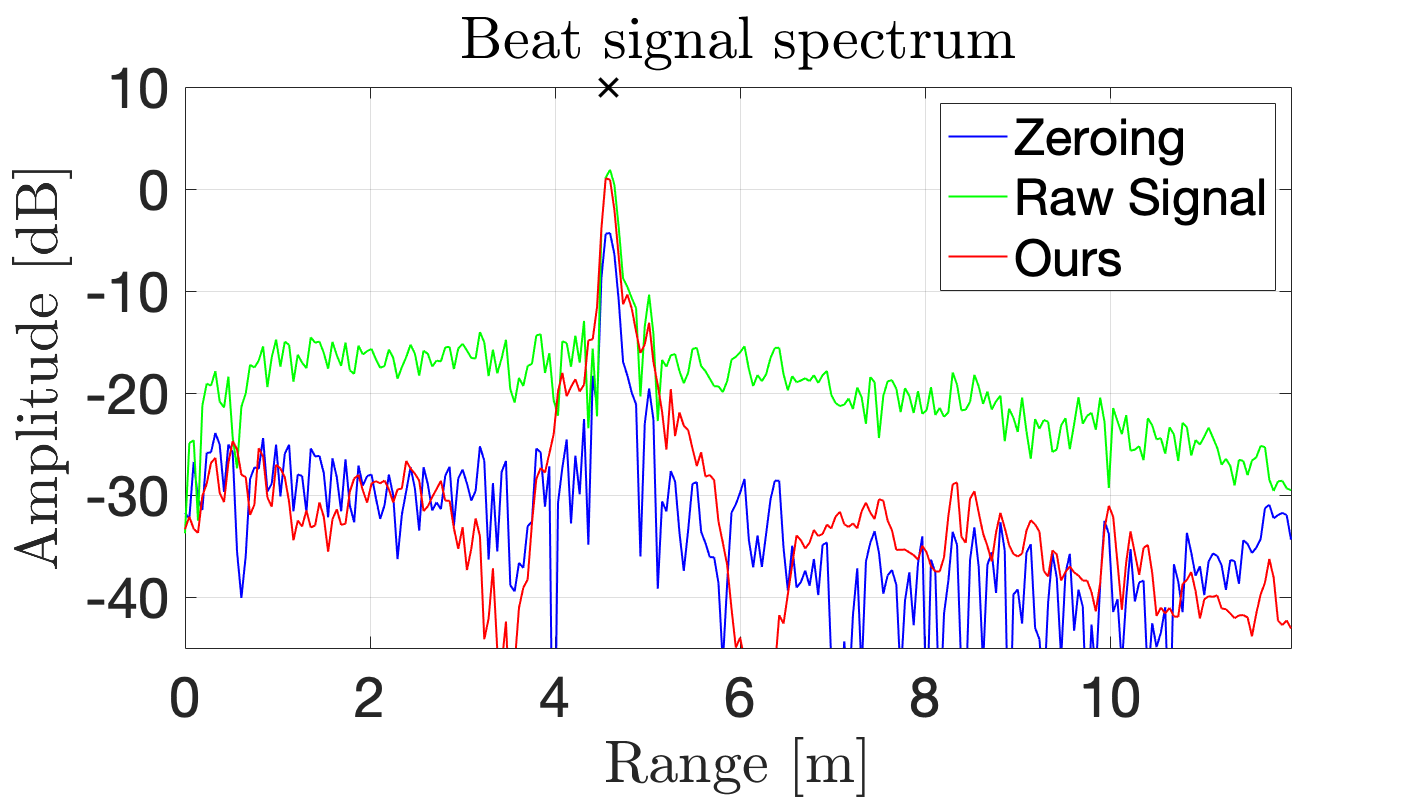} 
  \vspace{-0.55cm}
  \caption{$f_{0} = 76.75 GHz, B=1GHz$.}
  \label{real_subfig5}
\end{subfigure}
\begin{subfigure}{.33\textwidth}
  \centering
  \includegraphics[width=1.0\linewidth]{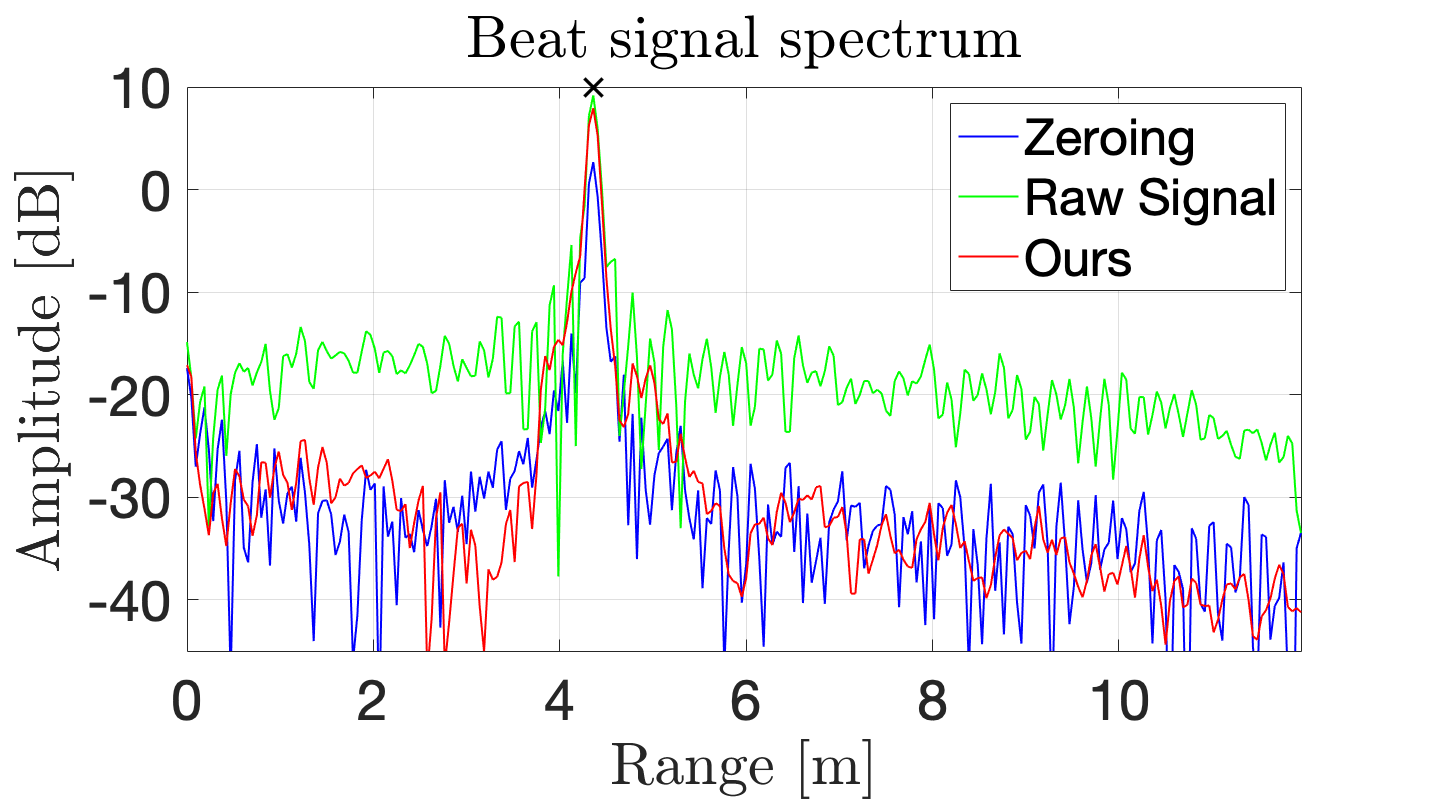} 
  \vspace{-0.55cm}
  \caption{$f_{0} = 76.25 GHz, B=1GHz$.}
  \label{real_subfig6}
\end{subfigure}
\begin{subfigure}{.33\textwidth}
  \centering
  \includegraphics[width=1.0\linewidth]{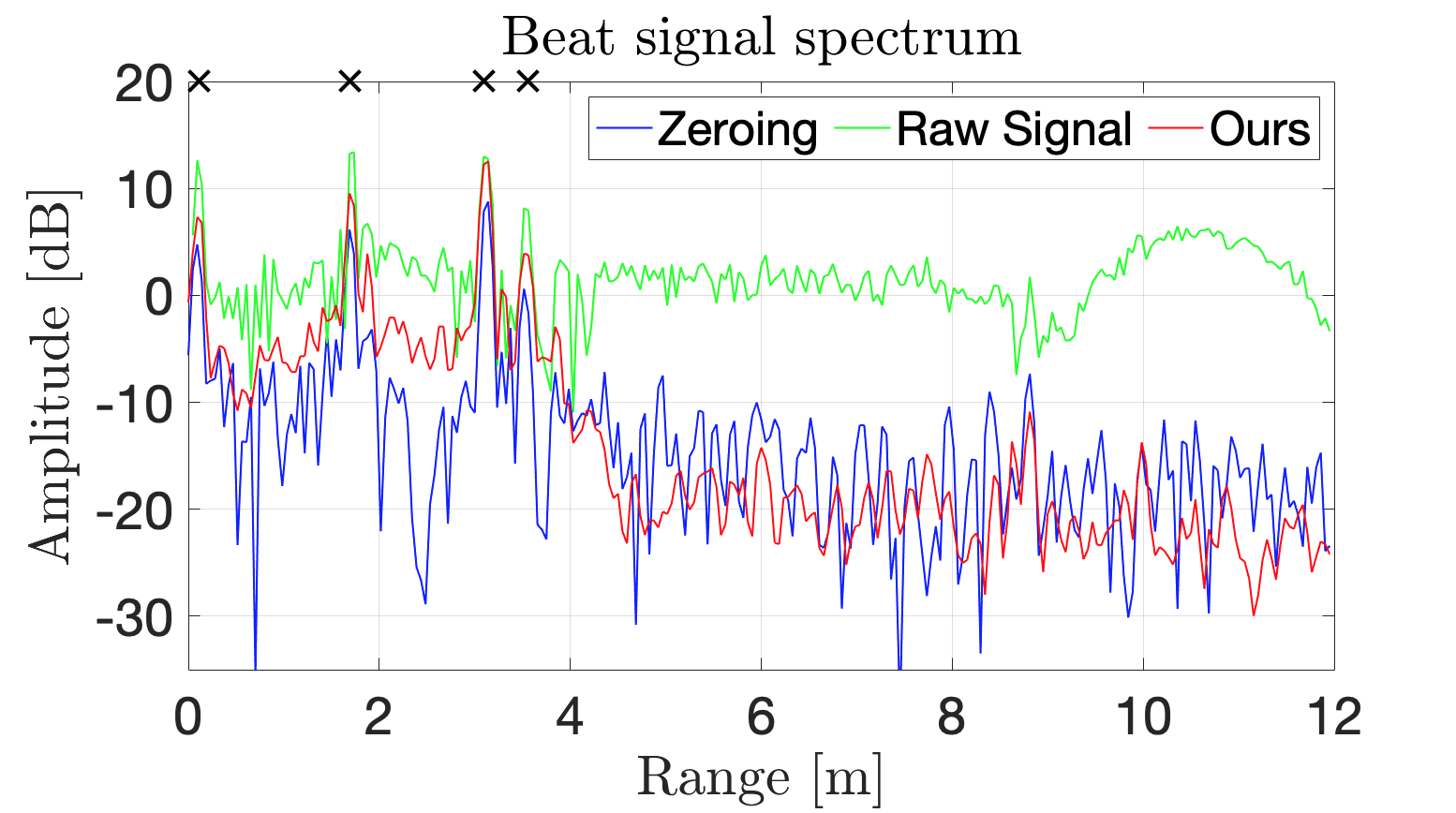}  
  \vspace{-0.55cm}
  \caption{$f_{0} = 77 GHz, B=1.6GHz$.}
  \label{real_subfig7}
\end{subfigure}
\begin{subfigure}{.33\textwidth}
  \centering
  \includegraphics[width=1.0\linewidth]{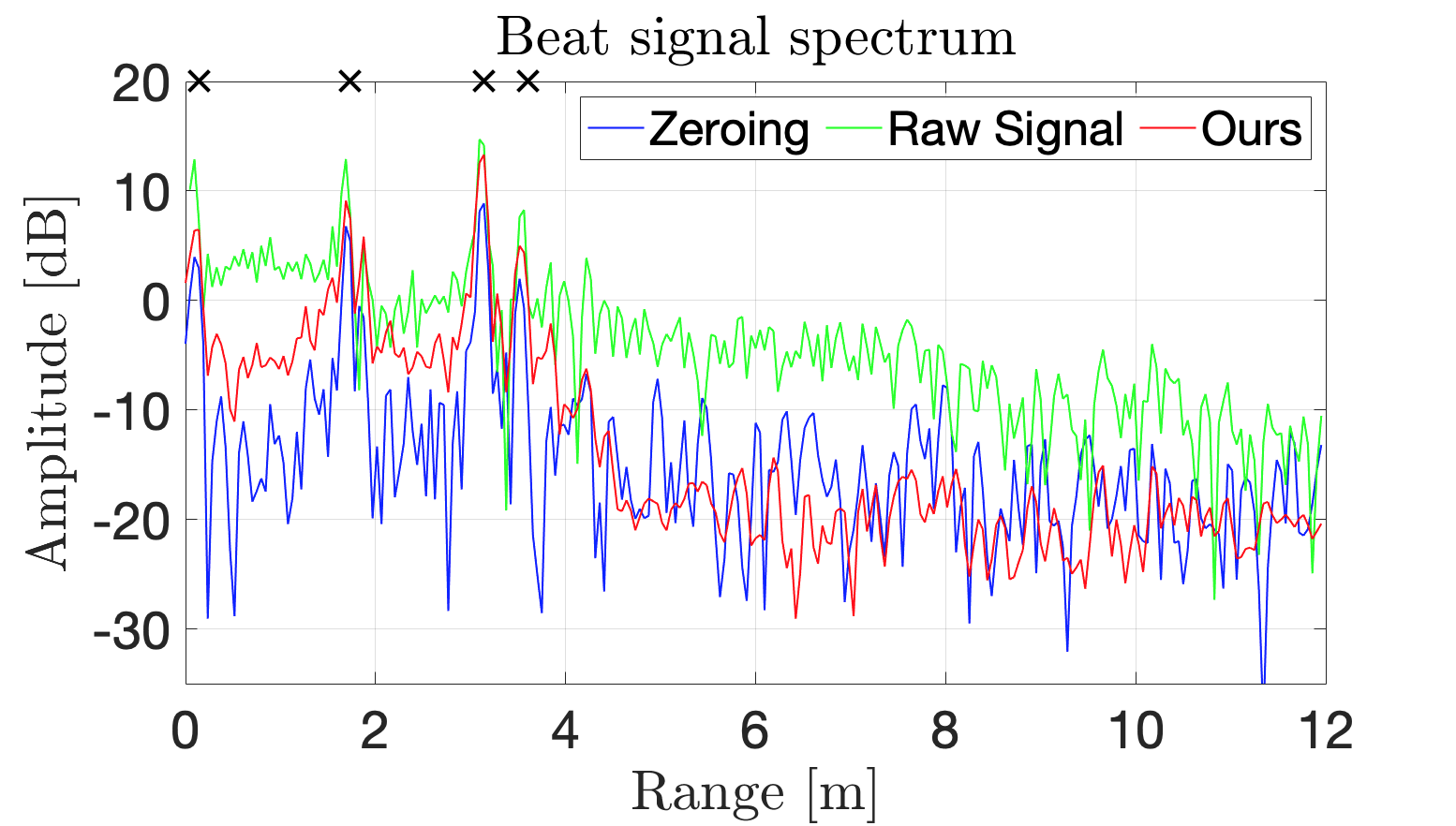}  
  \vspace{-0.55cm}
  \caption{$f_{0} = 77 GHz, B=1.6GHz$.}
  \label{real_subfig8}
\end{subfigure}
\begin{subfigure}{.33\textwidth}
  \centering
  \includegraphics[width=1.0\linewidth]{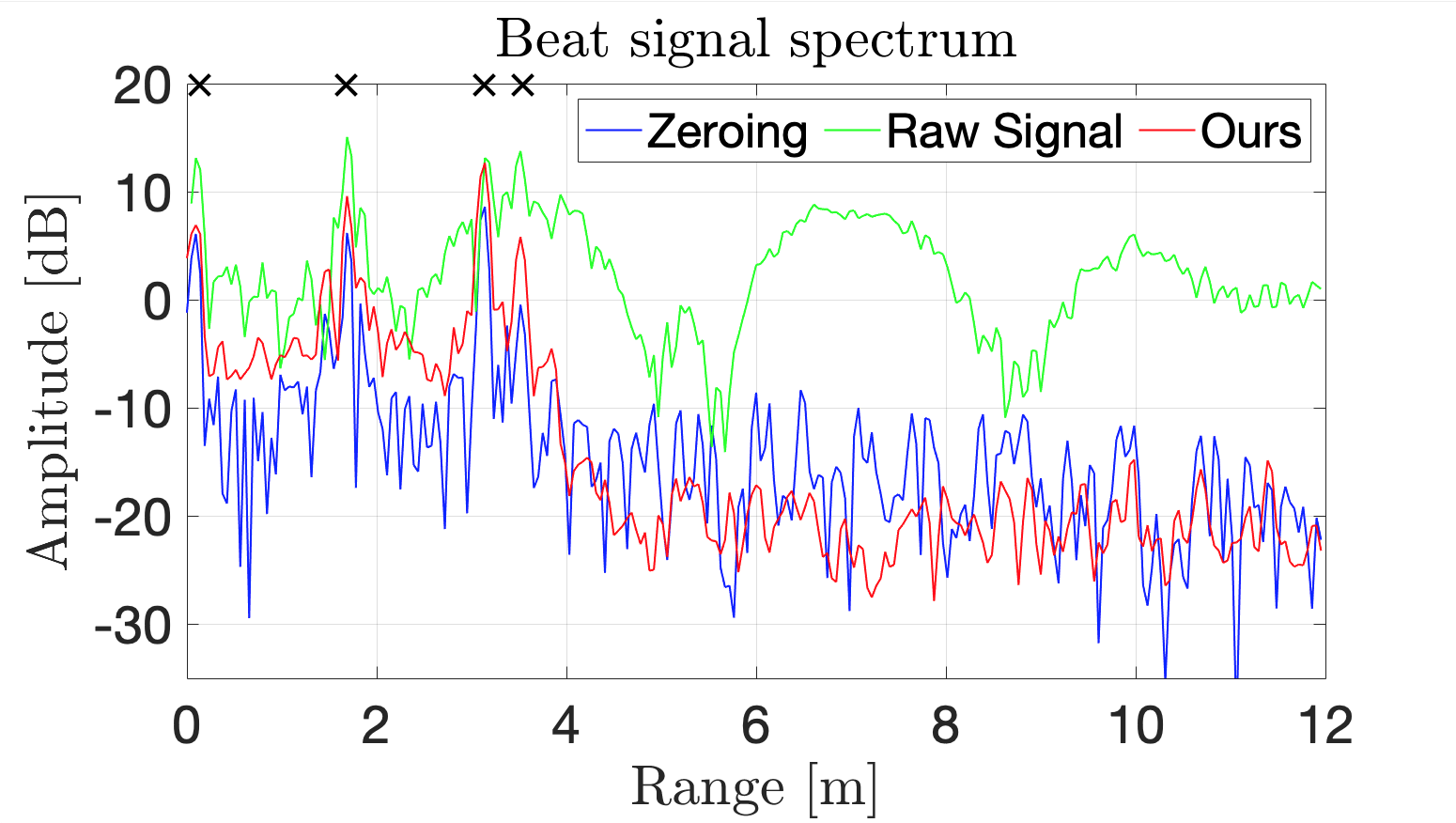}  
  \vspace{-0.55cm}
  \caption{$f_{0} = 77 GHz, B=1.6GHz$.}
  \label{real_subfig9}
\end{subfigure}
\vspace{-0.1cm}
\caption{Qualitative results provided by our FCN+pruning model in comparison with the zeroing method. The real signals with interference are acquired with two different automotive radar sensors. Data acquisition was made as follows: (a)-(f) with radar sensor from FAU and (g)-(i) with NXP TEF810X 77 GHz radar transceiver, respectively. The parameter $f_{0}$ refers to the interference central frequency and $B$ refers to the chirp bandwidth. Best viewed in color.}
\label{real_data_fig}
\vspace{-0.3cm}
\end{figure*}

%\vspace{-0.2cm}
\subsection{Results on ARIM-v2}

On the ARIM-v2 data set, which contains up to three interference sources per data sample, we compared our FCN (considering both conventional and weight pruning regimes) with the oracle (based on ground-truth labels), the zeroing baseline and the CNN of Rock et al.~\cite{rock2019complex}. The results reported in Table~\ref{tab_results_arim_v2} show that our approach provides superior results for all metrics, attaining performance levels quite close to the oracle. The differences between zeroing and our FCN on the ARIM data set become undoubtedly higher on the ARIM-v2 data set, because ARIM-v2 simulates a more difficult automotive scenario, in which a conventional method such as zeroing seems to fail to mitigate multiple sources of interference. Our FCN model attains almost half the error reached by zeroing, in terms of target phase MAE. Furthermore, our model estimates the amplitudes of targets with 0.86 dB better than zeroing on the test set. Another remarkable difference can be seen for the mean SNR improvement, where our network obtained a score of 15.36 dB, which is better than zeroing by 6.42 dB. In addition, we note that weight pruning leads to consistent improvements for all metrics, which seems to be considerably more important on ARIM-v2 than on ARIM. With or without pruning, our method also surpasses the state-of-the-art CNN of Rock et al.~\cite{rock2019complex}. In terms of the amplitude MAE, even the zeroing baseline outperforms the CNN of Rock et al.~\cite{rock2019complex}. In summary, the quantitative results demonstrate the superiority of our method.

In addition to the quantitative results presented so far, we illustrate a series of qualitative results on the ARIM-v2 test data set in Figure~\ref{arim-v2_data_fig}, comparing our approach against the zeroing method. Due to the fact that data samples are synthetically generated, we are able to compare the algorithms with the ground-truth signal without interference, allowing us to determine which method provides the desired result after interference mitigation. The plots depicted in Figure~\ref{arim-v2_data_fig} are vertically corespondent, meaning that, in the top plot on a column, the interference is mitigated by zeroing, while in the bottom plot, the same interference is mitigated by our FCN model. We handpicked three examples with multiple sources of interference, a type of incident that may occur in a real-life automotive scenario. We observe that, in this particular case, when there are multiple sources of interference, the zeroing approach fails to mitigate the interference and the targets can be barley observed because of the raised noise floor. Our model successfully mitigates the interference, providing an output very similar to the label. Although our model shows similar performance to baseline approaches when signals are affected by an interference with narrow length, in a difficult scenario, with multiple sources of interference or with wide-length interference, our approach clearly outperforms approaches such as zeroing, as it results from the plots presented in Figure~\ref{arim_data_fig} and Figure~\ref{arim-v2_data_fig}.

In order to provide a more detailed picture of our quantitative results, in Figure~\ref{statistics_fig}, we illustrate how our approach compares to zeroing in terms of three performance metrics (AUC, amplitude RMSE and phase RMSE) considering one, two and three sources of interference, from top to bottom, respectively. We observe that the differences between our FCN models and zeroing grows along with the number of interference sources, in favor of our approach, 
considering all performance measures.
We notice an important difference when we consider the RMSE on the phase of targets. The zeroing algorithm exhibits poor performance because, when there are multiple sources of interference, a substantial part of the signal is covered by interference. Therefore, we observe a substantial difference between our FCN models and zeroing. In Figure~{\ref{statistics_fig}}, we also illustrate the results of the deep FCN from our preliminary work~\mbox{\cite{Ristea-VTC-2020}}, excluding it from the graphs depicting the phase RMSE, since the deep FCN is not capable of recovering the phase. As the number of interference sources grows, we observe an increasing gap between our FCN+pruning and the deep FCN. Certainly, the gap is in favor of our method. We thus conclude that our current neural model is superior.

\begin{figure*}[!t]
\begin{center}
\centerline{\includegraphics[width=0.9\linewidth]{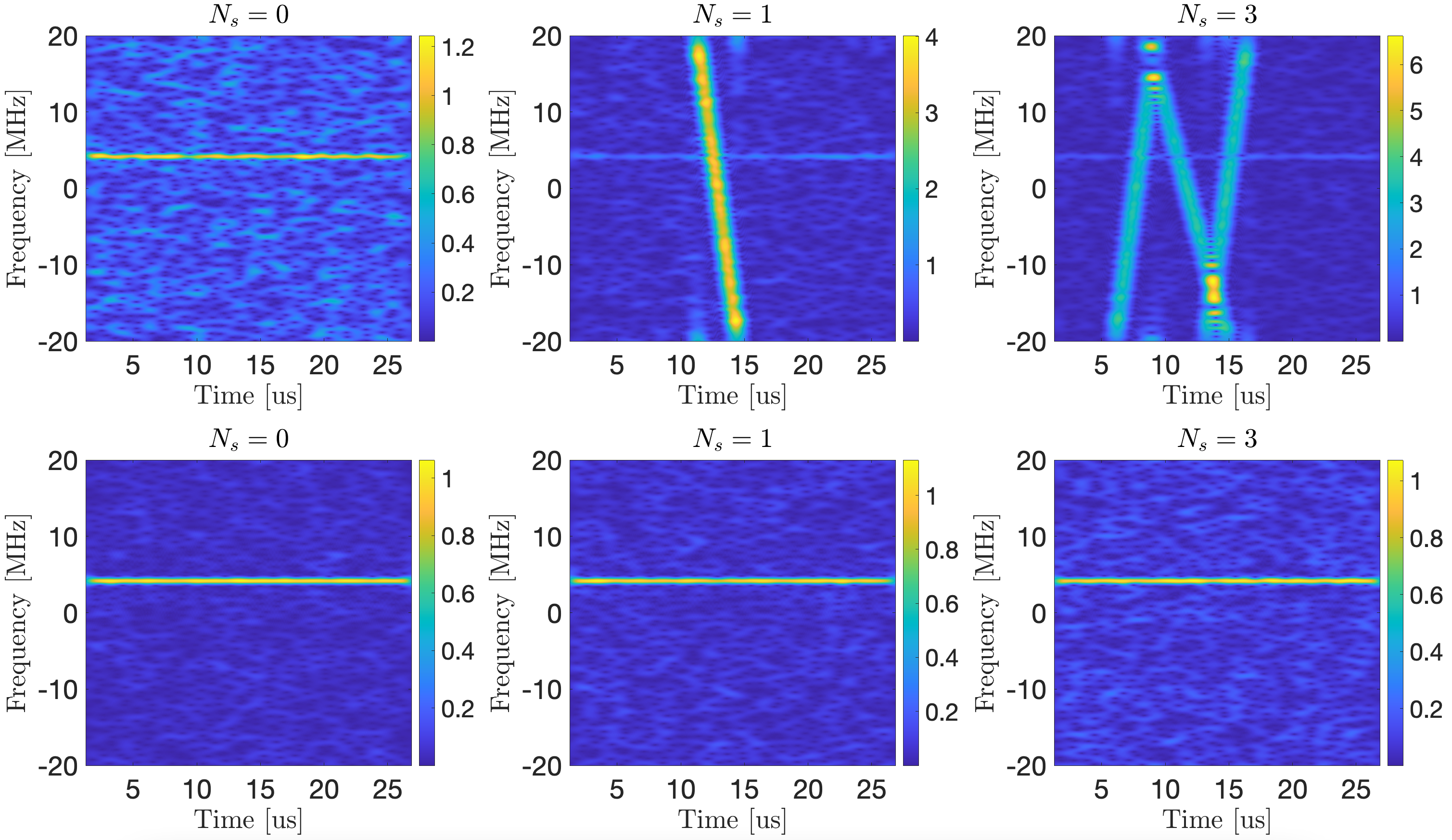}}
\caption{Qualitative results provided by our FCN+pruning model on ARIM-v2. We illustrate the input STFT (top row) and the output of the network processed with an inverse FFT and a STFT (bottom row), to have the same visualization. $N_s$ represents the number of interference sources. In all images, there is a single target (horizontal line) in the same position. Best viewed in color.}
\label{fig_correction}
\vspace{-0.5cm}
\end{center}
\end{figure*}

In Figure~\ref{fig_correction}, we added some visual results to observe the network's capacity to reduce noise and mitigate radar interference on ARIM-v2. In order to compare STFT data, we processed the output of the network by performing an inverse FFT, followed by a STFT. In this manner, we are able to reconstruct the STFT and compare it with the input STFT. We can observe that for no interference source ($N_s = 0$), the network acts like a denoising model and does not affect the target. When we feed STFT data affected by interference into the network, we observe that the interference is completely mitigated, even if input data is affected by multiple interference sources ($N_s=3$).

\begin{figure*}[!t]
\begin{center}
\centerline{\includegraphics[width=0.9\linewidth]{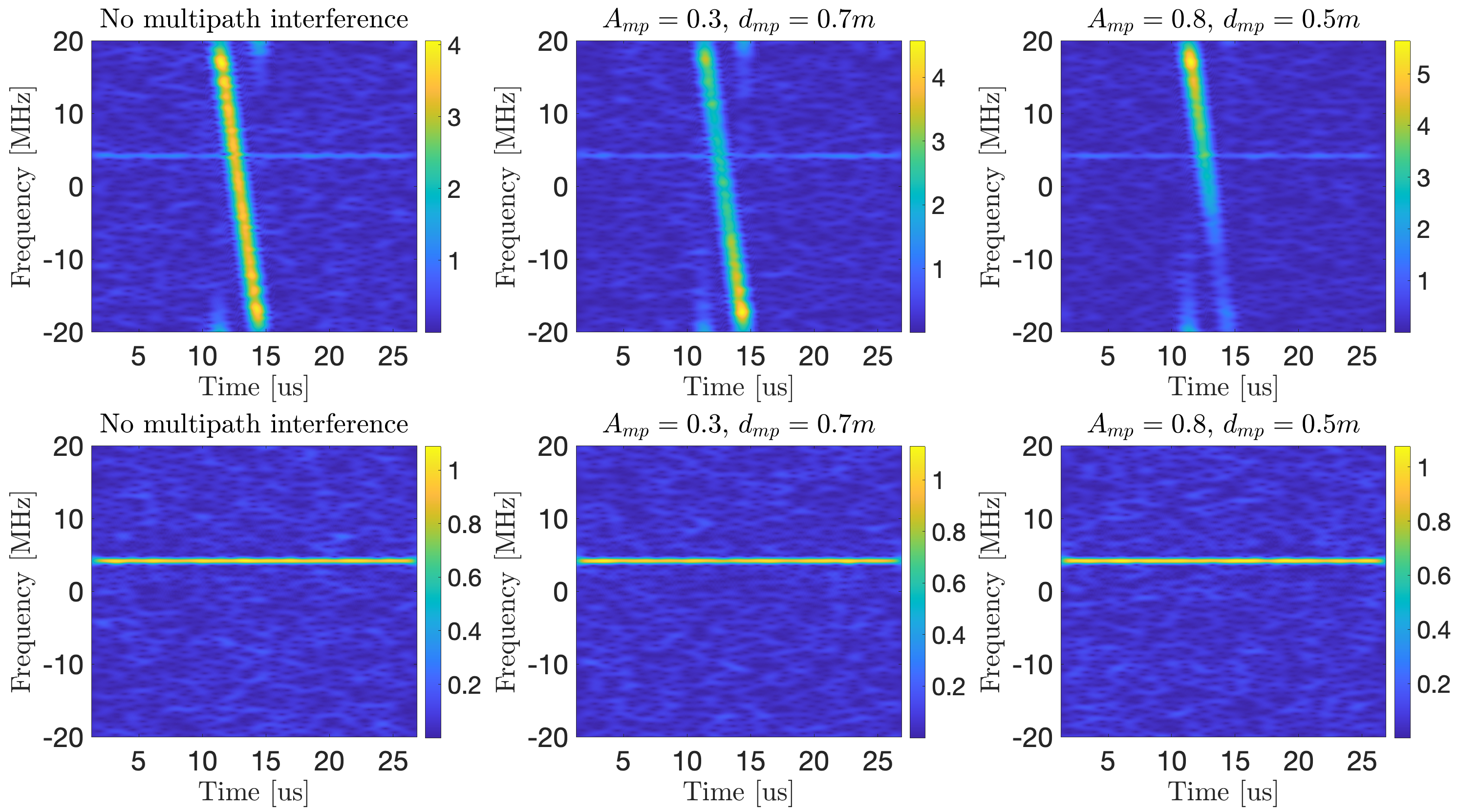}}
\caption{Qualitative results provided by our FCN+pruning model on synthetically generated data for multipath interference propagation. On the first row, there are signals affected by interference, and on the second row, the interference is mitigated with our network. $A_{mp}$ stands for the amplitude of the reflected interference and $d_{mp}$ stands for the path difference for the reflected interference with respect to the direct path. In all images, there is a single target (horizontal line) in the same position. Best viewed in color.}
\label{fig_multipath}
% \vspace{-0.5cm}
\end{center}
\end{figure*}

In order to analyze more complex interference scenarios, we synthetically generated samples with a single source of interference, while modeling the multipath propagation of the interference signal. The multipath simulation was performed by summing  the same interference signal to the signal affected by interference, with a delay and a different amplitude. The delay was chosen to correspond to $0.5$ and $0.7$ meters and the reflection amplitude was considered $0.3$ and $0.8$, in order to simulate both weak and strong reflectors. The results are shown in Figure~\ref{fig_multipath}. We can observe that our model successfully mitigates the interference, even if it corresponds to a multipath propagation.

%\vspace{-0.2cm}
\subsection{Generalization to Real Data}

\begin{figure*}[!t]
\begin{center}
\centerline{\includegraphics[width=0.9\linewidth]{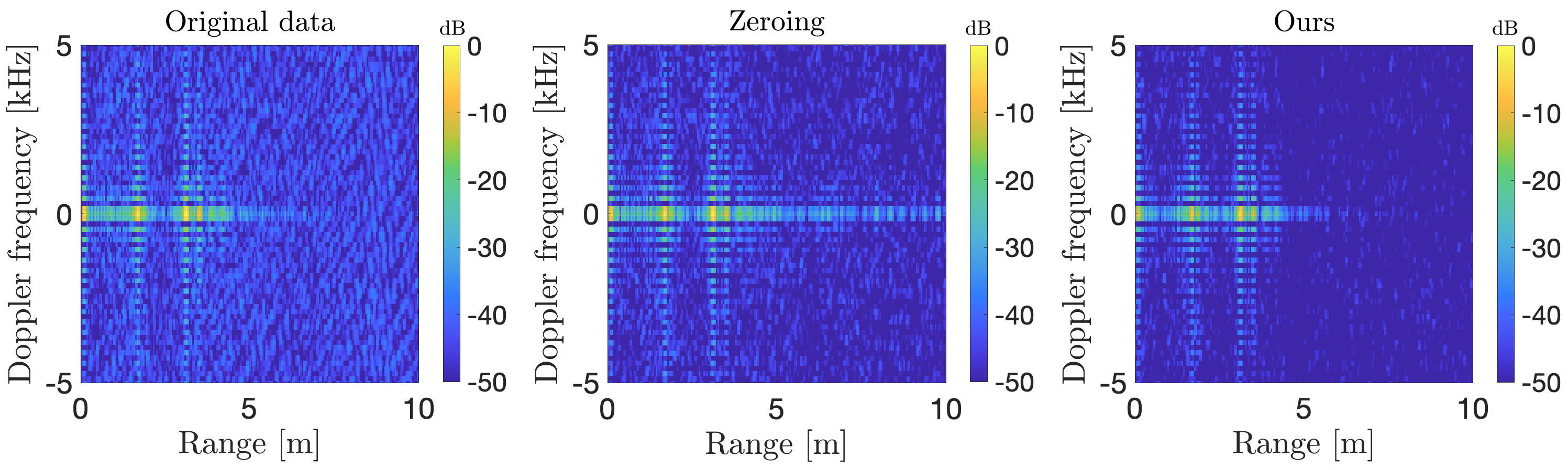}}
\caption{Range-Doppler maps for signals acquired with the NXP TEF810X 77 GHz radar transceiver. Left: the range-Doppler map for the original raw signals. Center: the range-Doppler map for signals processed with the zeroing method. Right: the range-Doppler map for signals processed with our FCN trained with pruning. Best viewed in color.}
\label{fig_doppler}
\vspace{-0.55cm}
\end{center}
\end{figure*}

The major concern regarding training a neural network on synthetically generated samples is the model's capacity to generalize to real data. Therefore, we evaluate the generalization capacity of our FCN on real data, by testing it on real samples collected with two different radar sensors. We underline that our FCN is never trained or fine-tuned on real data samples. In Figure~\ref{real_data_fig}, we present qualitative results on nine real samples with interference, comparing our method against zeroing. 

The first six plots, depicted in Figures~\ref{real_subfig1} to \ref{real_subfig6}, are generated with real data provided by FAU \cite{Fuchs2020}. We note that the targets are different among the presented signals, showcasing various scenarios. Moreover, the central frequency of the interference source is not always the same, having three distinct values: $76.25$ GHz, $76.5$ GHz and $76.75$ GHz. Looking at the results, it is clear that our network can provide more accurate estimations of the amplitude of targets, being able to mitigate the interference and to reduce the noise floor.

The last three plots, depicted in Figures~\ref{real_subfig7} to \ref{real_subfig9}, are made on data provided by the NXP company, which were captured with the NXP TEF810X 77 GHz radar transceiver in a couple of outdoor experiments on a two-lane road. The victim radar was mounted on the bumper of a car, while the interfering radar was mounted on a tripod in a fixed location outside the roadway. The main target was another moving car on the road. Besides the car, there were other reflections in the range profiles coming from surrounding targets (e.g., lighting poles, trees).
Even if the interference is more visible in these examples, our approach successfully mitigates the interference, providing better results in terms of amplitude of targets compared to the zeroing algorithm.

We highlight that the real data samples are collected with different radar sensors and have distinct central frequencies. Nevertheless, our model is able to mitigate the interference and surpass the baseline method, without any adjustment or fine-tuning. This demonstrates that our model has a good generalization capacity, being applicable to a wide range of radar sensors, without requiring any additional effort.

In addition to range profile processing, we tested the capacity of our network to clean real range-Doppler profiles, by processing separately each range profile and then concatenating them. We computed the range-Doppler experiment on data from the NXP company and tested our method against the zeroing baseline. The results are shown in Figure~\ref{fig_doppler}. We can observe that our FCN trained with pruning is able to better clean the range-Doppler map in comparison with the zeroing method.

\begin{table}
\centering
\noindent
\caption{Results provided by our FCN model (trained with both conventional and weight pruning regimes), on a generated test data set of radar signals with 4 to 6 interference sources, versus the oracle and the zeroing baseline. The best results (excluding the oracle) are highlighted in bold. The symbol $\uparrow$ means higher values are better and $\downarrow$ means lower values are better.}
\setlength\tabcolsep{4.5pt}
\begin{tabular}{|l|cccc|}
\hline
%\multicolumn{1}{N }{\textbf{}} & \multicolumn{4}{c }{\textbf{Evaluation}} & \multicolumn{4}{c }{} \\  
 & & & {MAE}$\downarrow$ & {MAE}$\downarrow$ \\
{Method} & $\overline{\Delta \mbox{SNR}}\uparrow$ & {AUC}$\uparrow$ & Amplitude & Phase \\
 &  & & {(dB)} & {(degrees)} \\ 
\hline
\hline
{Oracle (true labels)} & 20.34 & 0.970 & 0 & 0\\ 
\hline
{Zeroing} & 7.09 & 0.864 & 4.07 & 24.08 \\ 
\hline
{FCN (ours)} & 14.76 & 0.939 & 2.71 & 12.20\\ 
{FCN + pruning (ours)} & \textbf{15.13} & \textbf{0.942} & \textbf{2.55} & \textbf{11.27}\\ 
\hline
\end{tabular}
\label{tab_results_generalization}
\vspace{-0.3cm}
\end{table}

%\vspace{-0.2cm}
\subsection{Generalization to More Interference Sources}

In real automotive scenarios, a wide range of incidents may cause the radar sensor to fail during driving. A plausible situation could be that, in a specific moment, more interference sources affect the radar antenna. Therefore, we investigate the generalization capacity of our model to mitigate RFI from more sources than it was trained for. In this scope, we synthetically generated an additional test data set of 2,400 samples with four, five and six interference sources. We consider our FCN models trained on ARIM-v2 with both conventional and weight pruning regimes, resulting in an out-of-distribution evaluation setting. The results attained by our FCN models are compared with the oracle and the zeroing method. As shown in Table~{\ref{tab_results_generalization}}, our approach clearly outperforms the zeroing algorithm, being the closest method to the oracle. In terms of target phase MAE, our FCN based on weight pruning attains results with $12.81$ degrees better than zeroing. Moreover, the $\overline{\Delta \mbox{SNR}}$ is almost double for both FCN models compared to the zeroing baseline. Regarding the AUC, a measure which is very important in radar applications because it describes the ability to disentangle targets from noise, our best model has an improvement of $7.8\%$ compared to zeroing. In addition, we notice that weight pruning attains better performance compared to the conventional training regime, even when we test the generalization capacity on out-of-distribution data. This further supports our claim that weight pruning can act as a regularization method.

%\vspace{-0.2cm}
\subsection{Generalization to More Targets}

Another less expected situation that can occur in real automotive scenarios is generated by the presence of a multitude of targets in the same range profile. To demonstrate the capacity of our network to generalize to such situations, we generated an additional synthetic test set of 2,400 samples, such that each sample contains a randomly chosen number of targets between 5 to 10. Our network trained with weight pruning attains the best performance, as shown in Table~{\ref{tab_results_generalization_multiple_targets}}. Our best model has an improvement of 7.55 dB in terms of $\overline{\Delta SNR}$ in comparison with the zeroing method. Moreover, the MAE of the target's phase is reduced by half for our best model, when we take the zeroing baseline as reference. Once again, the efficiency of weight pruning is highlighted by the results, as it surpasses the conventional training method in terms of all metrics.

\begin{table}
\centering
\noindent
\caption{Results provided by our FCN model (trained with both conventional and weight pruning regimes), on a generated test data set of radar signals with a number of targets between 5 and 10, versus the oracle and the zeroing baseline. The best results (excluding the oracle) are highlighted in bold. The symbol $\uparrow$ means higher values are better and $\downarrow$ means lower values are better.}
\setlength\tabcolsep{4.5pt}
\begin{tabular}{|l|cccc|}
\hline
%\multicolumn{1}{N }{\textbf{}} & \multicolumn{4}{c }{\textbf{Evaluation}} & \multicolumn{4}{c }{} \\  
 & & & {MAE}$\downarrow$ & {MAE}$\downarrow$ \\
{Method} & $\overline{\Delta \mbox{SNR}}\uparrow$ & {AUC}$\uparrow$ & Amplitude & Phase \\
 &  & & {(dB)} & {(degrees)} \\ 
\hline
\hline
{Oracle (true labels)} &    19.32       &   0.954       & 0         & 0\\ 
\hline
{Zeroing}              &    9.99       &    0.897      &  2.40         & 15.23 \\ 
\hline
{FCN (ours)}           &    17.53       &   0.937       &  1.66         & 8.71 \\ 
{FCN + pruning (ours)} & \textbf{17.54} & \textbf{0.940} & \textbf{1.52} & \textbf{7.82}\\ 
\hline
\end{tabular}
\label{tab_results_generalization_multiple_targets}
\vspace{-0.3cm}
\end{table}

%\vspace{-0.2cm}
\section{Conclusion}
\label{sec_Conclusion}

In this paper, we proposed a novel fully convolutional network capable of estimating both magnitude and phase of automotive radar signals affected by multiple sources of interference. We also introduced a large-scale database of radar signals simulated in realistic and complex settings. We compared our FCN model with some state-of-the-art methods in a series of comprehensive experiments, showing that the proposed FCN provides superior results. We also released our novel data set to allow objective comparison in future work. To our knowledge, we are the first to establish a benchmark data set for automotive radar interference mitigation with multiple sources of interference. In future work, we aim to modify our FCN or to explore model distillation approaches in order to perform real-time processing on low-cost embedded devices. At the moment, real-time processing is only possible on expensive GPUs.

\section*{Acknowledgment}
The authors acknowledge all national funding authorities and the ECSEL Joint Undertaking, which funded the
PRYSTINE project under the grant agreement 783190. This work was co-funded through the Operational Program Competitiveness 2014-2020, Axis 1, contract no. 3/1.1.3H/24.04.2019, MySMIS ID: 121784. This article has also benefited from the support of the Romanian Young Academy, which is funded by Stiftung Mercator and the Alexander von Humboldt Foundation for the period 2020-2022.
Additionally, the authors would like to thank Jonas Fuchs and Anand Dubey from University of Erlangen-Nürnberg (FAU) and Lars van Meurs from NXP Semiconductors for providing real radar data under the PRYSTINE cooperation agreement.

\ifCLASSOPTIONcaptionsoff
  \newpage
\fi

\bibliographystyle{IEEEtran}
\bibliography{report}

\begin{thebibliography}{10}
\providecommand{\url}[1]{#1}
\csname url@samestyle\endcsname
\providecommand{\newblock}{\relax}
\providecommand{\bibinfo}[2]{#2}
\providecommand{\BIBentrySTDinterwordspacing}{\spaceskip=0pt\relax}
\providecommand{\BIBentryALTinterwordstretchfactor}{4}
\providecommand{\BIBentryALTinterwordspacing}{\spaceskip=\fontdimen2\font plus
\BIBentryALTinterwordstretchfactor\fontdimen3\font minus
  \fontdimen4\font\relax}
\providecommand{\BIBforeignlanguage}[2]{{%
\expandafter\ifx\csname l@#1\endcsname\relax
\typeout{** WARNING: IEEEtran.bst: No hyphenation pattern has been}%
\typeout{** loaded for the language `#1'. Using the pattern for}%
\typeout{** the default language instead.}%
\else
\language=\csname l@#1\endcsname
\fi
#2}}
\providecommand{\BIBdecl}{\relax}
\BIBdecl

\bibitem{Shibao-VTC-2019}
M.~{Shibao} and A.~{Kajiwara}, ``{Road Debris Detection Using 79GHz Radar},''
  in \emph{Proceedings of VTC2019-Fall}, 2019, pp. 1--4.

\bibitem{kunert2012eu}
M.~Kunert, ``{The EU project MOSARIM: A general overview of project objectives
  and conducted work},'' in \emph{Proceedings of EuRAD}, 2012.

\bibitem{brooker2007mutual}
G.~M. Brooker, ``Mutual interference of millimeter-wave radar systems,''
  \emph{IEEE Transactions on Electromagnetic Compatibility}, vol.~49, no.~1,
  pp. 170--181, 2007.

\bibitem{mosarim2010d15}
M.~Kunert, F.~Bodereau, M.~Goppelt, C.~Fischer, A.~John, T.~Wixforth,
  A.~Ossowska, T.~Schipper, and R.~Pietsch, ``D1.5 - {Study on the
  state-of-the-art interference mitigation technique, MOre Safety for All by
  Radar Interference Mitigation (MOSARIM) project},'' Robert Bosch GmbH, Tech.
  Rep., 2010.

\bibitem{uysal2019sync}
F.~{Uysal}, ``Synchronous and asynchronous radar interference mitigation,''
  \emph{IEEE Access}, vol.~7, pp. 5846--5852, 2019.

\bibitem{kim2018peer}
G.~{Kim}, J.~{Mun}, and J.~{Lee}, ``{A Peer-to-Peer Interference Analysis for
  Automotive Chirp Sequence Radars},'' \emph{IEEE Transactions on Vehicular
  Technology}, vol.~67, no.~9, pp. 8110--8117, Sep. 2018.

\bibitem{xu2018orthogonal_noise}
Z.~{Xu} and Q.~{Shi}, ``{Interference Mitigation for Automotive Radar Using
  Orthogonal Noise Waveforms},'' \emph{IEEE Geoscience and Remote Sensing
  Letters}, vol.~15, no.~1, pp. 137--141, Jan 2018.

\bibitem{bechter2017dbfmimo}
J.~{Bechter}, M.~{Rameez}, and C.~{Waldschmidt}, ``Analytical and experimental
  investigations on mitigation of interference in a dbf mimo radar,''
  \emph{IEEE Transactions on Microwave Theory and Techniques}, vol.~65, no.~5,
  pp. 1727--1734, May 2017.

\bibitem{laghezza2019nxp}
F.~{Laghezza}, F.~{Jansen}, and J.~{Overdevest}, ``{Enhanced Interference
  Detection Method in Automotive FMCW Radar Systems},'' in \emph{Proceedings of
  IRS}, 2019, pp. 1--7.

\bibitem{neemat2018interference}
S.~Neemat, O.~Krasnov, and A.~Yarovoy, ``{An interference mitigation technique
  for FMCW radar using beat-frequencies interpolation in the STFT domain},''
  \emph{IEEE Transactions on Microwave Theory and Techniques}, vol.~67, no.~3,
  pp. 1207--1220, 2018.

\bibitem{mun2018deep}
J.~Mun, H.~Kim, and J.~Lee, ``{A Deep Learning Approach for Automotive Radar
  Interference Mitigation},'' in \emph{Proceedings of VTC-Fall}, 2018.

\bibitem{Ristea-VTC-2020}
N.-C. Ristea, A.~Anghel, and R.~T. Ionescu, ``{Fully Convolutional Neural
  Networks for Automotive Radar Interference Mitigation},'' in
  \emph{Proceedings of VTC-Fall}, 2020.

\bibitem{rock2019complex}
J.~Rock, M.~Toth, E.~Messner, P.~Meissner, and F.~Pernkopf, ``{Complex Signal
  Denoising and Interference Mitigation for Automotive Radar Using
  Convolutional Neural Networks},'' in \emph{Proceedings of FUSION}, 2019.

\bibitem{fan2019interference}
W.~Fan, F.~Zhou, M.~Tao, X.~Bai, P.~Rong, S.~Yang, and T.~Tian, ``{Interference
  Mitigation for Synthetic Aperture Radar Based on Deep Residual Network},''
  \emph{Remote Sensing}, vol.~11, no.~14, p. 1654, 2019.

\bibitem{mun2020automotive}
J.~Mun, S.~Ha, and J.~Lee, ``{Automotive Radar Signal Interference Mitigation
  Using RNN with Self Attention},'' in \emph{Proceedings of ICASSP}, 2020, pp.
  3802--3806.

\bibitem{rock2021resource}
J.~Rock, W.~Roth, M.~Toth, P.~Meissner, and F.~Pernkopf, ``{Resource-Efficient
  Deep Neural Networks for Automotive Radar Interference Mitigation},''
  \emph{IEEE Journal of Selected Topics in Signal Processing}, vol.~15, no.~4,
  pp. 927--940, 2021.

\bibitem{Long-CVPR-2015}
J.~Long, E.~Shelhamer, and T.~Darrell, ``Fully convolutional networks for
  semantic segmentation,'' in \emph{Proceedings of CVPR}, 2015, pp. 3431--3440.

\bibitem{Fuchs2020}
J.~Fuchs, A.~Dubey, M.~Lübke, R.~Weigel, and F.~Lurz, ``{Automotive Radar
  Interference Mitigation using a Convolutional Autoencoder},'' in
  \emph{Proceedings of RADAR}, 04 2020.

\bibitem{han2015learning}
S.~Han, J.~Pool, J.~Tran, and W.~Dally, ``Learning both weights and connections
  for efficient neural network,'' \emph{Proceedings of NIPS}, vol.~28, pp.
  1135--1143, 2015.

\bibitem{liu2018rethinking}
Z.~Liu, M.~Sun, T.~Zhou, G.~Huang, and T.~Darrell, ``Rethinking the value of
  network pruning,'' \emph{Proceedings of ICLR}, 2019.

\bibitem{Srivastava-JMLR-2014}
N.~Srivastava, G.~Hinton, A.~Krizhevsky, I.~Sutskever, and R.~Salakhutdinov,
  ``{Dropout: A Simple Way to Prevent Neural Networks from Overfitting},''
  \emph{Journal of Machine Learning Research}, vol.~15, pp. 1929--1958, 2014.

\bibitem{alland2019spm}
S.~{Alland}, W.~{Stark}, M.~{Ali}, and M.~{Hegde}, ``{Interference in
  Automotive Radar Systems: Characteristics, Mitigation Techniques, and Current
  and Future Research},'' \emph{IEEE Signal Processing Magazine}, vol.~36,
  no.~5, pp. 45--59, 2019.

\bibitem{Ren-NIPS-2015}
S.~Ren, K.~He, R.~Girshick, and J.~Sun, ``{Faster R-CNN: Towards Real-Time
  Object Detection with Region Proposal Networks},'' in \emph{Proceedings of
  NIPS}, 2015, pp. 91--99.

\bibitem{Girshick-ICCV-2015}
R.~Girshick, ``{Fast R-CNN},'' in \emph{Proceedings of ICCV}, 2015, pp.
  1440--1448.

\bibitem{soviany2018optimizing}
P.~Soviany and R.~T. Ionescu, ``Optimizing the trade-off between single-stage
  and two-stage deep object detectors using image difficulty prediction,'' in
  \emph{Proceedings of SYNASC}, 2018, pp. 209--214.

\bibitem{wang2018supervised}
D.~Wang and J.~Chen, ``Supervised speech separation based on deep learning: An
  overview,'' \emph{IEEE/ACM Transactions on Audio, Speech, and Language
  Processing}, vol.~26, no.~10, pp. 1702--1726, 2018.

\bibitem{Chen-MICCAI-2018}
Y.~Chen, F.~Shi, A.~G. Christodoulou, Y.~Xie, Z.~Zhou, and D.~Li, ``{Efficient
  and accurate MRI super-resolution using a generative adversarial network and
  3D multi-level densely connected network},'' in \emph{Proceedings of MICCAI},
  2018, pp. 91--99.

\bibitem{georgescu2020convolutional}
M.-I. Georgescu, R.~T. Ionescu, and N.~Verga, ``{Convolutional Neural Networks
  with Intermediate Loss for 3D Super-Resolution of CT and MRI Scans},''
  \emph{IEEE Access}, vol.~8, no.~1, pp. 49\,112--49\,124, 2020.

\bibitem{Yu-ICIP-2017}
H.~Yu, D.~Liu, H.~Shi, H.~Yu, Z.~Wang, X.~Wang, B.~Cross, M.~Bramler, and T.~S.
  Huang, ``Computed tomography super-resolution using convolutional neural
  networks,'' in \emph{Proceedings of ICIP}, 2017, pp. 3944--3948.

\bibitem{bricman2018coconet}
P.~A. Bricman and R.~T. Ionescu, ``{CocoNet: A deep neural network for mapping
  pixel coordinates to color values},'' in \emph{Proceedings of ICONIP}, 2018,
  pp. 64--76.

\bibitem{ronneberger2015u}
O.~Ronneberger, P.~Fischer, and T.~Brox, ``{U-Net: Convolutional Networks for
  Biomedical Image Segmentation},'' in \emph{Proceedings of MICCAI}.\hskip 1em
  plus 0.5em minus 0.4em\relax Springer, 2015, pp. 234--241.

\bibitem{he2016deep}
K.~He, X.~Zhang, S.~Ren, and J.~Sun, ``Deep residual learning for image
  recognition,'' in \emph{Proceedings of CVPR}, 2016, pp. 770--778.

\bibitem{chung2014empirical}
J.~Chung, C.~Gulcehre, K.~Cho, and Y.~Bengio, ``{Empirical Evaluation of Gated
  Recurrent Neural Networks on Sequence Modeling},'' in \emph{Proceedings of
  DLRL Workshop}, 2014.

\bibitem{allen1977stft}
J.~B. {Allen} and L.~R. {Rabiner}, ``{A unified approach to short-time Fourier
  analysis and synthesis},'' \emph{Proceedings of the IEEE}, vol.~65, no.~11,
  pp. 1558--1564, Nov 1977.

\bibitem{LeCun-IEEE-1998}
Y.~LeCun, L.~Bottou, Y.~Bengio, and P.~Haffner, ``{Gradient-based Learning
  Applied to Document Recognition},'' \emph{Proceedings of the IEEE}, vol.~86,
  no.~11, pp. 2278--2324, 1998.

\bibitem{Krizhevsky-NIPS-2012}
A.~Krizhevsky, I.~Sutskever, and G.~E. Hinton, ``{ImageNet Classification with
  Deep Convolutional Neural Networks},'' in \emph{Proceedings of NIPS}, 2012,
  pp. 1097--1105.

\bibitem{georgescu2019recognizing}
M.-I. Georgescu and R.~T. Ionescu, ``{Recognizing Facial Expressions of
  Occluded Faces Using Convolutional Neural Networks},'' in \emph{Proceedings
  of ICONIP}, 2019, pp. 645--653.

\bibitem{Ristea-INTERSPEECH-2020}
N.-C. Ristea and R.~T. Ionescu, ``{Are you wearing a mask? Improving mask
  detection from speech using augmentation by cycle-consistent GANs},'' in
  \emph{Proceedings of INTERSPEECH}, 2020.

\bibitem{woo2018cbam}
S.~Woo, J.~Park, J.-Y. Lee, and I.~So~Kweon, ``{CBAM: Convolutional Block
  Attention Module},'' in \emph{Proceedings of ECCV}, 2018, pp. 3--19.

\bibitem{wang2017deep}
W.~Wang and J.~Shen, ``Deep visual attention prediction,'' \emph{IEEE
  Transactions on Image Processing}, vol.~27, no.~5, pp. 2368--2378, 2017.

\bibitem{Maas-WDLASL-2013}
A.~L. Maas, A.~Y. Hannun, and A.~Y. Ng, ``Rectifier nonlinearities improve
  neural network acoustic models,'' in \emph{Proceedings of WDLASL}, 2013.

\bibitem{Rumelhart-N-1986}
D.~E. Rumelhart, G.~E. Hinton, and R.~J. Williams, ``Learning representations
  by back-propagating errors,'' \emph{Nature}, vol. 323, no. 6088, pp.
  533--536, 1986.

\bibitem{Simonyan-ICLR-14}
K.~Simonyan and A.~Zisserman, ``{Very Deep Convolutional Networks for
  Large-Scale Image Recognition},'' in \emph{Proceedings of ICLR}, 2014.

\bibitem{Szegedy-CVPR-2015}
C.~Szegedy, W.~Liu, Y.~Jia, P.~Sermanet, S.~Reed, D.~Anguelov, D.~Erhan,
  V.~Vanhoucke, and A.~Rabinovich, ``{Going Deeper With Convolutions},'' in
  \emph{Proceedings of CVPR}, 2015, pp. 1--9.

\bibitem{He-CVPR-2016}
K.~He, X.~Zhang, S.~Ren, and J.~Sun, ``{Deep Residual Learning for Image
  Recognition},'' in \emph{Proceedings of CVPR}, 2016, pp. 770--778.

\bibitem{bentler2006digital}
R.~Bentler and L.-K. Chiou, ``Digital noise reduction: An overview,''
  \emph{Trends in amplification}, vol.~10, no.~2, pp. 67--82, 2006.

\bibitem{boudraa2004emd}
A.~Boudraa, J.~Cexus, and Z.~Saidi, ``{EMD-based signal noise reduction},''
  \emph{International Journal of Signal Processing}, vol.~1, no.~1, pp. 33--37,
  2004.

\bibitem{klasen2007binaural}
T.~J. Klasen, T.~Van~den Bogaert, M.~Moonen, and J.~Wouters, ``Binaural noise
  reduction algorithms for hearing aids that preserve interaural time delay
  cues,'' \emph{IEEE Transactions on Signal Processing}, vol.~55, no.~4, pp.
  1579--1585, 2007.

\bibitem{hu2007comparative}
Y.~Hu and P.~C. Loizou, ``A comparative intelligibility study of
  single-microphone noise reduction algorithms,'' \emph{The Journal of the
  Acoustical Society of America}, vol. 122, no.~3, pp. 1777--1786, 2007.

\bibitem{rudy2019deep}
S.~H. Rudy, J.~N. Kutz, and S.~L. Brunton, ``Deep learning of dynamics and
  signal-noise decomposition with time-stepping constraints,'' \emph{Journal of
  Computational Physics}, vol. 396, pp. 483--506, 2019.

\bibitem{chiang2019noise}
H.-T. Chiang, Y.-Y. Hsieh, S.-W. Fu, K.-H. Hung, Y.~Tsao, and S.-Y. Chien,
  ``{Noise reduction in ECG signals using fully convolutional denoising
  autoencoders},'' \emph{IEEE Access}, vol.~7, pp. 60\,806--60\,813, 2019.

\bibitem{zhang2013denoising}
X.-L. Zhang and J.~Wu, ``Denoising deep neural networks based voice activity
  detection,'' in \emph{Proceedings of ICASSP}, 2013, pp. 853--857.

\bibitem{Kingma-ICLR-2015}
D.~P. Kingma and J.~Ba, ``Adam: A method for stochastic optimization,'' in
  \emph{Proceedings of ICLR}, 2015.

\bibitem{xiao2019autoprune}
X.~Xiao and Z.~Wang, ``{AutoPrune: Automatic Network Pruning by Regularizing
  Auxiliary Parameters},'' in \emph{Proceedings of NeurIPS}, 2019, pp.
  13\,699--13\,709.

\bibitem{Russakovsky-IJCV-2015}
O.~Russakovsky, J.~Deng, H.~Su, J.~Krause, S.~Satheesh, S.~Ma, Z.~Huang,
  A.~Karpathy, A.~Khosla, M.~Bernstein, A.~C. Berg, and L.~Fei-Fei, ``{ImageNet
  Large Scale Visual Recognition Challenge},'' \emph{International Journal of
  Computer Vision}, vol. 115, no.~3, pp. 211--252, 2015.

\end{thebibliography}

\begin{IEEEbiography}[{\includegraphics[width=1in,height=1.25in,clip,keepaspectratio]{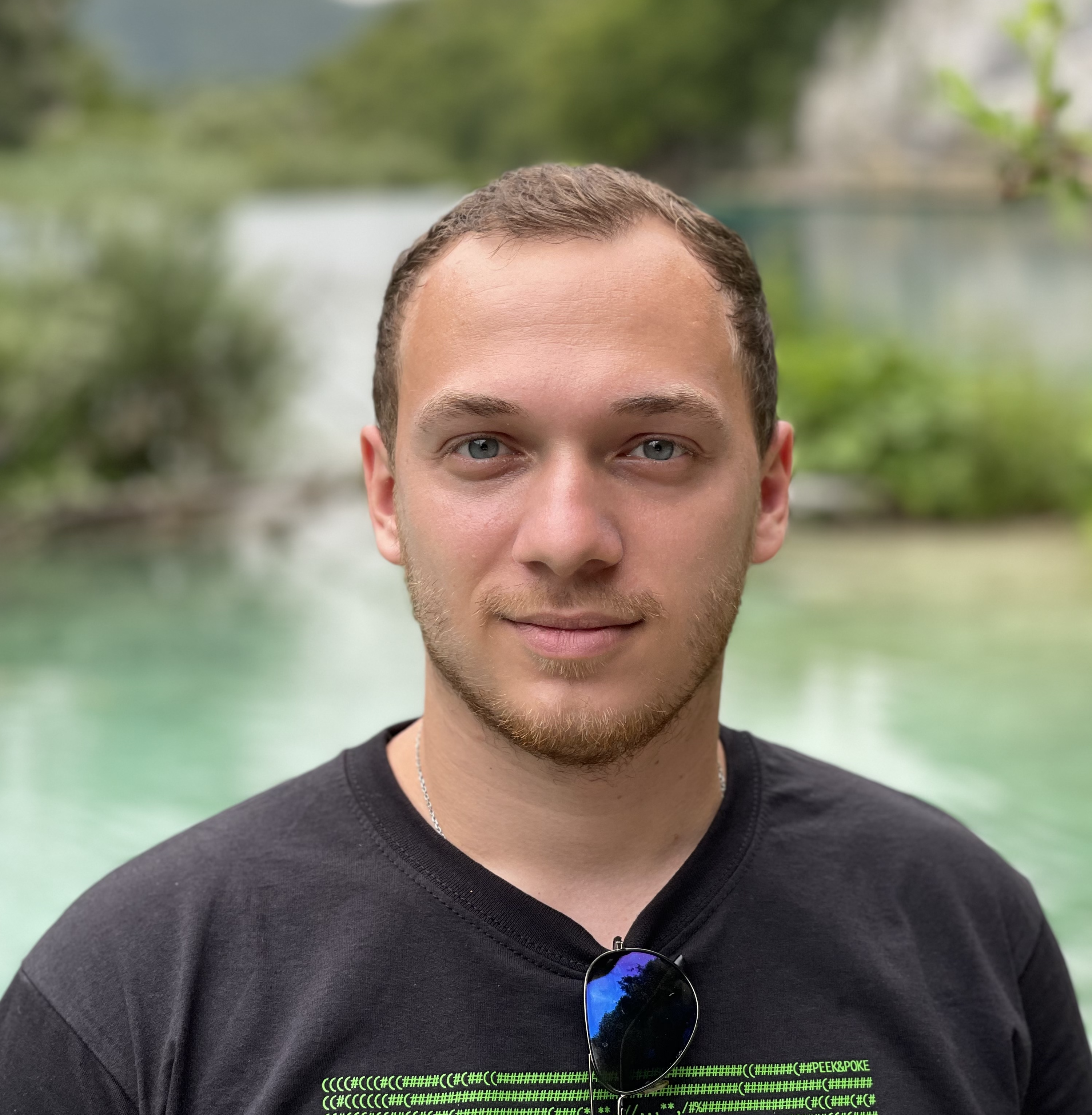}}]{Nicolae-C\u{a}t\u{a}lin Ristea} graduated as valedictorian from the Faculty of Electronics, Telecommunications and Information Technology, University Politehnica of Bucharest, in 2019. He received the M.S. degree in the image processing field from the same university. Nicolae is co-author of multiple papers accepted at top-tier conferences and journals, such as VTC, INTERSPEECH and Geoscience and Remote Sensing Letters. His research interests include artificial intelligence, computer vision, machine learning, signal processing and deep learning.
\end{IEEEbiography}

\begin{IEEEbiography}[{\includegraphics[width=1in,height=1.25in,clip,keepaspectratio]{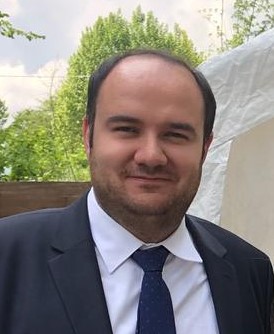}}]{Andrei Anghel} (S’11–M’16–SM’20) received the Engineering degree (as valedictorian) and the M.S. degree (with the highest grade) in electronic engineering and telecommunications from the University Politehnica of Bucharest, Bucharest, Romania, in 2010 and 2012, respectively. He received the joint Ph.D. degree in signal, image, speech, and telecoms, from the University of Grenoble Alpes, Grenoble, France, and in electronic engineering and telecommunications from the University Politehnica of Bucharest, Bucharest, Romania, in 2015 (awarded with the summa cum laude distinction). He received the habilitation degree in electronic engineering, telecommunications, and information technologies from the University Politehnica of Bucharest, Bucharest, Romania, in 2020.

Between 2012 and 2015, he worked as a Doctoral Researcher with Grenoble Image Speech Signal Automatics Laboratory (GIPSA-lab), Grenoble, France. In 2012, he joined the University Politehnica of Bucharest as a Teaching Assistant, where he is now an Associate Professor with the Telecommunications Department (Faculty of Electronics, Telecommunications and Information Technology) and researcher at the Research Centre for Spatial Information-CEOSpaceTech. His current research interests include remote sensing, radar, microwaves and signal processing. He is the author of more than 50 scientific publications, 2 textbooks, and a book about SAR signal processing for infrastructure monitoring.

Dr. Anghel regularly acts as a Reviewer for several IEEE and IET journals. He was the recipient of two gold medals (in 2005 and 2006) at the International Physics Olympiads.
\end{IEEEbiography}

\begin{IEEEbiography}[{\includegraphics[width=1in,height=1.25in,clip,keepaspectratio]{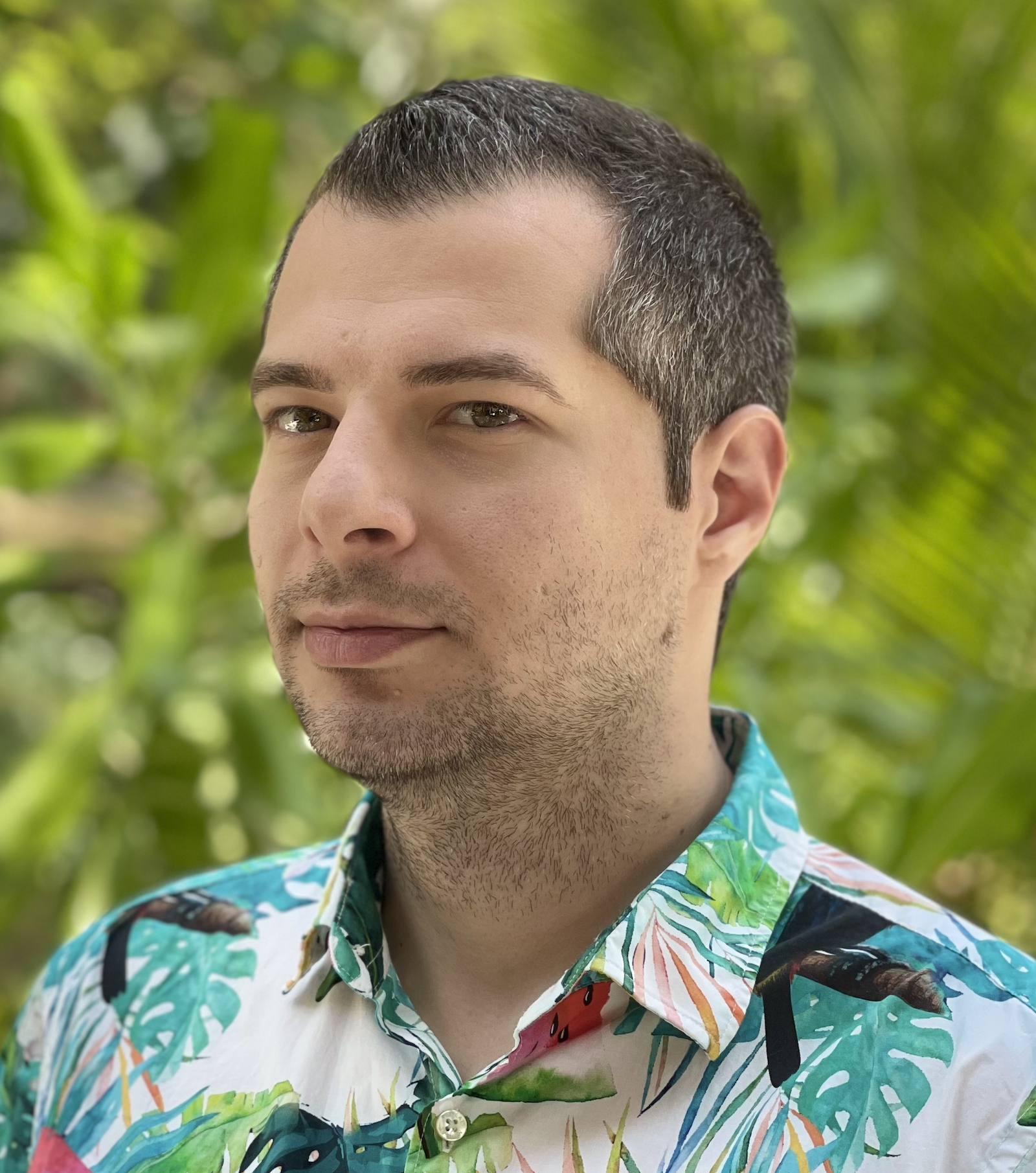}}]{Radu Tudor Ionescu} is Professor at the University of Bucharest, Romania. He completed his PhD at the University of Bucharest in 2013. He received the 2014 Award for Outstanding Doctoral Research in the field of Computer Science from the Romanian Ad Astra Association. His research interests include machine learning, text mining, computer vision, image processing and medical imaging. He published over 90 articles at international peer-reviewed conferences and journals, and a research monograph with Springer. Radu is editor of the journal Mathematics and served as an area chair at ICPR 2020. He received the "Caianiello Best Young Paper Award" at ICIAP 2013 for the paper entitled "Kernels for Visual Words Histograms". Radu also received the "Young Researchers in Science and Engineering" Prize from prof. Rada Mihalcea and the "Danubius Young Scientist Award 2018 for Romania" from the Austrian Federal Ministry of Education, Science and Research and by the Institute for the Danube Region and Central Europe. Together with other co-authors, he obtained good rankings at several international competitions: 4th place in the Facial Expression Recognition Challenge of WREPL 2013, 3rd place in the NLI Shared Task of BEA-8 2013, 2nd place in the ADI Shared Task of VarDial 2016, 1st place in the ADI Shared Task of VarDial 2017, 1st place in the NLI Shared Task of BEA-12 2017, 1st place in the ADI Shared Task of VarDial 2018.
\end{IEEEbiography}

\EOD

\end{document}